\documentclass
[11pt,tightenlines,eqsecnum,floats,aps,amsmath,amssymb,nofootinbib,prd,shownopacs,notitlepage]{revtex4}
\usepackage{graphicx, wrapfig}
\usepackage{amssymb}
\usepackage{color}
\usepackage{mathrsfs}
\usepackage{amsmath}
\usepackage{subfigure}
\usepackage{afterpage}

\usepackage{epstopdf}

\DeclareGraphicsExtensions{.pdf,.png,.jpg,.jpeg,.eps}

\def\f{\frac}

\def\rhopl{\rho_{{}_{\rm Pl}}}
\def\tpl{s_{{}_{\rm Pl}}}
\def\lpl{\ell_{{}_{\rm Pl}}}

\def\ig{\includegraphics}

\def\lp{\ell_{\rm Pl}}
\def\t{\tilde}
\def\h{\hat}

\def\R{\mathcal R }
\def\T{\mathcal T }

\def\Hp{\mathcal{H}_{\rm phy}}

\def\ab{\bar{a}}

\usepackage{enumerate}

\def\dd{{\rm d}}

\def\be{\nopagebreak[3]\begin{equation}}
\def\ee{\end{equation}}
\def\ba{\nopagebreak[3]\begin{eqnarray}}
\def\ea{\end{eqnarray}}

\def\dd{\rm d}
\def\Q{\mathcal{Q}}
\def\U{\mathcal{U}}
\def\B{\mathcal{B}}

\def\vk{\vec{k}}
\def\spl{s_{\rm Pl}}

\newcommand{\bfig}{\nopagebreak[3]\begin{figure}}
\newcommand{\efig}{\end{figure}}

\newcommand{\bmult}{\nopagebreak[3]\begin{multline}}
\newcommand{\emult}{\end{multline}}

\newcommand{\fref}[1]{Fig.\,\ref{#1}}
\newcommand{\tref}[1]{Table\,\ref{#1}}
\newcommand{\eref}[1]{eq.\,(\ref{#1})}

\newcommand{\sref}[1]{Sec.\,\ref{#1}}

\def\rcurv{r_{\rm curv}}
\def\rhosup{\rho_{\rm sup}}

\def\phid{\dot\phi}
\def\phib{\phi_{{}_{\rm B}}}
\def\phidb{\phid_{{}_{\rm B}}}
\def\ab{a_{{}_{\rm B}}}
\def\Hb{H_{{}_{\rm B}}}

\def\half{\f{1}{2}}

\def\RB{{\mathfrak R}_{\rm B}}

\def\Rcmb\star{{\mathfrak R}_{\rm CMB}}

\def\as{A_s}
\def\ns{n_s}
\def\camb{\texttt{CAMB}~}
\def\cosmomc{\texttt{COSMOMC}~}
\def\efolds{$e$-folds~}

\def\mpc{{\rm Mpc^{-1}}}
\def\mpl{m_{{}_{\rm Pl}}}
\def\celltt{C^{\rm TT}_\ell}
\def\cellee{C^{\rm EE}_\ell}
\def\cellte{C^{\rm TE}_\ell}

\begin{document}

\title{Quantum Gravity in the Sky: \\ Interplay between fundamental theory and observations}
\author{Abhay Ashtekar${}^{1,2}\,$}
\email{ashtekar@gravity.psu.edu} 
\author{Brajesh Gupt${}^{1}\,$}
\email{bgupt@gravity.psu.edu}
\affiliation{
${}^{1}$ Institute for Gravitation and the Cosmos \& Physics Department, The Pennsylvania State University, University Park, PA 16802 U.S.A.\\
\\
${}^{2}$ Perimeter Institute for Theoretical Physics, 31 Caroline St N, Waterloo, Ontario, Canada N2L 2Y5
}

\pacs{}

\begin{abstract}

Observational missions have provided us with a reliable model of the evolution of the universe starting from the last scattering surface all the way to future infinity. Furthermore given a specific model of inflation, using quantum field theory on curved space-times this history can be pushed \emph{back in time} to the epoch when space-time curvature was some $10^{62}$ times that at the horizon of a solar mass black hole! However, to extend the history further back to the Planck regime requires input from quantum gravity. An important aspect of this input is the choice of the background quantum geometry and of the Heisenberg state of cosmological perturbations thereon, motivated by Planck scale physics. This paper introduces first steps in that direction. Specifically we propose two principles that link quantum geometry and Heisenberg uncertainties in the Planck epoch with late time physics and explore in detail the observational consequences of the initial conditions they select. We find that the predicted temperature-temperature (T-T) correlations for scalar modes are indistinguishable from standard inflation at small angular scales even though the initial conditions are now set in the deep Planck regime. However, \emph{there is a specific power suppression at large angular scales}. As a result, the predicted spectrum provides a better fit to the PLANCK mission data than standard inflation, where the initial conditions are set in the general relativity regime. Thus, our proposal brings out a deep interplay between the ultraviolet and the infrared. Finally, the proposal also leads to specific predictions for power suppression at large angular scales also for the (T-E and E-E) correlations involving electric polarization. The PLANCK team is expected to release this data in the coming year.  

\end{abstract}

\maketitle

\tableofcontents

\section{Introduction}
\label{s1}

Thanks to a powerful confluence of observations and theory, our understanding of the very early universe has deepened tremendously over the past two decades. In particular, we have learned that the observed temperature fluctuations in the cosmic microwave background (CMB)  provide seeds for formation of the large scale structure (LSS) of the universe. A natural avenue to account for the CMB fluctuations themselves is provided by the theory of  \emph{quantum} cosmological perturbations on a classical Friedman, Lema\^{i}tre, Robertson, Walker (FLRW) background space-time. In this setting, the inflationary scenario has emerged as the leading phenomenological paradigm. In particular, detailed calculations carried out using a single inflaton in the Starobinsky potential provides an excellent fit to the PLANCK mission observations \cite{planck2013}. This quantitative fit is impressive because it traces the origin of the observed LSS back to the onset of inflation, when the scale factor was some $10^{-52}$ times that in the CMB epoch and the matter density was some $10^{68}$ times nuclear density!  

But inflation is not a fundamental theory like, say, general relativity. Rather, as emphasized by Peebles \cite{peebles},  it serves as a ``a framework on which to hang a theory.''  This view does not diminish the importance of the underlying ideas; indeed, as the history of the first two decades of quantum mechanics vividly illustrates, frameworks that provide scaffoldings can play a paramount role in the development of the fundamental theory. Rather, this viewpoint provides a better vantage point to examine inflation. In particular, since many of the criticisms of inflation are rooted in the implicit assumption that it represents a fundamental theory, they now lose their force.  We will adopt Peeble's viewpoint and regard inflation as a powerful working model that, together with the CMB observations, can provide guidance for construction of the desired fundamental theory. The goal of this and the accompanying paper \cite{ag2} is to provide the first steps in the construction of the desired theory by building on some recent work in loop quantum cosmology (LQC) \cite{aan1,aan2,aan3,agullomorris,agulloassym}.%
\footnote{Within LQC, Other observational aspects have also been examined using complementary approaches in \cite{bcgmrev,lcbg,madrid,bg2}.}

A key question in any model of the early universe is the choice of initial conditions. In the current cosmological paradigms this corresponds to the choice of the background FLRW space-time and the (Heisenberg) state of quantum fields representing the scalar and tensor modes of cosmological perturbations. In the standard inflationary scenario, the choice of the FLRW solution is left largely open and the Heisenberg state of perturbations is chosen using two considerations. First, during the slow-roll phase of inflation, space-time geometry is well-approximated by de Sitter space-time (since the change in the Hubble parameter is slow) and quantum fields in de Sitter space-time admit a unique, maximally symmetric, regular Heisenberg state, called the Bunch-Davies (BD) vacuum. Second, since the co-moving wave numbers of observable modes range roughly from $10^{-1}\,k_{\star}$ to some $300\, k_{\star}$  --where $k_{\star} = 2\times 10^{-3} {\rm Mpc}^{-1}$ is the pivot mode--  it suffices to focus just on these modes.  Therefore, a few \efolds before $k_{\star}$ exits the Hubble horizon, the Heisenberg state is assumed to be in the BD vacuum for all these modes. The procedure has a degree of ambiguity because, even though the Hubble parameter decreases only slowly, it \emph{is} time dependent, whence the Heisenberg state so chosen varies, depending on the precise instant at which the BD condition is imposed. Thus we have a small class of Heisenberg states --rather than a unique state-- to work with. But one can perform explicit calculations to show that this ambiguity is too small to have observable effects, e.g., in calculating the power spectrum.

However, conceptually, the procedure seems awkward because it asks us to choose the Heisenberg state by making appeal to the geometry of the universe at an ad-hoc intermediate time. Furthermore, in the \emph{backward} Heisenberg evolution, the physical wavelength of all  observable modes exceeds the curvature radius at sufficiently early times (see the left  panel of \fref{fig:rcurv}). When this occurs, interaction with the background curvature  excites them. Thus, standard inflation asks that the Heisenberg state of perturbations be so chosen that it has `just the right amount of excitations' \emph{at very early times,} waiting to be removed by the subsequent evolution (forward in time), so that all the observable modes end-up in the BD ground state at the onset of slow-roll. From a conceptual perspective, this is an unnatural past hypothesis. 

A more important limitation arises from the assumption that the FLRW geometry satisfies Einstein's equations all the way back to the big bang. There is every expectation that quantum gravity effects would intervene in the Planck regime, significantly modify Einstein's equations, thereby resolving the big bang singularity. Therefore one needs to first understand the \emph{quantum} FLRW geometry at very early times and evolve cosmological perturbations on this quantum geometry from some well-defined initial conditions in the Planck regime. A more satisfactory past hypothesis could then be formulated using fundamental properties of the quantum geometry at Planck scale and basic tenets of quantum mechanics such as the uncertainty principle. Such an approach would also squarely address the standard `trans-Planckian issues' associated with perturbations.%
\footnote{Since the background FLRW geometry is not specified in standard inflation, there can be a very large number of \efolds prior to the onset of slow-roll. Therefore, the physical wavelength of observable modes can be an arbitrarily small fraction of the Planck length at early times, making the use of QFT on a classical FLRW space-time untenable (see, e.g., \cite{brandentrans}). LQC faces these trans-Planckian issues directly by replacing the classical FLRW geometry by an appropriate quantum state $\Psi_{o}(a,\phi)$. We will also see that the physical principles introduced in this paper severely limit the pre-inflationary \efolds. As a result, all observable modes have frequencies compatible with the underlying quantum geometry in the Planck regime.}
If the initial conditions are so chosen in the quantum gravity regime, quantum fields may well get excited during evolution from this epoch and therefore \emph{not} be in the BD vacuum at the onset of inflation. A priori, then, it is not clear whether such a more fundamental proposal to narrow down the quantum state of the universe would be compatible with observations. In this precise sense, there is a possibility of testing quantum gravity proposals using CMB. Conversely, observations of the very early universe can guide us in our search of basic theoretical principles that dictate Planck scale physics.  Sections \ref{s2}-\ref{s4} (and the accompanying paper \cite{ag2}) provide a concrete proposal to illustrate this interplay between fundamental theory and observations. 

Let us summarize the general setting. We will use LQC where the big bang singularity has been shown to be resolved in a wide variety of cosmological models, and replaced by a big bounce 
(for reviews, see, e.g., \cite{asrev,ps3,agullocorichi}). The basic mechanism is provided by \emph{quantum Riemannian geometry} that underlies Loop Quantum Gravity (LQG) \cite{ashtekarloopsstatus,rovelliqg,thiemannbook}. Corrections to Einstein's equations reveal a new repulsive force originating in quantum geometry. The force is negligible if space-time curvature is less than $\sim 10^{-3}$ times Planck curvature, making classical cosmology an excellent approximation to LQC during and after inflation. However, in the Planck regime the force grows \emph{extremely} rapidly, overwhelms the classical attraction, and causes the universe to bounce. Thus, in the Planck regime, there is a well-defined quantum theory in which \emph{all physical observables remain bounded.} In particular, there is an absolute upper bound on the eigenvalues of operators representing the matter density and the Hubble rate. 

To consistently handle the propagation of scalar and tensor modes near the bounce, one needs an extension of quantum field theory (QFT) on classical FLRW space-times to that on \emph{quantum} FLRW geometries, represented by wave functions $\Psi_{o} (a,\phi)$ in the LQC physical Hilbert space (where $a$ denotes the scale factor and $\phi$ the inflaton field). At first the task seems extremely difficult. However, there is an unexpected simplification \cite{akl,aan2,aan3}: So long as the back reaction of the scalar and tensor perturbations $\h{\Q}, \h{\T}^{(I)}$ (with $I=1,2$) can be neglected on the background quantum geometry $\Psi_{o}$, dynamics of quantum fields $\h{\Q}, \h{\T}^{(I)}$ on $\Psi_{o}$ is \emph{identical} to that of quantum fields  $\h{\Q}, \h{\T}^{(I)}$ propagating on a smooth FLRW metric $\t{g}_{ab}$ which is constructed in a precise manner from $\Psi_{o}$.%
\footnote{In the pre-inflationary epoch, the curvature perturbation $\h\R$ for scalar modes become ill-defined at the turn-around point where $\dot\phi =0$. Therefore, in the LQC literature, one uses the Mukhanov-Sasaki gauge invariant scalar perturbation $\h{Q}$ in the pre-inflationary dynamics and converts the result to $\h\R$  at the end of inflation.}
Since this result sets up an \emph{exact} correspondence, it enables one to `lift' various mathematical techniques of regularization and renormalization from QFT on FLRW space-times to that on quantum FLRW space-times. The smooth tensor field $\t{g}_{ab}$ is referred to as the \emph{dressed, effective metric}. It captures information not only about where $\Psi_{o}$ is peaked but also about those fluctuations in $\Psi_{o}$ that the propagation of fields $\h{\Q}, \h{\T}^{(I)}$ is sensitive to. This is why the correspondence is exact. The idea is to first choose  a Heisenberg state $\psi$ of perturbations $\h{\Q}, \h{\T}^{(I)}$ and evolve the fields assuming that the back reaction can be neglected, and then check at the end that the initial assumption is satisfied. The final result is then a self-consistent solution with Heisenberg operators $\h{\Q}, \h{\T}^{(I)}$ in a state $\psi$, propagating on a FLRW quantum geometry $\Psi_{o}$ \cite{aan1,aan3}. Away from the Planck regime, the description rapidly reduces to the standard one, involving quantum fields propagating on a FLRW solution $g_{ab}$ to Einstein's equations. 

The final physical predictions that can be compared with observations --such as the Temperature-Temperature (T-T) power spectrum \cite{aan1,aan2,aan3}, non-gaussianities \cite{agullomorris} and hemispherical asymmetry \cite{agulloassym}-- depend, of course, not only on the general LQC framework, but also on the choices of the dressed effective metric $\t{g}_{ab}$ of the background FLRW geometry and the quantum state $\psi$ of scalar and tensor perturbations. In the remainder of this paper, we will propose physical principles that will tremendously narrow down  choices of $\t{g}_{ab}$ and $\psi$, and test it against observations. We will find that, at the onset of slow-roll,  the state $\psi$ so selected is indistinguishable from the BD vacuum for modes $\ell \gtrsim30$ but differs from it for the long wavelength modes with $\ell \lesssim 30$, \emph{leading to power suppression only at large angular scales}, thereby providing a better agreement with the data than standard inflation. This suggests that the principles may be capturing some essential features of a deeper theory and may provide guidance for further work, e.g., aimed at a `derivation' of our postulates from more fundamental considerations.

The paper is organized as follows. In section \ref{s2} we use the rich information provided by the PLANCK mission to construct a summary of the dynamics of the universe starting from the onset of inflation to infinite future using Einstein's equations. The portion of the history between the time $t_{\star}$ when the pivot mode $k_{\star}$ exits the Hubble horizon to the end of inflation depends on the choice of inflationary potential. The Starobinsky potential is preferred phenomenologically by the PLANCK data and is also attractive from conceptual considerations since it could arise from the gravitational sector of a more fundamental theory without having to introduce to new scalar field or a potential by hand. Therefore in this paper we will focus on Starobinsky potential. However, to test the robustness of our results, throughout the paper we also discuss the quadratic potential side by side.%
\footnote{The predicted scalar power spectrum for the quadratic potential is compatible with observations but the predicted value of the ratio $r$ of tensor to scalar power is disfavored. In this paper we focus only on scalar modes.}
In section \ref{s3} we introduce new principles to greatly narrow down the background FLRW quantum geometry $\t{g}_{ab}$  and the Heisenberg state $\psi$ of perturbations. The selection of $\t{g}_{ab}$ is dictated by fundamental features of the quantum Riemannian geometry, particularly by the eigenvalues of the area operator \cite{alarea}. The selection of $\psi$ is made by a quantum generalization of Penrose's \cite{rp-weyl} Weyl curvature hypothesis developed in \cite{ag2}. In section \ref{s4}, we use these choices of $\t{g}_{ab}$  and $\psi$ together with the well-developed procedures from LQC \cite{akl,aan1,aan2,aan3}, summarized above, to carry out detailed calculations for the temperature-temperature (T-T), the temperature and electric polarization (T-E) and electric polarization - electric polarization (E-E) correlation functions. As noted above, predictions from LQC supplemented with the two principles, agree with those of standard inflation for modes at small angular scales, $\ell \gtrsim 30$ although the Heisenberg state $\psi$ is now chosen in the Planck regime rather than at the onset of the slow-roll phase. However the situation is different for the long wavelength modes $\ell \lesssim 30$ because of an unforeseen interplay \cite{aan3} between the ultraviolet (UV) properties of the quantum FLRW geometry, and the infrared (IR) properties of cosmological perturbations $\h{\Q},\h{\T}^{(I)}$ (discussed in section \ref{s4.1.2}). \emph{In our state $\psi$, these modes are \emph{not} in the BD vacuum at the onset of slow-roll.}%
\footnote{There is sizable literature in cosmology on interesting consequences of using non-BD vacua (see especially \cite{nishant1,nishant2}). The difference is that, in our case, the departure from the BD vacuum is not postulated but results from dynamics in the deep Planck regime. Previous investigations were entirely in the general relativity regime.}
Consequently, now the predictions differ from those of standard inflation. We show that our $\psi$ leads to a suppression of power for $\ell \lesssim 30$ for the T-T correlation functions and that this suppression provides a better $\chi^{2}$-fit to the data compared to the prediction of standard inflation. We have predictions of power suppression also for the T-E and E-E correlation functions. If future observations were to clearly falsify them, we would have to abandon at least one of the two principles. In section \ref{s5} we summarize our results and discuss several issues they raise. In particular, we emphasize two points: (i) while the power suppression at the largest angular scales can arise from other mechanisms (see, e.g., \cite{ContaldiFastRoll,ClineFastRoll,JainFastRoll, PedroFastRoll, LelloFastRoll,cai3,cai2,cai1}), these ideas are driven by phenomenological, rather than fundamental considerations. Our
goal is to investigate if CMB observations can provide us new clues about fundamental physics in the Planck epoch; and, (ii) nonetheless, our specific proposal for initial conditions should be regarded only a first stab at the problem. Observations may rule it out and, even if it survives future observational tests, one would have to replace it with a deeper principle that arises from more systematic quantum gravity considerations.

Our conventions are as follows. We use signature -,+,+,+ and set $c=1$.  But keep $G$ and $\hbar$ explicitly in various equations to facilitate the distinction between classical and quantum effects. As is usual, we will set $\kappa = 8\pi G$ and, following the convention that is common in cosmology literature we set the scale factor today to one; $a_{0} =1$. For the area gap $\Delta$ we use the value $\Delta = 5.17$ that comes from the black hole entropy calculations \cite{Domagala_gamma,Meissner_gamma}. Finally, we will use Planck units that are common in the quantum gravity literature rather than the reduced Planck units often used in cosmology. (Thus, our Planck mass $\mpl= \sqrt{\hbar/G}$ is related to the reduced Planck mass $M_{\rm Pl}$ via $\mpl = \sqrt{8\pi} M_{\rm Pl}$.) Unless otherwise stated, \emph{numerical values} of all quantities are given in dimensionless Planck units $\lpl = \mpl = t_{\rm Pl} =1$.

\section{Dynamics of the universe from observations and theory}\label{s2}

This section is divided into three parts. Einstein's equation together with the data provided by the PLANCK mission suffice to determine the space-time geometry from the surface of last scattering --which we will refer to as the CMB surface-- all the way to infinite future. In the first part of this section we discuss salient features of this geometry. Understanding these features is important because, not only the CMB observations and the ongoing surveys of the large scale structure (LSS), but also such observations that could \emph{ever} be performed in the future by an `eternal observer' over her \emph{infinite} lifetime refer only to this portion of space-time. We will therefore refer to this portion (depicted in the left panel of Fig. 1) as \emph{the  observationally relevant} patch of the universe.

Can one extend this portion of space-time to the past of CMB surface where new physics is likely to play an important role? For this task, one has to develop theoretical scenarios for the early universe and work out the consequences that could be observed in CMB and surveys probing the LSS. In the second part of this section we discuss this extension in the framework of inflation, using the Starobinsky and quadratic potentials. This analysis enables one to extend the space-time structure all the way to the very early time $t_{\star}$ at which the pivot mode $k_{\star} = 0.002 \mpc$ exits the Hubble horizon. Although this is some 117 \efolds before $t_{\rm CMB}$, space-time curvature is still some 11-12 orders of magnitude below the Planck scale. Therefore, one can still trust general relativity (GR) during this entire phase, whence  this extension of space-time (shown in the right panel of Fig. 1) is based on GR.

What about a further extension to the past of $t=t_{\star}$? Now one quickly reaches the Planck regime where GR can no longer serve as a viable approximation. Therefore, predictions about the salient features of the universe to the past of the onset of the slow-roll inflation will generally depend on the specific quantum gravity theory that is contemplated. In the third part of this section we will summarize the relevant ideas from LQG that are needed for this extension over the 11-12 orders of magnitude in curvature from $t=t_{\star}$ to the Planck regime.\\

\subsection{Future of the universe according to PLANCK and GR}
\label{s2.1}

\bfig
 \ig[width=0.47\textwidth]{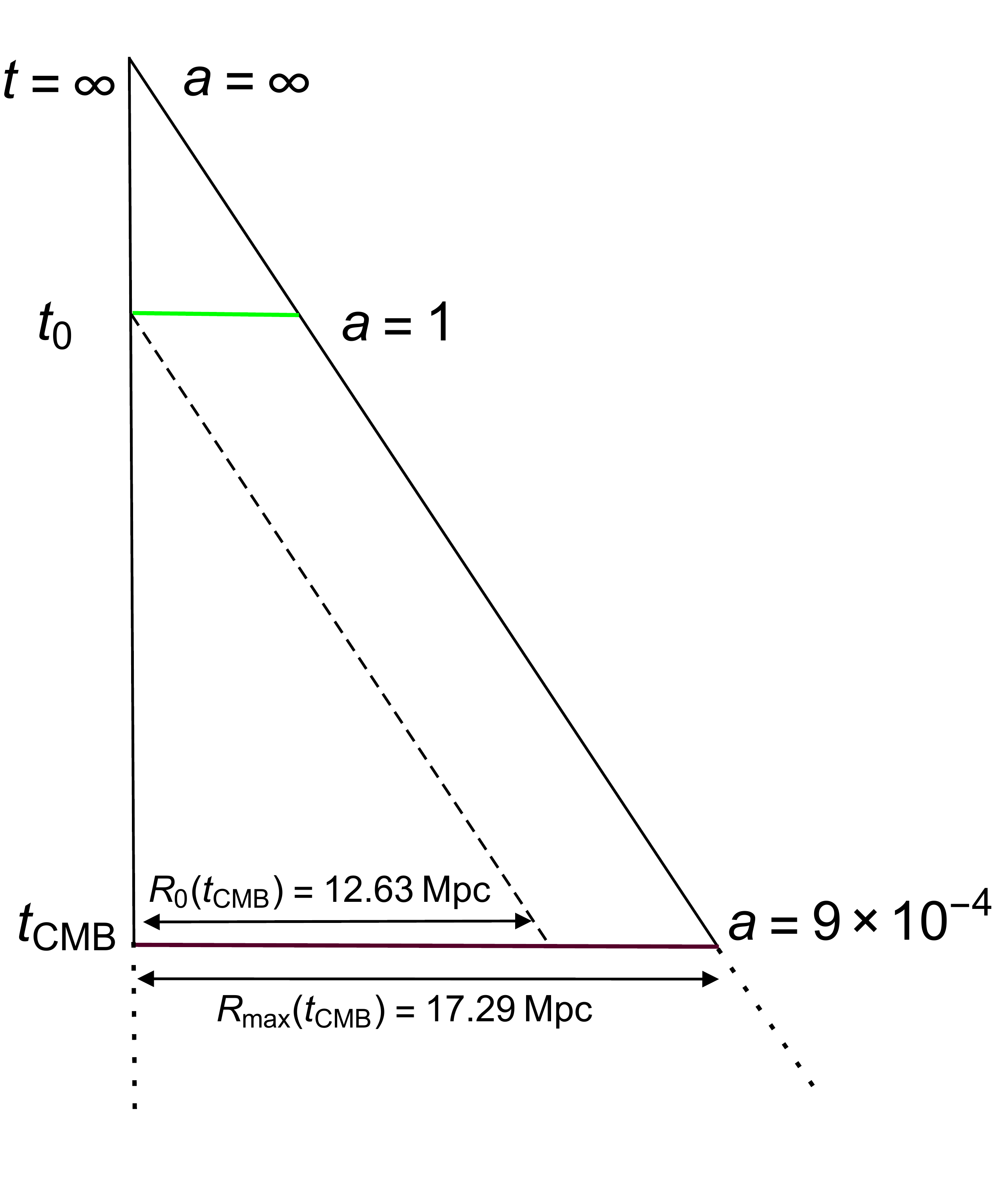}
 \hskip0.5cm
 \ig[width=0.47\textwidth]{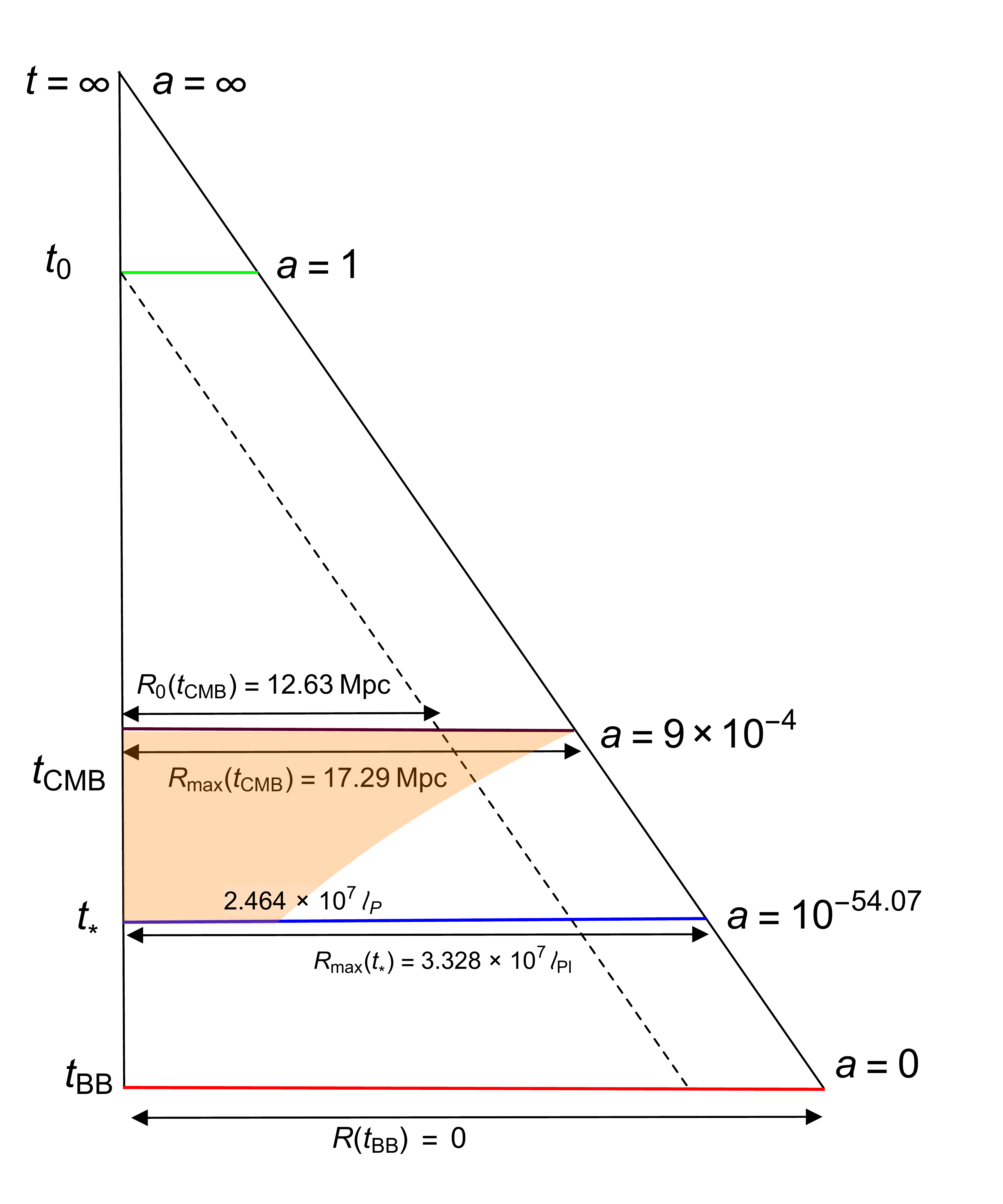}\caption{Observationally relevant patch of the FLRW universe.\\ {\it Left panel:} Conformal diagram of the FLRW geometry of our universe. The portion to the future of the CMB epoch is determined by the PLANCK data \cite{planck15xx} and Einstein's equations. The Cauchy data on the portion of the CMB surface bounded by a sphere of radius $R_{\rm max} = 17.29$ Mpc determine the evolution in the full interior of the Cauchy horizon of the eternal observer.
\\ 
{\it Right panel:} The history of the universe, extended to the past of the CMB surface using Starobinsky inflation, discussed in section \ref{s2.2}. $t_{\star}$ is the proper time at which the pivot mode $k_{\star}$ exits the Hubble horizon and $t_{\rm BB}$ refers to the big bang singularity. Because $\sim 117$ \efolds separate $t_{\star}$ from $t_{\rm CMB}$, the ball of radius of $\sim 10^{-26}$ cm at time $t_{\star}$ expands out to fill the ball of radius $R_{\rm max} \sim 17.29~{\rm Mpc}$ at $t_{\rm CMB}$ that covers the entire universe accessible to an eternal observer. As in standard inflation, we have extended space-time to the past of $t= t_{\star}$ using Einstein's equations with inflaton as matter source.}
\label{figs:history}
\efig

The PLANCK mission has provided us with values of parameters that are needed in Einstein's equations to determine a specific FLRW solution to the future of the CMB surface. More precisely, throughout this paper, we will use the values reported in the PLANCK XX paper \cite{planck15xx} that puts constraints on inflation%
\footnote{These numbers come from the so-called TT, TE, EE +lowP data, which corresponds ``to the combination of the likelihood at $\ell \geq 30$ using TT, TE, and EE spectra and the low-$\ell$ multipole likelihood''. See Table 3 and footnote 4 in \cite{planck15xx}.}   
(the error bars signify 68\% confidence level):
\be \label{parameters}
H_{0} = 67.27 \pm\ 0.66\, {\rm km\,\,s^{-1}\, Mpc^{-1}}; 
\quad\quad {\rm and} \quad \Omega_{m} = 0.3156\pm 0.0091\, .  \ee 
For simplicity, we will work just with the best fit values. \emph{Therefore, one has to keep in mind that our final results can be trusted only within  appropriate error bars.} Since the radiation contribution $\Omega_{r}$ is negligible for the significant figures quoted here, we will set $\Omega_{\Lambda} = 1- \Omega_{m} = 0.6844$. Then, setting the scale factor today to one ($a_{0}=1$), and taking the CMB surface to be at $z=1100$ (see, e.g., \cite{planck15xiii}, section 6.7.3), using Einstein's equation we reconstructed the physical FLRW geometry of the universe to the future of the CMB surface. To understand the causal structure, it is simplest to cast the space-time metric in a manifestly conformally flat form:
\be \label{metric}
{\dd} s^{2} = a^{2}(\eta) \big(- {\dd}\eta^{2} + {\dd} {\vec{x}}^{2}\big), \ee
where the conformal time $\eta$ is given, up to a constant, by
\be \label{eta}
\eta = \int \f{{\dd}{t}}{a(t)}\,\, =\int \f{{\dd} a}{Ha^{2}}  
= \int\, \f{{\dd} a}{ a^2H_{0} \big(\Omega_{m} ({a_{0}^{3}}/{a^{3}})\, +\, \Omega_{\Lambda} \big)^{\f{1}{2}} } \quad. \ee
For definiteness, we will set $\eta= 0$ today (i.e. at $t=t_{0}$).
As we now show, the most striking feature of the FLRW space-time that results from the PLANCK values of $H_{0}$ and $\Omega_{m}$ and $\Omega_{\Lambda}$ is that its global structure is profoundly altered by the presence of a positive cosmological constant $\Lambda$. 

\vskip0.4cm
{\footnotesize{
\begin{table} 
\caption{History of the universe. The first three rows refer only to the Planck data \cite{planck15xx} and GR. The last row refers to Starobinsky inflation discussed in the next subsection. $t_{\star}$ is the proper time at which the pivot mode $k_{\star}$ exits the Hubble horizon. The second column provides the values of conformal time $\eta$ at various epochs (with $\eta=0$ today); the third gives the number of \efolds measured from today; the fourth provides the physical radius of the universe at various as seen today; and the last column gives the maximum radius of the universe at various epochs, as seen by the eternal observer.} \label{history} \vskip0.3cm                                                                   
 \begin{tabular}{ccccc}
\hline
Epoch &  $\eta $ (Mpc)  &  $n_e$ & $R_0$  & $R_{\rm max}$ \\
\hline
\hline
$t_\infty$  &  $5,129$ & $\infty$ &  NA & $5,391$ Mpc \\
\\
\color{black} {$t_0$}  & \color{black} {$0$} & \color{black} {$0$} &
\color{black} {$0$} & \color{black} {$5,129$ Mpc} \\
\\
 $t_{\rm CMB}$  &  $-13,890$ & 7.003 & 12.63 Mpc & 17.29 Mpc \\
\\
$t_{*}$  & $-14,630$ & 124.5 & $2.464\times10^7~\ell_{\rm pl}$ &
$3.328\times10^7~\ell_{\rm pl}$ \\
\\
\hline
  \end{tabular}
\end{table}
}}%
Because of the positive $\Lambda$, the space-time  can be conformally completed such that it admits a smooth future infinity $\mathcal{I}^{+}$ which is now \emph{space-like} \cite{abk1}. This surface corresponds to infinite proper time $t$. Every cosmic observer (whose world line is orthogonal to the $t={\rm const}$ surfaces) will intersect $\mathcal{I}^{+}$ in her infinite future. The key consequence of a positive $\Lambda$ is that the causal past of this observer contains only a \emph{finite portion} of any $t={\rm const}$ surface: \emph{every eternal observer} has a past horizon. If --as was widely believed until the 1990s-- our universe had a vanishing $\Lambda$, the situation would have been drastically different. Then the universe would have been matter dominated in the future and the scale factor would have grown only as $t^{2/3}$ rather than exponentially, whence an eternal observer would have been able to see the entire (infinite) hypersurfaces $t={\rm const}$. Thus, because $\Lambda$ is positive, every eternal (cosmic) observer will have access to only a finite spatial portion of the universe at any time, and in particular at the CMB time. Therefore, it suffices to restrict oneself to this \emph{observationally relevant} patch of the universe, whose boundary is the cosmological horizon of an eternal observer. We will do so. The boundary of this patch at the CMB surface consists of the interior of the 2-sphere of radius $17.29$ Mpc. This 2-sphere is special in that the Cauchy data specified on this part of the CMB surface determines fields in the interior of the entire patch.

The structure of this portion of space-time is summarized in (the left panel of) Fig \ref{figs:history} and (first three rows of) Table \ref{history}, where $\eta$ denotes the conformal time (with $\eta=0$ today); $n_{e} = |\ln (a(\eta)/a_{0})|$ denotes the number of \efolds between now and the conformal time $\eta$; $R_{0}(\eta)$ is the proper radius of the portion of the universe at time $\eta$ that is visible to us \emph{today}; and $R_{\rm max}(\eta)$ is the maximum radius of the universe at time $\eta$ that is visible to the eternal observer.

We will use these features of space-time geometry in section \ref{s4} in working out observational consequences of the two new Principles  introduced out in section \ref{s3}.

\subsection{Constraints on inflation from the PLANCK data}
\label{s2.2}

Inflationary framework provides an avenue to extend the space-time provided by the PLANCK data to the \emph{past} of the CMB surface. However, this can only be done within specific models of inflation since different models lead to different extensions. We will work with a single inflaton in either the Starobinsky or the quadratic potential.

Let us begin by recalling the notions that play an important role in the discussion of the slow-roll phase of inflation for \emph{any} potential. First, there are two sets of slow-roll parameters (see, e.g., \cite{Liddle:1994dx}). The fundamental parameters, associated with space-time dynamics are: 
\be
  \epsilon := -\f{\dot H}{H^2},\quad {\rm and}\quad \delta := \f{\ddot H}{\dot H H}\,\,\, .\ee
During the slow-roll phase these dimensionless parameters are small, whence one generally ignores quantities that are quadratic or higher order in them. A second set of slow-roll parameters refer to the inflaton potential: 
\be
  \epsilon_V := \f{1}{16\pi G} \left(\f {V^\prime}{V}\right)^2,\quad {\rm and} \quad
    \delta_V := \f{1}{8\pi G} \left(\f{V^{\prime\prime}}{V}\right)
\ee
where `prime' denotes the derivative with respect to $\phi$. The smallness of these parameters captures the idea that the potential energy of the scalar field is changing very slowly. The two sets of parameters can be related to each other using equations governing dynamics during the slow-roll phase if one neglects terms which are quadratic and higher order:
\be \epsilon \approx \epsilon_V, \qquad {\rm and } \qquad \delta \approx \delta_V -\epsilon_V. \ee
Observations provide the values of the amplitude of the scalar power spectrum $\as$  and the spectral index $\ns$ of scalar perturbations at the pivot 
mode $k=k_{\star}$. Given any inflaton potential $V(\phi)$ leading to a slow-roll phase of dynamics, and assuming that the quantum field $\h{\R}$ representing the scalar (curvature) perturbation is in the BD vacuum a few \efolds before $k_{\star}$ exists the Hubble horizon%
\footnote{The dynamical equation satisfied by each mode features the curvature radius $\rcurv = [6\,(\ddot{a}/a + (\dot{a}/a)^{2})]^{-\f{1}{2}}$ --rather than the Hubble radius $R_{H} = (a/\dot{a})$. Therefore, what matters from fundamental considerations is when the pivot mode exists the curvature radius. However, during the slow-roll phase, curvature radius is \emph{very} close to the Hubble radius $R_{H} = 1/H$ and one speaks of the mode exiting the Hubble horizon.}
during the slow-roll, one can express $\as$ and $\ns$ in terms of the Hubble rate and slow-roll parameters as follows:
\ba
\label{eq:as}
  \as(k_{\star}) &=& \f{G\hbar\, H_\star^2}{\pi \epsilon_\star}\,\, \simeq \,\,\left(\f{G \hbar\, H_\star^2}{\pi
             \epsilon_V}\right)\, \Big{|}_{\phi=\phi_\star}\,\,\,\, {\rm and} \\
\label{eq:ns}
  \ns(k_{\star}) &=& 1- 4 \epsilon_\star + 2\delta_\star \,\,\simeq \,\,\left(1 - 6~\epsilon_{V} + 2 ~
              \delta_{V}\right)\big{|}{\phi=\phi_\star}.
\ea
Here the subscript `$\star$' denotes that the quantities are evaluated at the time $t_{\star}$ when the pivot mode $k_\star$ exist the Horizon during inflation and $\simeq$ denotes the equality in the slow-roll approximation. Given an inflaton potential $V(\phi)$, (\ref{eq:as}) and (\ref{eq:ns}) can be regarded as algebraic equations for two unknowns, $H_\star$ and $\phi_\star$. Thus, using the values of $\as$ and $\ns$ from observations we can solve these equations for $H_{\star}$ and $\phi_{\star}$. For a general potential, the solutions could constitute a discrete family but for the Starobinsky and quadratic potentials, the solutions are in fact unique. These values of $H_\star$ and $\phi_\star$ then directly determine the numerical values of the slow role parameters $\epsilon_{V}$ and $\delta_{V}$ at time $t_{\star}$, and hence also of $\epsilon_{\star}$ and $\delta_{\star}$ within the slow-roll approximation. Next, at time $t_{\star}$ we have the Friedmann equation 
\be H_\star^2 = \f{8\pi G}{3} \left(\half \phid_\star^2 + V(\phi_\star)\right) \ee     
and equation of motion of the inflaton $\phi$ 
\be 3 H_\star \phid_\star + V^\prime(\phi_\star) \simeq 0,
\ee
where we have neglected $\ddot\phi$ using the slow-roll approximation. These two equations enable one to determine the free parameter --the mass-- in both these potentials, as well as value $\dot\phi_\star$ of the time derivative of the inflaton at time $t_{\star}$.

The PLANCK data provides the following values for $\as$ and $\ns$ at the pivot scale%
\footnote{\label{pivot} The PLANCK mission uses $\t{k}_{\star} = 5\times 10^{-2}~\mpc$ as the pivot scale. We use the WMAP pivot scale $k_{\star} = 2\times 10^{-3}~{\rm Mpc}^{-1}$ to facilitate comparison with other papers in LQC. For this,  as is common, we calculated $A_{s}(k_{\star})$ from the PLANCK values $A_{s}(\t{k}_{\star}), \, n_{s}(\t{k}_{\star})$ using their  ansatz $\mathcal{P}_{\R}(k) = A_{s}(\t{k}_{\star}) (k/\t{k}_{\star})^{n_{s}(\t{k}_{\star})-1}$. }
$k_\star =0.002~\mpc$  \cite{planck15xiii,planck15xx}
\be \label{eq:asns}
 \as = (2.474\pm0.116) \times10^{-9} \qquad {\rm and} \qquad \ns = 0.9645\pm0.0062\, .
\ee
Using the best fit values, the procedure described above led us to the following values of various parameters:
\begin{itemize}
\item for Starobinsky potential $V(\phi) = \textstyle{\f{3\, m^2}{32\pi G}\,\, \big(1-e^{-\sqrt{\f{16\pi G}{3}}\phi}\big)^2}$
\ba 
 H_\star &=& 1.321\times10^{-6} \mpl, \qquad \phi_\star =  1.064~\mpl\nonumber, \\
 \epsilon_\star &=& 2.244\times10^{-4}  ,~   \qquad\qquad m = 2.676 \times10^{-6} ~\mpl.
\label{eq:paramQ}
\ea
\item  for the quadratic potential $V(\phi) = \textstyle{\f{1}{2}}m^2\phi^2$
\ba
 H_\star &=&  8.304\times10^{-6}~\mpl, \qquad \phi_\star = 2.994~\mpl, \nonumber \\
 \epsilon_\star &=& 8.875\times10^{-3},  \qquad\qquad  m = 1.353 \times10^{-6}~\mpl
\label{eq:paramS}
\ea
\end{itemize}
The fact that the values of these parameters are noticeably  different brings out the fact that the dynamics prior to $t_{\rm CMB}$ in the two cases is quite different. However, for each of the two potentials, analytical considerations imply that these data are sufficient to determine a unique FLRW solution to Einstein's equation with the inflaton as matter source between $t_{\star}$ and $t_{\rm CMB}$. Thus, while the inflationary scenario does not have any a priori constraints on the choice of the FLRW background, given a potential, one does obtain an unique solution once precise values of $\as(k_{\star})$ and $\ns(k_{\star})$ are given. 
This extension from $t= t_{\rm CMB}$ to $t= t_{\star}$ is enormous both in terms of space-time geometry (since it covers $\sim 117$ \efolds) and in terms of dynamics of matter (since the density increases by a factor of $\sim 10^{103}$!) Salient properties of the space-time geometry of the two extensions are shown in the right panel of Fig \ref{figs:history} and are listed in the last row of Table \ref{history}.\\ 

{\it Remark}: In the last para we have described an idealized situation in the sense that we assumed that $\as (k_{\star})$ and $\ns (k_{\star})$ are known exactly. Observations provide values of $\as(k_{\star})$ and $n_s(k_{\star})$ only within error bars. Using them, one obtains a `pencil' of solutions --rather than a unique solution-- to the past of $t=t_{\star}$. Furthermore, because the inflationary trajectories are attractive, as we will see in section \ref{s4}, this ``pencil'' has a counter-intuitive feature.  

\subsection{Elements of LQG and LQC}
\label{s2.3}

As noted in the beginning of this section, if one further wishes to extend cosmology to an even earlier epoch to the past of $t=t_{\star}$, we need a quantum theory of gravity. LQG provides a natural arena for this task because, as explained in section \ref{s1}, its underlying quantum geometry naturally introduces modifications in Einstein's equations which become dominant in the Planck regime, leading to  singularity resolution in a variety of cosmological models \cite{bojowaldprl01,aps3,ps09,asrev,ps3,agullocorichi}.
To make this article self-contained we will now recall the salient features of LQG and LQC that are relevant for our analysis (although they have been discussed extensively in the LQC literature).

In GR, gravity is encoded in space-time geometry. Therefore, it is natural to expect that a quantum theory of gravity will require or lead to an appropriate quantum generalization of this Riemannian geometry. The LQG community adopted this viewpoint seriously and systematically developed a rigorous mathematical theory of quantum  Riemannian geometry in the 1990s (for reviews, see, e.g.,\cite{ashtekarloopsstatus,rovelliqg,thiemannbook}). This theory has a natural in-built discreteness at the Planck scale. In particular, one can  define self-adjoint operators representing geometric observables --such as areas of physical surfaces and volumes of physical regions \cite{alarea,alvol}. It has been shown that geometry is quantized in the direct sense that eigenvalues of geometric operators are discrete \cite{rs,alarea,alvol,length}. In particular, there is a smallest non-zero eigenvalue  of the area operator, $\Delta\,\lp^{2}$. Thus the theory introduces a new fundamental, dimensionless parameter $\Delta$, called the \emph{area gap}. This \emph{microscopic} parameter sets the scale for new phenomena.%
\footnote{This is the fundamental \emph{physical parameter} of LQG, related to the mathematical Barbero-Immirzi parameter $\gamma$ via $\gamma = \Delta/(4\sqrt{3}\,\pi)$ \cite{ag1,aa-ijmpd}. In this paper we will use the commonly used value $\Delta = 5.170$ of the area gap  from black hole entropy calculations. It could be modified by more sophisticated considerations. For comparisons with observations, this would slightly alter the number of pre-inflationary \efolds (between the LQC bounce and the onset of the slow-roll phase) but not affect any of the main conclusions.}

In all cosmological models considered in LQC, there is a well-defined Hilbert space $\Hp$ of physical states that satisfy the quantum Hamiltonian constraint. This is a \emph{difference} equation (in scale factor) whose step-size is dictated by the area-gap $\Delta$; it goes over to the Wheeler-DeWitt equation of geometrodynamics only in the limit $\Delta \to 0$. Thus, the area gap plays a key role in the LQC dynamics. Here we will focus on spatially homogeneous, isotropic (i.e. FLRW) models \cite{aps3}. Then, the role played by the area gap can be made explicit: Because $\Delta \not=0$, the operator $\h\rho$ representing matter density acquires an universal upper-bound on the entire Hilbert space $\Hp$, given by 
\be \rhosup\, = \,  \frac{{\rm 18\pi}}{G^{2}\hbar\, \Delta^{3}}\,\, \approx \,\,0.4092\, \rho_{\rm Pl}.\ee
Note that in the limit in which the area gap goes to zero, i.e., if quantum geometry effects are ignored, we obtain the classical GR result that the energy density can be arbitrarily large. (This is analogous to the fact that the ground state energy of the hydrogen atom diverges in the limit $\hbar \to 0$). The boundedness of $\h{\rho}$ immediately implies that all physical states $\Psi_{o}$ of the FLRW geometry undergo a bounce. A common strategy used in LQC is the following. One starts with a wave function $\Psi_{o}$ that is sharply peaked at a classical FLRW trajectory \emph{at late times}, when GR is an excellent approximation, and `evolves' it using the LQC Hamiltonian constraint. One finds that it remains sharply peaked also in the Planck regime but now on the \emph{effective} trajectory satisfying (\ref{effeq}) and bounces when the matter density $\langle \Psi_{o} |\h{\rho}| \Psi_{o} \rangle$ reaches $\rhosup$, resolving the big bang singularity and replacing it with a regular quantum bounce.

The dynamics of these sharply peaked states is well described by certain \emph{effective equations}.%
\footnote{As in the original discussion of the Starobinsky potential, an attractive strategy is to begin with the $R + R^{2}/(6m^{2})$ action so that all degrees of freedom are encoded in geometry and the inflaton field $\phi$ emerges in the passage from the Jordon to the Einstein frame. Then the fundamental description is in the Jordan frame. Normally, the $R + R^{2}/(6m^{2})$ theory would be regarded as an effective theory that incorporates certain quantum corrections, encoded in the second term. If, instead, one were to replace general relativity with this theory at the classical level and then use loop quantum cosmology techniques, one would obtain effective equations discussed in \cite{ma1,ma2}.}
More precisely, the Friedmann and Raychaudhuri equations of GR  
\be H^{2} \equiv (\dot{a}/a)^{2} = \f{8\pi G}{3} \rho \qquad{\rm and}\qquad \dot H = -4 \, \pi G\, (\rho\,+p)\,, \ee 
where $\rho$ and $p$ respectively are the energy density and pressure of the matter field, are replaced by the LQC effective equations  that incorporate leading order quantum corrections:
\be H^{2} = \frac{8\pi G}{3} \rho\,\, \big(1 -\f{\rho}{\rho_{\rm sup}}\
\big)\, \qquad{\rm and} \qquad
\dot H =  -4 \, \pi G\, (\rho\,+p)\, \Big(1- 2\f{\rho}{\rho_{\rm sup}}\Big).\,  \label{effeq}\ee
Because of the terms involving $\rho_{\rm sup}$, the leading order corrections are already sufficient to resolve the singularity. Eqs (\ref{effeq}) also bring out the fact that GR is an excellent approximation to  LQC so long as the matter density $\rho$ is much smaller than $\rho_{\rm sup}$, and in particular at the onset of inflation, $t=t_{\star}$, when $\rho \sim 10^{-11} \times \rhopl$. 
At the bounce, the Hubble parameter of the effective trajectory \emph{vanishes} in stark contrast to general relativity where it is very large in the Planck regime and diverges at the big bang. The scalar curvature of the effective metric reaches its universal upper bound $R_{\rm sup} = 62\, \lp^{-2}$ at the bounce. (For details, see, e.g., \cite{asrev}.)

Next, let us summarize the situation with respect to the scalar and tensor perturbations $\h{\Q},\, \h{\T}^{I}$ where $\h{Q}$ is the Mukhanov-Sasaki scalar perturbation and $I =1,2$ denote the two polarizations of the tensor perturbation. In LQC these quantum fields propagate not on a classical FLRW background geometry $g_{ab}$ but on the \emph{quantum} FLRW geometry of $\Psi_{o}$. Because the quantum geometry is completely regular, one can trust the LQC evolution also in the Planck regime. A priori, development of the QFT on a quantum geometry seems like a daunting task. However, as noted in section \ref{s1}, a key simplification arises because of the following result \cite{akl}. Suppose the back reaction of the perturbations $\h{\Q},\, \h{\T}^{I}$ on the background quantum geometry $\Psi_{o}$ can be neglected. Then dynamics of $\h{\Q},\, \h{\T}^{I}$ on the quantum geometry $\Psi_{o}$ is \emph{completely equivalent} to that of quantum fields $\h{\Q},\, \h{\T}^{I}$ propagating on a quantum corrected, dressed, effective, smooth FLRW geometry $\t{g}_{ab}$ constructed from $\Psi_{o}$ as follows: 
\be \label{qcg} \t{g}_{ab} dx^a dx^b \equiv {\dd}\t{s}^2 =
\tilde{a}^{2} (-{\dd}\t{\eta}^{2}\, + \,  {\dd}{\vec{x}^2} )\, .\ee
where
\be \label{qpara} 
\tilde{a}^4 = \f{\langle \hat{H}_o^{-\f{1}{2}}\,
\hat{a}^4(\phi)\, \hat{H}_o^{-\f{1}{2}}\rangle}{\langle
\hat{H}_o^{-1}\rangle};\quad \quad
d\tilde{\eta} = \langle \h{H}_{o}^{-1/2}\rangle\, (\langle \h{H}_{o}^{-1/2}\, \h{a}^{4}(\phi)\, \h{H}_{o}^{-1/2} \rangle)^{1/2}\,\, d\phi\,\,. \ee
Here, all operators and their expectation values refer to the Hilbert space of the background FLRW quantum geometry: the expectation values are taken in the state $\Psi_o$,\, $\h{H}_{o}$ is the `free' Hamiltonian in absence of the inflaton potential,\, and $\h{a}(\phi)$ is the (Heisenberg) scale factor operator \cite{akl,aan3}. 

Thus, when the test field approximation holds, in LQC the tensor modes simply satisfy $\t{\Box} \h{\T}_{I}=0$. The dynamical equation of the scalar perturbation $\h\Q$ is slightly more complicated. In QFT on classical FLRW space-times, it satisfies  $(\Box + \U/a^{2}) \h{\Q} =0$  where $\U$ is a potential constructed from the background FLRW solution. In the LQC dynamics, this $\U$ is also dressed by quantum corrections and is replaced by \cite{aan3}
\be \label{qpot} \t{\U}(\phi) = \f{\langle \h{H}_o^{-\f{1}{2}}\,
\h{a}^2(\phi)\, \h{\U}(\phi) \h{a}^2(\phi)\, \h{H}_o^{-\f{1}{2}}
\rangle}{\langle \hat{H}_o^{-\f{1}{2}}\, \hat{a}^4(\phi)\,
\hat{H}_o^{-\f{1}{2}}\rangle}\,\,\, . \ee
Thus, the evolution equation of the scalar mode $\h\Q$ is now given by $(\t\Box + \t\U)\, \h{Q} = 0$. Note that the scalar modes `experience' the same dressed metric $\t{g}_{ab}$ as the tensor modes \cite{aan3}. It is evident from (\ref{qpara}) and (\ref{qpot}) that the expressions of the dressed metric and the dressed potential could not have been guessed a priori. They resulted from explicit, detailed calculations \cite{akl,aan3}. They `know' not only the effective trajectory (of Eq. (\ref{effeq})) on which $\Psi_{o}$ is peaked but also certain fluctuations in $\Psi_{o}$. Finally note that the equivalence is \emph{exact} in the test field approximation; it does not involve any additional assumptions, e.g., on the wavelengths of the modes. 

Because of this exact correspondence, given an inflaton potential, the task of narrowing down the choice of the quantum geometry state $\Psi_{o}$ by invoking a past hypothesis reduces to that of narrowing down the choices of the dressed effective metrics $\t{g}_{ab}$. We use this fact in the next section.\\ 

\emph{Remark:} Note that, in the deep Planck era, one needs full LQG to describe quantum geometry. This geometry is distributional. For example, physical areas of 2-surfaces are quantized and change discontinuously. The wave function $\Psi_{o}$ captures only the coarse grained properties that refer to the scale factor. Use of the dressed effective metric $\t{g}_{ab}$ corresponds to a further simplification: $\t{g}_{ab}$ is a smooth tensor field that captures only those properties of $\Psi_{o}$ that are relevant for the dynamics of cosmological perturbations. Thus $\t{g}_{ab}$ is a mathematical construct that enables us to ignore a variety of subtle features of the physical quantum geometry in the Planck regime in calculations of $n$-point functions of scalar and tensor modes.


\section{Physical principles}
\label{s3}

We will now introduce two principles to narrow down the Heisenberg states of the quantum corrected --or dressed, effective-- metric $\t{g}_{ab}$ and the state $\psi$ of the scalar mode $\h\Q$ propagating on  $\t{g}_{ab}$.  The proposal is based on fundamental considerations: properties of quantum geometry in LQG, the Heisenberg uncertainties satisfied by $\h\Q, \h{\T}^{I}$ and emergence of classical behavior at late times. Nonetheless, \emph{it should be viewed only as a first step that may guide us to a deeper understanding of initial conditions,} perhaps one that may lead to their derivation from dynamical considerations of quantum gravity. As mentioned in section \ref{s1}, initial conditions selected by the proposal also lead to predictions that are supported by observations, suggesting that the proposal could be pointing us in the right direction if there is indeed an interplay between CMB observations and quantum gravity.
%

\subsection{Narrowing down the quantum corrected background geometry} 
\label{s3.1}

Let us begin with the motivation behind this principle. The dressed effective metric $\t{g}_{ab}$ satisfies Einstein's equation to great accuracy outside the Planck regime --i.e., when the matter density falls below $\sim 10^{-4} \rhopl$.%
\footnote{One does find departures in our numerical simulations which are accurate to 10 decimal figures, but not within the first four significant figures used in this paper. Note that the true physical geometry in LQG is distributional \cite{ashtekarloopsstatus,rovelliqg,thiemannbook} also in the low curvature regime. But it is very well approximated by a smooth metric satisfying Einstein's equations when appropriately coarse-grained.} 
In the classical regime, then, we are interested in solutions to FLRW equations with a cosmological constant. Qualitatively, the causal structure of all these space-times is the same as that depicted in the left panel of Fig \ref{figs:history}. An eternal observer, in particular, has access only to a finite portion of this space-time. Furthermore, CMB or astronomical surveys measuring the LSS cannot probe space-time to the past of the CMB surface. Therefore we have referred to the portion of the universe within the cosmological horizon that lies to the future of the CMB surface as the observationally relevant patch of the universe. \emph{Our goal is to extend this patch to the past of the CMB surface via theoretical considerations} that do not refer to the
details of the matter sources in this extension. From causal structure considerations it is clear that in any specific model there would be a very large number of solutions all of which have the same space-time structure to the future of the CMB surface shown in  Fig. \ref{figs:history}. For example, even within the extension provided by standard inflation with a specific potential, the solutions can feature an arbitrarily large number of \efolds to the past of the $t=t_{\star}$ surface. All these solutions will exhibit the desired slow-roll after $t=t_{\star}$ providing viable extensions of the space-time specified only to the future of $t=t_{\rm CMB}$. Our task is to significantly narrow down this freedom by adding a new physical input.

\bfig \vskip-0.4cm
 \hskip0.2cm 
 \ig[width=0.49\textwidth]{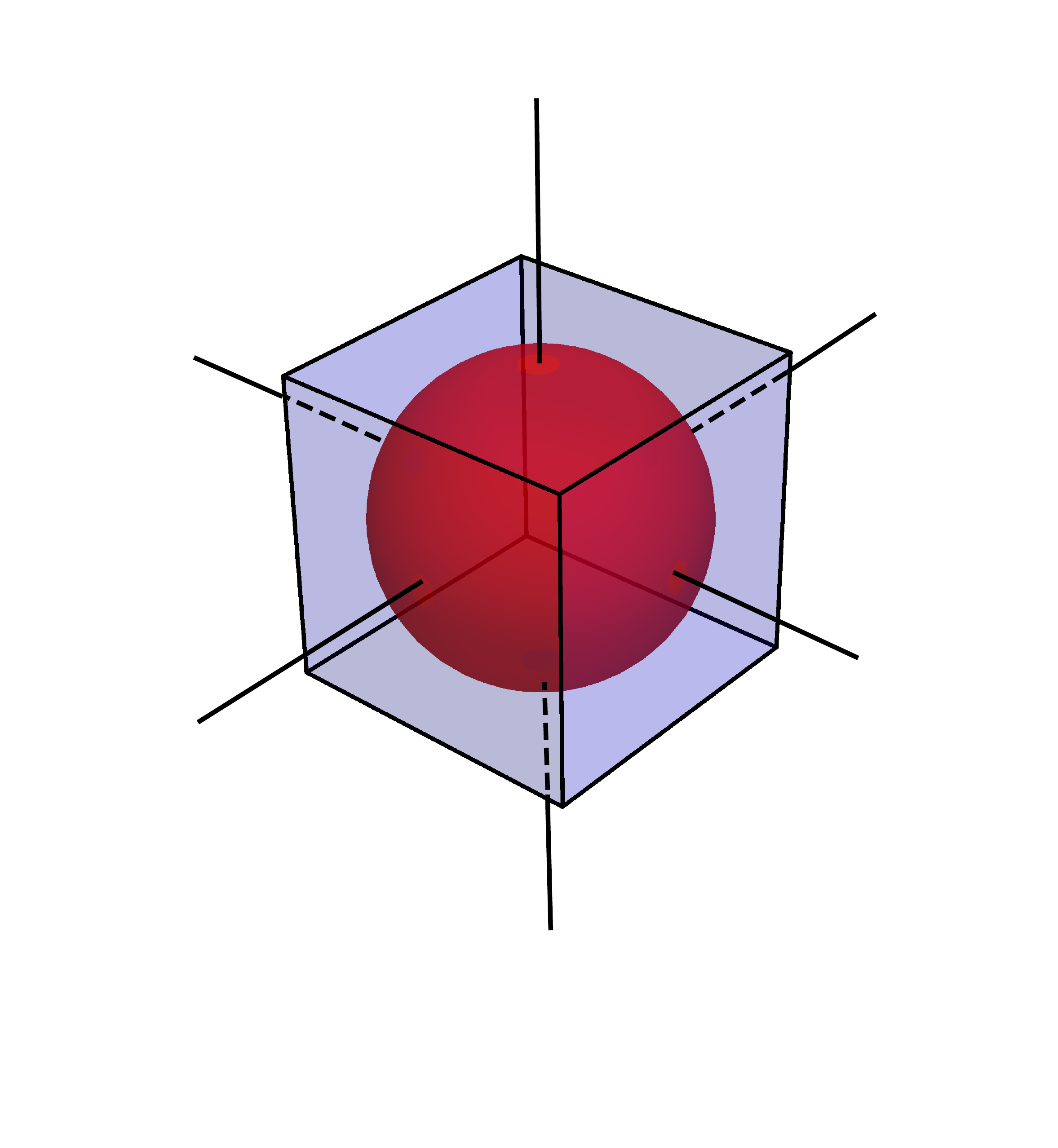}
 \ig[width=0.45\textwidth]{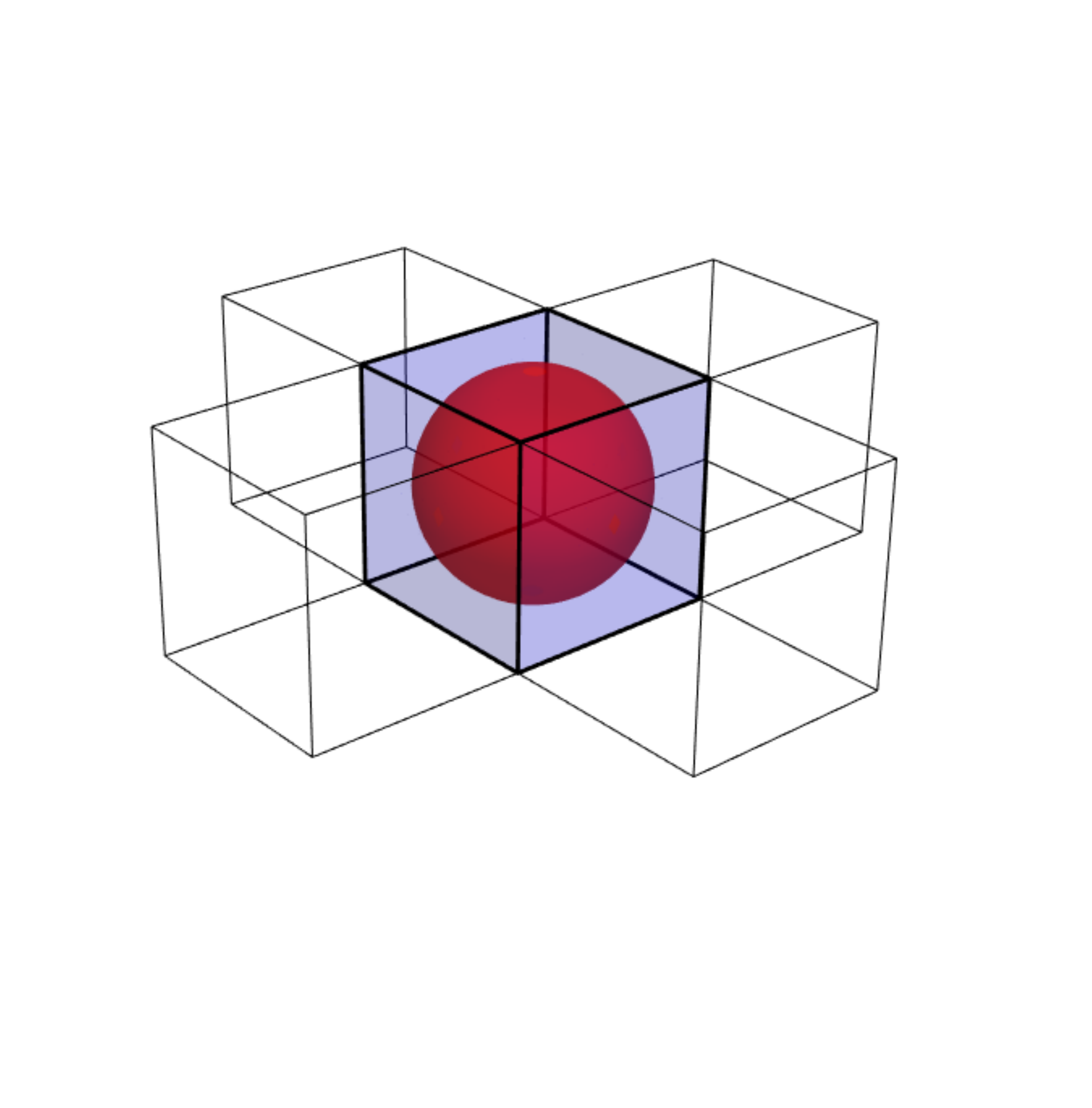}
 \vskip-1.1cm
 \caption{A simple depiction of the LQG geometry at the bounce surface.\\
 {\it Left panel:} The `elementary cell' and its dual graph. The edges of the graph intersect the elementary 2-sphere in 6 points, and each intersection deposits a quantum of area $\Delta\, \lp^{2}$ on the 2-sphere. Principle 1 states that the 2-sphere (in red) enclosed in an `elementary cell' (dual to a single vertex) at the bounce surface expands out to fill the largest 2-sphere on the CMB-surface that is visible to an eternal observer. (See the left panel of Fig.\ref{figs:history}.)\\
{\it Right panel:} The quantum FLRW geometry is defined by a graph with edges along the $x,y,z$ axes meeting in 6-valent nodes. The figure shows the dual cellular decomposition of the spatial 3-manifold at $t=t_{\rm B}$ where each cubical cell is dual to a node of the graph, and each face of the cell is dual to an edge of the graph.}
\label{figs:bounce}
\efig

Every solution $\t{g}_{ab}$ under consideration defines the largest 2-sphere $\mathbb{S}^{2}_{\rm CMB}$ at $t_{\rm CMB}$ with radius  $R_{\rm max}$\,  --the intersection of the $t=t_{\rm CMB}$ surface with the cosmological horizon (see Fig. \ref{figs:history}). The entire observationally relevant patch of the universe is contained within this comoving 2-sphere for $t \ge t_{\rm CMB}$. In any one extension of this space-time to the \emph{past} of the CMB surface, the physical area of this comoving 2-sphere would decrease. The question is: What would it be at the LQC bounce surface $t=t_{\rm B}$? In any given solution, the physical area of this comoving 2-sphere is smallest at the bounce. \emph{The idea is to severely restrict permissible solutions by demanding that the area of this 2-sphere at  $t=t_{\rm B}$ be minimum possible, allowed by the area operator in the underlying LQG quantum geometry.}

Let us recall the nature of this quantum geometry from \cite{awe2}. Given the isometries of $\t{g}_{ab}$, the simplest spin-network underlying this geometry is a  `cubical' graph with edges along the $x,y,z$ directions that intersect in six-valent nodes all of whose edges carry the spin label $j=1/2$. The dual of this graph provides a decomposition of the spatial manifold with cubical cells (see the right panel of Fig. \ref{figs:bounce}), where the nodes of the graph lie at the centers of cubes and edges of the graph intersect the face of the cube. The smallest sphere $\mathbb{S}^{2}_{\rm B}$ in this quantum geometry is the one contained on just one cell%
\footnote{Because quantum geometry is distributional, it does not matter whether the 2-sphere is contained in the cube or surrounds the cube. In either case, it would have 6 intersections with the edges of the graph that defines the quantum geometry, whence the LQG area would be the same.} 
(see the right panel of Fig. \ref{figs:bounce}). Since $\mathbb{S}^{2}_{\rm B}$ has 6 intersections with the edges, each depositing an area $\Delta\, \lp^{2}$, its the total quantum area is $6 \Delta\, \lp^{2} \approx 31.02\,\lp^{2}$. Thus, our first principle to limit the allowed extensions is to demand:
\begin{quote}
\textbf{Principle 1:} \emph{The dressed effective metric $\t{g}_{ab}$ should be such that the past evolution of the 2-sphere $\mathbb{S}^{2}_{\rm CMB}$ yields the `elementary' 2-sphere $\mathbb{S}^{2}_{\rm B}$ at the bounce with total area $(6\Delta)\,\, \lp^{2}$.} 
\end{quote}
This condition is a significant restriction on the allowed $\t{g}_{ab}$ \emph{to the past of the CMB surface.} 

As noted at the end of section \ref{s2.3}, effects of quantum geometry appear at three levels: discreteness of eigenvalues of geometric operators in full LQG, coarse graining tailored to our FLRW setting captured in $\Psi_{o}$, and the dressed effective metric $\t{g}_{ab}$ that encodes all the information in $\Psi_{o}$ that is relevant for the dynamics of scalar and tensor modes. Principle 1 bridges the first and the third levels: one tracks the backward-in-time evolution 2-sphere $\mathbb{S}^{2}_{\rm CMB}$ all the way to $\mathbb{S}^{2}_{\rm B}$ using $\t{g}_{ab}$ and relates the physical radius of $\mathbb{S}^{2}_{\rm CMB}$ determined by $\t{g}_{ab}$ to that of $\mathbb{S}^{2}_{\rm B}$ determined in full LQG.

In section \ref{s4} we will use the value of $R_{\rm max}$ provided by the PLANCK data (see section \ref{s2.1}) and the constraints they provide on inflation with Starobinsky and quadratic potentials (see section \ref{s2.2}) to analyze the consequences of Principle 1. We will find that for each of the two potentials, Principle 1 selects only two dressed effective metrics $\t{g}_{ab}$ that extend the space-time geometry from $t_{\rm CMB}$ all the way to the bounce, $t_{\rm B}$. Furthermore, the phenomenological consequences of these two solutions are indistinguishable. In particular, for each potential, one obtains a definite (and small) number of pre-inflationary \efolds between $t_{\rm B}$ and the time $t_{\star}$ when the pivot mode $k_{\star}$ exits the Hubble horizon during inflation. \\ 

\emph{Remarks:}

1. Note that Principle 1 does not refer to inflation at all. The only ingredients are: (i) \emph{quantum geometry} at the bounce that dictates the minimum area of a 2-sphere; and (ii) the area of the 2-sphere at $t_{\rm CMB}$ defined by the past horizon of an eternal observer.  Heuristically, by choosing the minimum area at the bounce, we are asking that the FLRW quantum geometry of the observationally relevant universe be in the `ground state' of the area operator (the state that would assign the `elementary sphere' zero area does not belong to the physical Hilbert space \cite{aps3}). Interestingly, as discussed in section \ref{s4.4}, there are tight observational bounds on how large $\mathbb{S}^{2}_{\rm BB}$ can be. Increasing its area at the bounce by even a factor of 10 is ruled out at 95\% confidence level. However, so far the observational viability of Principle 1 has been investigated \emph{only in the context of the Starobinsky and quadratic potentials in the inflationary paradigm.}

2. There is some ambiguity in the precise choice of the CMB surface. However, this has negligible effect in terms of observational consequences. For example, of one chooses the CMB surface to be at $z=1110$ rather than $z=1100$, the number $N_{\rm B-CMB}$ of \efolds from the bounce to the CMB surface would change by $- 0.0090$, while, as we will see in section \ref{s4}, for the Starobinsky potential in the inflationary scenario $N_{\rm B-CMB} = {134.3}$ for $z=1100$.

3. From the full LQG perspective, Principle 1 asks us to start from a cell that has a single node. As a manifold, the \emph{entire portion of the CMB surface inside $\mathbb{S}^{2}_{\rm CMB}$ is contained in this cell} at $t_{\rm CMB}$. However, during evolution, more and more nodes would be created and the labels on the links and nodes of the graph would also evolve, leading to a vast expansion of \emph{physical} areas of 2-surfaces and volumes of 3-dimensional regions, making the physical geometry correspondingly richer. This paradigm leads to interesting questions. For example, one can ask: What radius does a 2-sphere of, say, half the size as $\mathbb{S}^{2}_{\rm CMB}$ have when evolved back to the bounce surface? In the LQG quantum geometry, the answer is that it would again be of area $(6\Delta)\,\, \lp^{2}$! In fact in the LQG quantum geometry, \emph{every} 2-sphere contained in $\mathbb{S}^{2}_{\rm CMB}$ has the same area at the bounce surface so long as it envelops the node. At the bounce surface, geometry defined by $\t{g}_{ab}$ is a poor approximation to this true quantum geometry since areas of 2-spheres as measured by $\t{g}_{ab}$ change continuously from $0$ to $(6\Delta)\,\,\lp^{2}$, although it is fully adequate to describe the dynamics of scalar and tensor modes all the way to the bounce. Thus the Planck regime is very subtle.

Another interesting issue concerns the physical wave lengths of observable modes. Recall that in CMB we observe modes in the co-moving wave number range $\sim \,\, (0.1 k_{\star}, 300 k_{\star})$. The dressed effective metric $\t{g}_{ab}$ has no problem with accommodating all these modes. But what is the situation from the fundamental LQG perspective? Can the elementary cell $\mathbb{S}^{2}_{\rm B}$ at the bounce accommodate all these modes from a fundamental perspective of quantum geometry that has an inbuilt UV cut-off? A definitive answer must await a much more complete understanding of quantum dynamics in full LQG. But at this stage we can discuss this issue semi-heuristically. If the area eigenvalues were equally spaced (as the energy levels of a harmonic oscillator), the natural UV cut-off could imply that a vast majority of these modes simply don't exist in the Planck regime. However, the LQG quantum geometry is subtle. The eigenvalues of the area operator \emph{crowd exponentially} for large areas \cite{Domagala_gamma,Meissner_gamma,area2,area1} and, since perturbed geometry is inhomogeneous, to analyze perturbations we are led to consider \emph{all} area eigenvectors, not just those that are adapted to homogeneous isotropic situations, depicted in \fref{figs:bounce}. And this structure is already rich enough to allow a few thousand eigenvectors of the area operator with eigenvalues  $a_{n} \le (6\Delta)\,\,\lp^{2}$.%
\footnote{We thank Ivan Agullo and Fernando Barbero for a first estimate of this number.}  
Therefore, heuristically, one can argue that the observed modes would be accommodated in the elementary cell when one replaces the dressed effective description provided by $\t{g}_{ab}$ by one based in full LQG. Moreover, it is likely that the precise quantum geometry of full LQG will tell us that modes with $k \gg 300k_{\star}$ do not exist because of the UV behavior of quantum gravity. Observationally, this restriction will not be directly relevant because the structure of the universe corresponding to these forbidden scales has origin in local astrophysics rather than cosmology. But conceptually, the issue is important, e.g., to obtain more fundamental regularization and renormalization schemes. All these considerations are beyond the scope of this paper but constitute an important aspect of ongoing investigation into implications of the UV regularity of LQC.

4. In full LQG, it may perhaps be more natural to consider a tetrahedral decomposition of the spatial 3-manifold than a cubical one depicted in Fig. \ref{figs:bounce}. Since faces of the elementary tetrahedron will have 4 intersections with the links of the graph rather than 6, the physical area of the 2-sphere $\mathbb{S}^{2}_{\rm B}$ at the bounce would now be $4\Delta\,\, \lp^{2}$. It turns out that this new number provides an even better fit to the constraints from observations discussed in section \ref{s4.4}.

5. Finally, we wish to emphasize that Principle 1 is meant only as a stepping stone to more fundamental considerations. Since observationally we know the large scale structure of the universe quite well to the future of $t_{\rm CMB}$, it seems natural to pose the question: How does the theory constrain the extensions to the past of $t_{\rm CMB}$? On the other hand, from quantum gravity considerations, it is the quantum geometry in the Planck regime that offers a natural point of departure and we are led to ask: How does the elementary cell at the bounce evolve? The two issues are intertwined in Principle 1: As remarked above, Principle 1 interpolates the geometry defined by the dressed effective metric $\t{g}_{ab}$ at $t_{\rm CMB}$ and that defined by full LQG at $t_{\rm B}$. In this respect, it is qualitatively similar to Bohr-Sommerfeld considerations in quantum mechanics which go back and forth between classical phase space and quantum mechanics. The viewpoint is that Principle 1 is a shadow left on our current understanding of LQC by full LQG which, like the Bohr-Sommerfeld principle, can nonetheless lead to interesting consequences and also provide hints to uncover more fundamental principles. It only offers a point of departure; it does \emph{not} provide the final picture.

\subsection{Narrowing down quantum states $\psi$ of perturbations}
\label{s3.2}

We will now assume that we are given a dressed effective geometry $\t{g}_{ab}$ (satisfying Principle 1) and consider quantum fields $\h{\Q}, \h{\T}^{I}$ representing cosmological perturbations thereon. We wish to introduce a second principle  to narrow down the Heisenberg state of \emph{perturbations} by incorporating certain aspects of the physics of the Planck regime. 

The introduction of this second principle is a two-step process. The first extends Penrose's Weyl curvature hypothesis to LQC and is discussed in detail in \cite{ag2} for both tensor and scalar modes. Here we will only summarize the main ideas and results --without proofs-- for scalar modes. This step enables one to select a small ball $\B$ in the space of all homogeneous, isotropic, quasi-free vacua, by making appeal to the fundamental uncertainty principles, applied in the Planck regime. This step is not tied to inflation. The second step will enable us to select a single state within the ball $\B$. It is tied to the inflationary paradigm. In section \ref{s4} we will see that at $t=t_{\star}$, this state is indistinguishable from the BD vacuum for Fourier modes of the scalar perturbations for $\ell \gtrsim 30$ but differs from the BD vacuum for modes at large angular scales $\ell \lesssim 30$.

\subsubsection*{Step 1: Quantum generalization of Penrose's Weyl curvature hypothesis}
\label{s3.2.1}

Penrose's hypothesis \cite{rp-weyl} was formulated in the context of GR 
and the intent was to quantify the sense in which the big bang represents an extremely rare initial state: It posits that, in spite of the strong curvature singularity, the Weyl curvature should vanish at the big bang. This was meant as an appropriate `past hypothesis' which makes the initial conditions very special, thereby providing an avenue to argue that the universe is initially in a low entropy state. The hope has been  that such an initial condition would naturally emerge from an appropriate quantum gravity theory. 

Now, one expects that big bang singularity will be resolved by quantum gravity. Therefore the appropriate quantum gravity extension of this hypothesis has to refer, instead, to the Planck regime in which physics is extreme but regular. Since Weyl curvature is now an operator satisfying non-trivial commutation relations, there will be no state on which it vanishes identically. The best one can do is to ask that the expectation values of observables that constitute the Weyl tensor should vanish and have as small dispersions around zero as are allowed by the uncertainty principle. As discussed in \cite{ag2}, the analysis has several subtle elements. In particular, one has to first consider tensor modes, recast the Weyl tensor considerations in terms of metric perturbations and then carry over the final results to scalar modes. This quantum generalization of the Weyl curvature hypothesis ensures that in the Planck regime the universe is as isotropic and homogeneous as quantum dynamics and Heisenberg uncertainties permit. Therefore, the generalization was called the \emph{quantum homogeneity and isotropy hypothesis} (QHIH). Since this paper focuses only on scalar modes, we will now discuss the restrictions imposed by the QHIH on the Heisenberg states of the gauge invariant Mukhanov-Sasaki scalar perturbations $\h\Q$.

Let $(M, \t{g}_{ab})$ denote the dressed effective space-time of LQC. To avoid inessential infrared issues, we will assume that the spatial manifold is a 3-torus $\mathbb{T}^{3}$ with comoving volume $V_{o}$.%
\footnote{If the topology is $\mathbb{R}^{3}$ rather than $\mathbb{T}^{3}$, Fourier modes of the field $\h{\Q}$ are operator-valued distributions rather than operators. Therefore, the extension of our construction becomes technically cumbersome, although it is conceptually straightforward.} 
If the co-moving circumference of each of the circles along the $x,y,z$ axes are taken to be, say $2/k_{\star}$, then the observationally relevant universe will always be within this $\mathbb{T}^{3}$.  We are interested in the canonical algebra generated by $\h\Q$ and their conjugate momenta $\h\Pi$, and the class $\mathcal{C}$ of all regular `vacua' that are compatible with the underlying symmetries. These conditions imply that the states are homogeneous, isotropic, quasi-free (i.e. Gaussian). Even with these restrictions, the class $\mathcal{C}$ is large. The QHIH restricts them severely. One begins by fixing a time $t=t_{0}$ in the Planck regimes and looks for states in $\mathcal{C}$ that satisfy the following three conditions \emph{at that time}:\\
(i) Expectation values of the canonically conjugate operators vanish:
\be \langle \psi_{0}|\,\h{\Q}_{\vk}\,| \psi_{0} \rangle =0 \quad {\rm and}\quad\langle \psi_{0}|\,\h{\Pi}_{\vk}\,| \psi_{0} \rangle =0 \ee
(ii) The product of uncertainties are minimized:
\be \Delta (\h{\Q}_{\vk})\,\, \Delta (\h{\Pi}_{\vk}) = \f{\hbar}{2}\,  V_{o} \ee
(iii) The uncertainties are equally distributed among the canonically conjugate variables. Using (i), one can show \cite{ag2} that this requirement that is succinctly captured in the equation:
\be\sigma_{k}^{2}\,:=\, k\,[\Delta (\h{\Q}_{\vk})]^{2} +\f{1}{k}\, [\Delta (\h{\Pi}_{\vk})]^{2} = \f{\hbar}{2}\, V_{o}. \label{dispersion}
\ee
Note that, because of condition (i), the uncertainty $[\Delta (\h{\Q}_{\vk})]^{2}$ is given by just the expectation value of  $\f{1}{2} \big(\h{\Q}_{\vk}\h{\Q}_{\vk}^{\dag}+ \h{\Q}_{\vk}^{\dag} \h{\Q}_{\vk}\big)$, and similarly for $[\Delta (\h{\Pi}_{\vk})]^{2}$.

Somewhat surprisingly, imposition of conditions (i)\! -\! (iii) at $t=t_{0}$, suffices to select a unique Heisenberg state $|0_{t_{0}}\rangle$ \cite{ag2}.%
\footnote{This is not surprising because the quasi-free states are completely determined by 2-point functions. Note also that the potential $\t{\mathcal{U}}$ of section \ref{s2.3} has negligible effect on dynamics under consideration because $\t{\mathcal{U}}/k^{2} <10^{-5}$ for observable modes. Therefore, although in principle condition (iii) should be modified by replacing the explicit factors of $k$ with $\sqrt{{k}^{2} + \t{\mathcal{U}}}$ for simplicity we did not do this.}
However, while in flat --or, more generally, stationary-- space-times the resulting state is  independent of the choice of $t_{0}$, this is no longer the case now because the underlying geometry defined by $\t{g}_{ab}$ is itself time-dependent. Thus, as we pass from $t=t_{0}$ to another time $t= t_{1}$, say, the dispersion $\sigma_{k}^{2}$ in the Heisenberg state $|0_{t_{0}} \rangle$ is no longer the minimum possible; it grows. Conceptually, this is completely analogous to the fact that, in standard inflation, if we select the BD vacuum $|0_{t_{0}}\rangle$ at a time $t=t_{0}$ it is not the BD vacuum at a slightly different $t=t_{1}$. Our idea is to use dynamics in the \emph{Planck regime} to restrict states in the class $\mathcal{C}$ to a small ball $\B$.

By \emph{Planck regime}, we will mean a `space-time slab' $S_{I}$ defined by the time interval $I= [t_{-}, \, t_{+}]$ around the bounce time $t_{\rm B}$,\, \emph{outside} of which the energy density of matter satisfies $\rho \le 10^{-4}\, \rhopl$.%
\footnote{We call $S_{I}$ the Planck regime because outside this slab   general relativity cannot be distinguished from LQC within the 4 significant figures precision of this paper.} 
We are then led to consider states $|0_{t}\rangle$ for all $t \in I$. 
Consider the dispersion in the Heisenberg state $|0_{t_{1}} \rangle$ evaluated at time $t_{2}$:
\be \sigma_{k}^{2} ({t_{1}}|{t_{2}}) := k\, \langle 0_{t_{1}} |\, \f{1}{2} \big(\h{\Q}_{\vk}\h{\Q}_{\vk}^{\dag}+ \h{\Q}_{\vk}^{\dag} \h{\Q}_{\vk}\big)(t_{2})\, | 0_{t_{1}} \rangle +\f{1}{k}\, \langle 0_{t_{1}} |\, \f{1}{2} \big(\h{\Pi}_{\vk}\h{\Pi}_{\vk}^{\dag}+ \h{\Pi}_{\vk}^{\dag} \h{\Pi}_{\vk}\big)(t_{2})\, |0_{t_{1}} \rangle \ee
Then 
\be \label{doublesup} s_{k}^{2}\, :=\,  \sup_{t_{1},t_{2}\in I}  \sigma_{k}^{2} \, ({t_{1}}|{t_{2}})\, , \ee
obtained by taking the supremum with respect to both $t_{1}$ and $t_{2}$, provides us a definite measure of minimum dispersions we must allow if we wish to characterize states that are preferred by uncertainty considerations in the full Planck regime $S_{I}$. Therefore, \emph{we define our preferred ball $\B$ as consisting \emph{all} states $|\psi\rangle$ in $\mathcal{C}$ in which the uncertainty $\sigma^{2}_{k}$ is bounded by $s_{k}^{2}$ in the Planck regime:}
\be \label{ball} 
k\, \langle \psi |\,\f{1}{2} \big(\h{\Q}_{\vk}\h{\Q}_{\vk}^{\dag}+ \h{\Q}_{\vk}^{\dag} \h{\Q}_{\vk}\big) (t)\,| \psi\rangle +\f{1}{k}\, \langle \psi |\,\f{1}{2}\big(\h{\Pi}_{\vk}\h{\Pi}_{\vk}^{\dag}+ \h{\Pi}_{\vk}^{\dag} \h{\Pi}_{\vk}\big) (t) \,| \psi \rangle\,\, \le\,\, s_{k}^{2} \ee 
for all $k$ and for all $t \in I$. The initial condition imposed by the QHIH is that we should restrict ourselves only to the Heisenberg states in this ball.

 This quantum generalization of the Weyl curvature hypothesis can be thought of as follows. The ball $\B$ contains just those states in $\mathcal{C}$ which are singled out by making appeal, \emph{in the Planck regime}, to the same fundamental conceptual considerations that single out the standard Fock vacuum in stationary space-times \cite{ag2}. The additional steps in the construction are forced upon us simply because the geometry is time-dependent in cosmology; indeed, if geometry were stationary in the Planck regime, the ball $\B$ would collapse to a single state --the vacuum associated with that stationary slab. In any state in the ball $\B$ the quantum universe -- described by the background $\Psi_{o}$ and quantum scalar perturbations thereon-- is as homogeneous and isotropic in the Planck regime as Heisenberg uncertainties and quantum dynamics allow it to be.\\

\emph{Remarks:}

1. The final ball $\B$ contains many more states than the 1-parameter family of vacua $|0_{t}\rangle$ with $t \in I$. In these additional states the uncertainties are not minimized at any one instant $t \in I$. However, when we consider the \emph{full} Planck regime $S_{I}$, they are on the same footing as the states $|0_{t}\rangle$ because the upper bound on the dispersion $\sigma^{2}_{k}$ in the full Planck regime $S_{I}$ for any one $|\psi\rangle$ in $\B$ is the same as that for the $|0_{t}\rangle$ we began with. This is why we regard $\B$ as the appropriate generalization to the Planck regime of LQC of the `ground' state in stationary space-times, picked out by the uncertainty relations.

2. In this subsection and Ref \cite{ag2} we only provide a quantum generalization of the Weyl curvature hypothesis. No attempt is made in `deriving' this condition from dynamical considerations of quantum gravity. But we hope that the availability of a precise statement in the quantum regime brings us one step closer in this task.

\subsubsection*{Step 2: Maximal classical behavior at the end of inflation}
\label{s3.2.2} 

We will now select a single state from the ball $\B$ by imposing an additional condition. This condition assumes that we are in the inflationary paradigm. One of the interesting effects of inflation is that the quantum dynamics of fields representing cosmological perturbations is well-approximated by classical dynamics after just a few \efolds (for a summary, see, e.g. \cite{kps}). This emergence of classicality has several different facets, each with its own conceptual subtleties \cite{ack}. However, for our purposes, it suffices to note just one aspect. During inflation, the field $\h{\Q}_{\vk}$ gets highly squeezed and the dispersion in its conjugate momentum $\h{\Pi}_{\vk}$ increases correspondingly. In non-relativistic quantum mechanics of familiar microscopic systems, the dynamics of such highly squeezed quantum states is not well approximated by classical dynamics because the large dispersion in the conjugate momentum causes the system to have a large dispersion in the configuration variable rather quickly. But the situation is different in inflation because the time derivative is give by $\partial_{t}{\h\Q}_{\vk}(t) = a^{-3}(t) \h\Pi_{\vk}(t)$ and the scale factor increases exponentially during inflation. As a result, the state remains sharply peaked in $\h{\Q}_{\vk}$ as time evolves and the evolution of the quantum $n$-point functions of the field $\h{\Q}_{\vk}(t)$ are well-approximated by those of their classical analogs in a precise sense \cite{ack}. 

In this paradigm, then, classical behavior is most enhanced at the end of inflation if the state is maximally peaked in $\h{\Q}_{\vk}$. Therefore we will select a preferred state $\psi$ in the ball $\B$ by demanding:
\begin{quote}
\textbf{Principle 2:} \emph{The Heisenberg state $\psi$ of the scalar mode $\h{\Q}_{\vk}$ should be the one that minimizes the dispersion in $\h{\Q}_{\vk}$} at the end of inflation \emph{within the ball $\B$ singled out by QHIH, the quantum generalization of the Weyl curvature hypothesis.}
\end{quote}
Although the ball $\B$ is in an infinite dimensional space of states $\mathcal{C}$, since it is compact, one would expect the minimum to exist. We will see in section \ref{s4} that this is indeed the case: for each of the two inflaton potentials, dynamics singles out a unique state $\psi$ in the ball $\B$. Using the dressed effective metric $\t{g}_{ab}$ selected by principle 1 and the Heisenberg state $\psi$ of the quantum perturbation $\h{\Q}_{\vk}$ selected by principle 2, we will work out the dynamics from the deep Planck regime to the end of inflation and calculate the $n$-point functions and compare these theoretical results with observations.\\ \goodbreak

\emph{Remarks:}

1. Note that in Principle 2, it is important that the minimization of uncertainty in $\h\Q$ is carried out within the ball $\B$. Had we attempted the minimization process without any reference to any such restriction, we would not have found a state minimizing the uncertainty.

2. In practice, the task of implementing Principle 2 is best carried out as follows. First, one calculates the dispersions
\be \sigma^{2}_{k}\,(t|t_{\rm B}) = k\, \langle 0_{t} |\, \f{1}{2} \big(\h{\Q}_{\vk}\h{\Q}_{\vk}^{\dag}+ \h{\Q}_{\vk}^{\dag} \h{\Q}_{\vk}\big)(t_{\rm B})\, | 0_{t} \rangle +\f{1}{k}\, \langle 0_{t} |\, \f{1}{2} \big(\h{\Pi}_{\vk}\h{\Pi}_{\vk}^{\dag}+ \h{\Pi}_{\vk}^{\dag} \h{\Pi}_{\vk}\big)(t_{\rm B})\, |0_{t} \rangle \ee
at the bounce time $t_{\rm B}$ in the 1-parameter family of vacua $|0_{t}\rangle$, with $t\in I$. Then one considers the ball $\B_{(t_{\rm B})}$ of states $\t\psi$ in $\mathcal{C}$ in which the dispersion at time $t_{B}$ is less than equal to 
\be  (s^{2})_{(t_{\rm B})} := \sup_{t\,\in\, I} \sigma^{2}_{k}\,(t|t_{\rm B})\, ,\ee
so that the ball is given by:
\be \B_{(t_{\rm B})} := \{\,\,\t\psi\in \mathcal{C}\,\Big{|}\, \big[\, k\, \langle \t\psi |\, \f{1}{2} \big(\h{\Q}_{\vk}\h{\Q}_{\vk}^{\dag}+ \h{\Q}_{\vk}^{\dag} \h{\Q}_{\vk}\big) (t_{\rm B})\, |\t\psi \rangle +\f{1}{k}\, \langle \t\psi|\, \f{1}{2} \big(\h{\Pi}_{\vk}\h{\Pi}_{\vk}^{\dag}+ \h{\Pi}_{\vk}^{\dag} \h{\Pi}_{\vk}\big)(t_{\rm B})\, |\t\psi\rangle \,\big] \le (s^{2})_{(t_{\rm B})}\,\}\, . \ee
The ball $\B_{(t_{\rm B})}$ is tied to the bounce because the states $\t\psi$ are all `on the same footing' as the 1-parameter family of vacua $|0_{t}\rangle$ as far as the dispersions \emph{evaluated at} $t_{\rm B}$ are concerned. (For a general $|\psi\rangle \in \B$, the dispersions may not assume the minimum value $\hbar V_{o}/2$ at any time $t$ in the Planck regime). One then minimizes the uncertainty in $\h{\Q}(t)$   \emph{at the end of inflation} $t=t_{\rm end}$. This selects a unique state $\psi_{(t_{\rm B})}$ in the ball $\B_{(t_{\rm B})}$. In the last step one replaces the bounce time $t_{\rm B}$ in the above construction by an arbitrary time $t \in I$ and minimizes the uncertainty in $\h{\Q} (t_{\rm end})$ among the resulting states $\psi_{(t)}$. \emph{This is the state $\psi$ selected by Principle 2.} In essence this procedure divides the task of taking the double supremum in (\ref{doublesup}) into two steps, first taking the sup in $t_{1}$ and then in $t_{2}$. We could have stopped after the first step and just worked with the ball $\B_{(t_{\rm B})}$ since the states $\t\psi$ in this ball can also be regarded as the set that generalizes the notion of the `ground state' in stationary space-times. {Our final results on power spectra would have remained essentially the same.} We chose $\B$ in place of $\B_{(t_{\rm B})}$ only to avoid having to give preference to the bounce time in evaluating dispersions.

3. Given any inflaton potential, Principles 1 and 2 enable us to select $\t{g}_{ab}$ and $\psi$ uniquely (within error bars discussed in section \ref{s2}). Because the selection of $\t{g}_{ab}$ refers to the radius $R_{\rm max}$ at CMB time and that of $\psi$ refers to the dispersion in $\h{\Q}_{\vk}$ at the end of inflation, they incorporate certain `final conditions'. But note that these Principles interpolate between the Planck regime and late time. That is, the Heisenberg states have been selected by asking that the physics they encode in the Planck regime be  correlated in a very specific way to the physics that emerges from them at late times. The required correlation refers to natural features of the universe at both ends: quantum geometry and Heisenberg uncertainties in the Planck regime and  intrinsic features of the observationally relevant universe and emergence of classicality and at late times. However, in other areas of physics it is customary to restrict states by their properties only at an initial time. Principles 1 and 2 represent a significant departure from this practice. We do not have a strong view on whether this shift is essential because of issues that are unique to cosmology --there is but one universe we can observe!-- or if it is only a short-gap measure that will be transcended with a deeper understanding of quantum gravity. Our current emphasis is only on understanding the interplay between theoretical proposals that use fundamental features of Planck scale physics and what observations tell us about the late time universe. The viewpoint is that one should first confront preliminary ideas to observations and, if they turn out to be viable, use them as stepping stones to a deeper and more fundamental formulation.

\section{Observational Consequences}
\label{s4}

This section is divided into four parts. In the first, we begin with a discussion of the evolution of the dressed effective metric $\t{g}_{ab}$ selected by Principle 1 for two inflaton potentials: the Starobinsky and the quadratic potentials discussed in section \ref{s2}. Using $\t{g}_{ab}$, we then investigate the evolution of the physical wavelengths of observable modes from the bounce to the end of inflation and contrast it with the evolution in GR from the big bang to the end of inflation in GR. This analysis will reveal that only the observable modes with the \emph{longest} wavelengths encounter curvature in the Planck regime get exited and are not in the BD vacuum at the onset of inflation (i.e. at time $t_{\star}$). In the second part, we use Principle 2 to determine the preferred Heisenberg state $\psi$ of perturbations. In the third subsection we turn to the computation of the primordial power spectrum using the preferred state. The scalar power spectrum at the end of inflation provides the initial conditions for the Boltzmann equations which govern the evolution of density perturbation during radiation and matter domination era in the post inflationary phase. We use \camb \cite{camb} to solve the Boltzmann equations and obtain the corresponding temperature anisotropy spectrum $\celltt$, the predicted E-mode polarization spectrum $\cellee$ and cross correlation between the temperature anisotropy and E-mode polarization spectrum $\cellte$. In the fourth subsection, we perform Markov Chain Monte Carlo (MCMC) simulations to verify the `goodness' of the Principle 1 using the publicly available code \texttt{COSMOMC} \cite{cosmomc}.


\subsection{Dynamics of the background fields}
\label{s4.1}

\subsubsection{Selection of the dressed effective background solution}
\label{s4.1.1}

In section \ref{s2.1} we discussed dynamical properties of the space-time metric to the future of the CMB surface using general relativity and the data provided by the PLANCK mission. In this subsection we will discuss properties of the dressed effective metric $\t{g}_{ab}$ from the bounce to the CMB surface. We will assume that the state $\Psi_{o}$ of the quantum FLRW geometry is sharply peaked. This ensures that $\t{g}_{ab}$ satisfies the effective equations (\ref{effeq}) to the accuracy required in our numerical simulations. We will constraint $\t{g}_{ab}$ using Principle 1 and three inputs distilled from the PLANCK data (and GR, away from the Planck regime): (i) the inflationary scenario with either the Starobinsky or the quadratic potential, where the value of the mass parameter $m$ is determined in each case from the PLANCK data; (ii) the value $R_{\rm max} = 17.29$ Mpc of the maximum radius of the universe at the CMB time in the observationally relevant universe; and, (iii) the number of \efolds between $t_{\rm CMB}$ and $t=t_{\star}$ (when the pivot mode exits the Hubble horizon during inflation).

Let $\mathring{\RB}$ denote the `area-radius' of the elementary sphere at the bounce surface that enters Principle 1. Then, since the area of the elementary sphere is\,\, $6\,\Delta\, \lp^{2} \approx 31.01\,\lp^2$, we have $\mathring{\RB} = \big(6\Delta/4\pi\big)^{\f{1}{2}}\, \lp\, \, \approx 1.57~\lp$. Principle 1 requires that this sphere should expand out to the largest sphere  in the observationally relevant universe at $t_{\rm CMB}$, which has radius $R_{\rm max}= 17.29$ Mpc. This implies that the metric $\t{g}_{ab}$ is such that there are 134 \efolds between $t_{\rm B}$ and $t_{\rm CMB}$. Next, constraints on inflation discussed in section \ref{s2.2} provide the number $N_{\star\,-\,\rm{CMB}}$ of \efolds between $t=t_{\star}$, to the CMB surface. For the Starobinsky potential we have $N_{\star\,-\,\rm{CMB}} \approx 117.47$ and for the quadratic potential we have $N_{\star\,-\,\rm{CMB}} \approx 119.30$. Therefore, we know that $\t{g}_{ab}$ is such that the number $N_{\rm{B}\,-\,\star}$ of \efolds between the bounce and $t=t_{\star}$ is given by  $N_{\rm{B}\,-\,\star} \approx16.83$ for the Starobinsky potential and $N_{\rm{B}\,-\,\star} \approx 15.00$ for the quadratic potential.%
\footnote{As explained in footnote \ref{pivot}, we first used the PLANCK values  $A_{s}(\t{k}_{\star}), n_{s}(\t{k}_{\star})$ to calculate the amplitude $A_{s}$ at $k=k_{\star}$, and then used this value of $A_{s}(k_{\star})$ to calculate various parameters at $t=t_{\star}$ in section \ref{s2.2}. An alternate procedure would be to use the PLANCK pivot scale $\t{k}_{\star}$ directly, calculate various parameters at the time $\t{t}_{\star}$ when the mode $\t{k}_{\star}$ exits the horizon during inflation, evolve the resulting initial data, and calculate the number of \efolds $N_{\rm B\, -\, \star}$ between $t_{B}$ and $t_{\star}$. For the Starobinsky potential this procedure yields $N_{\rm B\, -\, \star} =16.89$ and for the quadratic potential it yields $N_{\rm B\, -\, \star} = 15.03$. Therefore the final results are quite insensitive to which of the two procedures one uses.}
We will now use this constraint to narrow down the initial conditions for the dressed, effective background solution at the bounce. 

Recall that the space of initial conditions at the bounce for the background effective solution is 4-dimensional, parameterized by $(\ab,\Hb,\phib,\phidb)$. We know that $\Hb =0$ at the bounce and the discussion of the last paragraph implies that there have been 141.30 \efolds from $t_{\rm B}$ until today, $t=t_{0}$. Since we have followed the standard convention and set $a_{0} =1$, we have $\ab = e^{-141.3}$. Values of $\phi_{\rm B}$ and $\dot\phi_{\rm B}$ are subject to the Hamiltonian constraint $\phidb^2/2+V(\phib) = \rhosup \equiv 0.4092 \rhopl$. Therefore, for any given value of $\phi_{B}$ we obtain two 
initial data sets at the bounce. Principle 1 says that only those data sets are permissible whose evolution yields the appropriate number of \efolds $N_{\rm B\, -\, \star}$ between $t_{B}$ and $t_{\star}$. It turns out that for each potential we obtain precisely two initial data sets:
\footnote{For both potentials, the value of $N_{\rm{B}\,-\,\star}$ up to four significant figures is sensitive to the value of $\phib$ up to 7 decimal places. Our calculations were carried out keeping 10 decimal figures. But to conform to the rest of the paper, we report $\phib$ only up to four significant figures.}
\begin{itemize}
\item  For the Starobinsky potential: 
\be
       \phib=-1.421~\mpl,\qquad\phidb=0.9046~\mpl^2 \label{eq:iniS}, 
\ee
for which the kinetic and potential energy density at the bounce are  ${\rm KE}\approx0.4092~\rhopl$ and ${\rm PE}=2.387\times10^{-8}~\rhopl$ and%
\footnote{{Note that since the potential energy is very small, the kinetic energy and the total energy density at the bounce are the same up to four significant figures reported here. Differences arise after eight decimal places.}}
\be
       \phib=3.296~\mpl,\quad\qquad\phidb=-0.9046~\mpl^2 \label{eq:iniS2},
\ee
with ${\rm KE}\approx0.4092~\rhopl$ and ${\rm PE}=2.136\times10^{-13}~\rhopl$.

\item  For the quadratic potential:
\be 
    \phib=1.033~\mpl,\quad\qquad \phidb=0.9046~\mpl^2 \label{eq:iniQ} 
\ee 
with ${\rm KE}\approx0.4092~\rhopl$ and ${\rm PE}=9.750\times10^{-13}~\rhopl$ and 
\be 
    \phib=5.482~\mpl,\quad\qquad \phidb=-0.9046~\mpl^2 \label{eq:iniQ2} 
\ee 
with ${\rm KE}\approx0.4092~\rhopl$ and ${\rm PE}=2.751\times10^{-11}~\rhopl$.

\end{itemize}
\begin{table}
\caption{Evolution time lines for the Starobinsky and quadratic potential (with initial conditions given by Eqs.(\ref{eq:iniS}) and (\ref{eq:iniQ}) respectively). {For both potentials there are sufficient \efolds ($>60$) during inflation.} All quantities are reported in Planck units. $N_{\rm B}$ is the number of \efolds from the bounce up to the event that labels each column.}
\label{tab:timeline}
\begin{tabular}{ccccccc}
\hline
Event       &   & Bounce & $\rho=10^{-4}~\rhopl$ & $KE=PE$  & $t=t_\star$  & $t=t_{\rm end}$ \\
\hline
           & $t$      & 0       & 11.52             & $2.231\times10^5$       & $9.340\times10^6$      & $5.293\times10^7$ \\
           & $\phi$   &-1.421   & -0.630            & 0.9670                 & 1.065                  & 0.1875 \\
Starobinsky& $\rho$   & 0.4092  & 0.001            & $4.111\times 10^{-13}$  & $2.082\times 10^{-13}$  & $7.356\times 10^{-14}$ \\
           & $H$      & 0       & 0.029             & $1.855\times 10^{-6}$   & $1.321\times 10^{-6}$   & $7.850\times 10^{-7}$ \\
           & $N_B$    & 0       & 1.386             & 4.717                 & 16.83                 & 72.29 \\
\hline
            & $t$     & 0       & 11.52             & $3.423\times10^4$       & $1.287\times10^6$       & $1.375\times10^7$ \\
            & $\phi$  & 1.033   & 1.823             & 3.118                 & 2.994                 & 0.2820 \\
Quadratic   & $\rho$  & 0.4092  & 0.001             & $1.779\times 10^{-11}$  & $8.233\times 10^{-12}$  & $9.307\times 10^{-14}$ \\
            & $H$     & 0       & 0.029             & $1.194\times 10^{-5}$   & $8.304\times 10^{-6}$   & $8.831\times 10^{-7}$ \\
            & $N_B$   & 0       & 1.386             & 4.085                 & 15.00                 & 71.56 \\
\hline
\end{tabular}
\vskip0.5cm

\end{table}

For each potential, space-time geometry to the future of the bounce starting from the two sets of initial data turned out to be essentially indistinguishable {(within the accuracy of results reported here.).}%
\footnote{Interestingly, the evolution to the past of the bounce of the two solutions is quite different. This behavior is likely to have interesting implications, e.g., to the issue of the arrow of time, mentioned in section \ref{s5}. However, we will not discuss this point in detail since it is not directly relevant to the focus of this paper.}
For brevity we will present results using only the first initial data set in each case. The salient features of the solutions are summarized in Table \ref{tab:timeline} and figures \fref{figs:NbRhoStarobinsky} and \fref{figs:NbRhoQuadratic}. The Table shows values of various parameters at the bounce time $t_{\rm B}$; at the time when the Planck regime ends ($\rho = 10^{-4}\rhopl$); at the time when the inflaton has lost enough kinetic energy to make it equal to the potential energy, at $t=t_{\star}$ when the pivot mode exits the Hubble horizon during slow-roll; and at $t=t_{\rm end}$ when the inflation ends (i.e. the slow-roll parameter $\epsilon$ becomes $1$). The figures show the number of \efolds and energy density as a function of time for the two potentials.\\

\bfig
 \ig[width=0.47\textwidth]{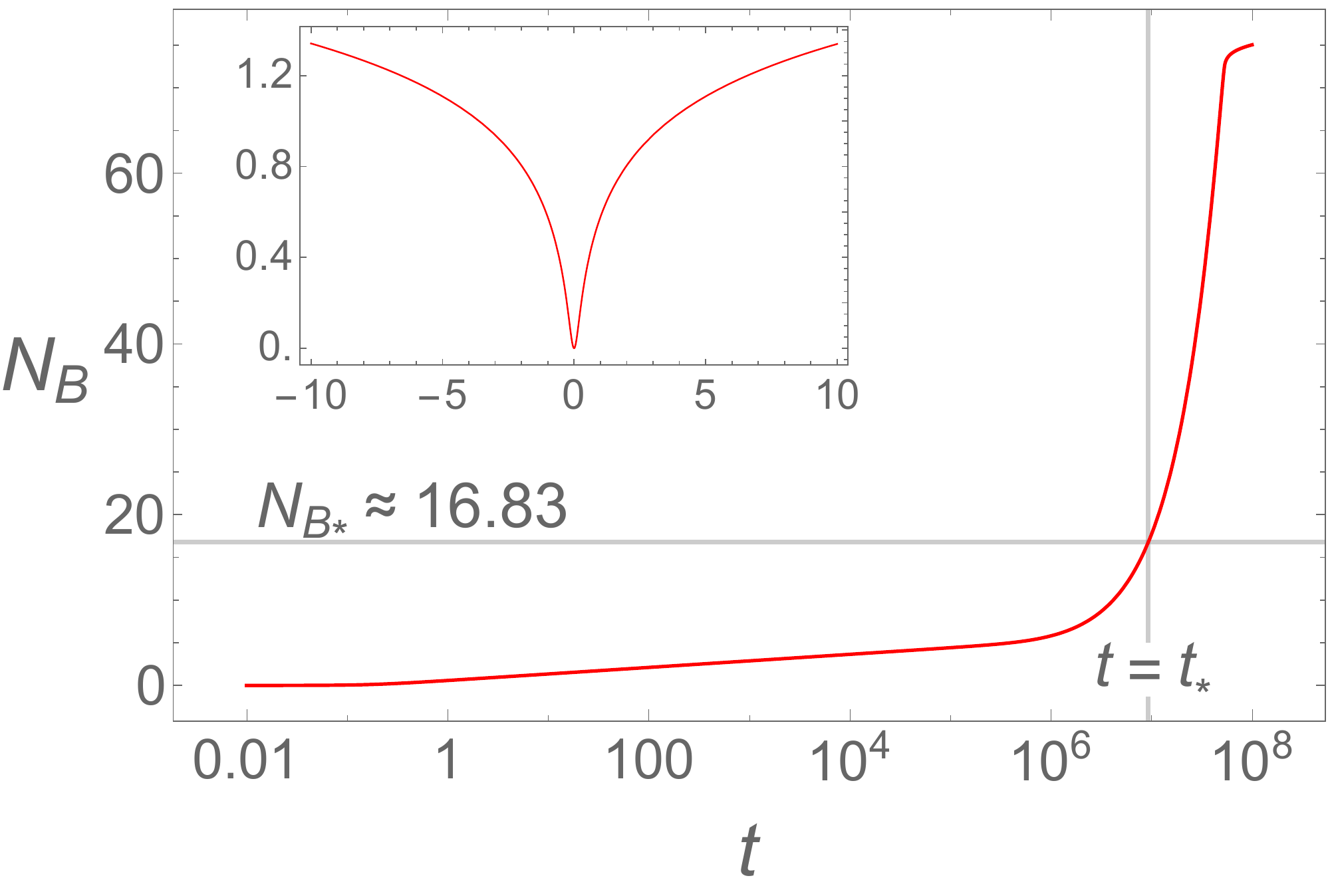}
 \hskip0.5cm
 \ig[width=0.47\textwidth]{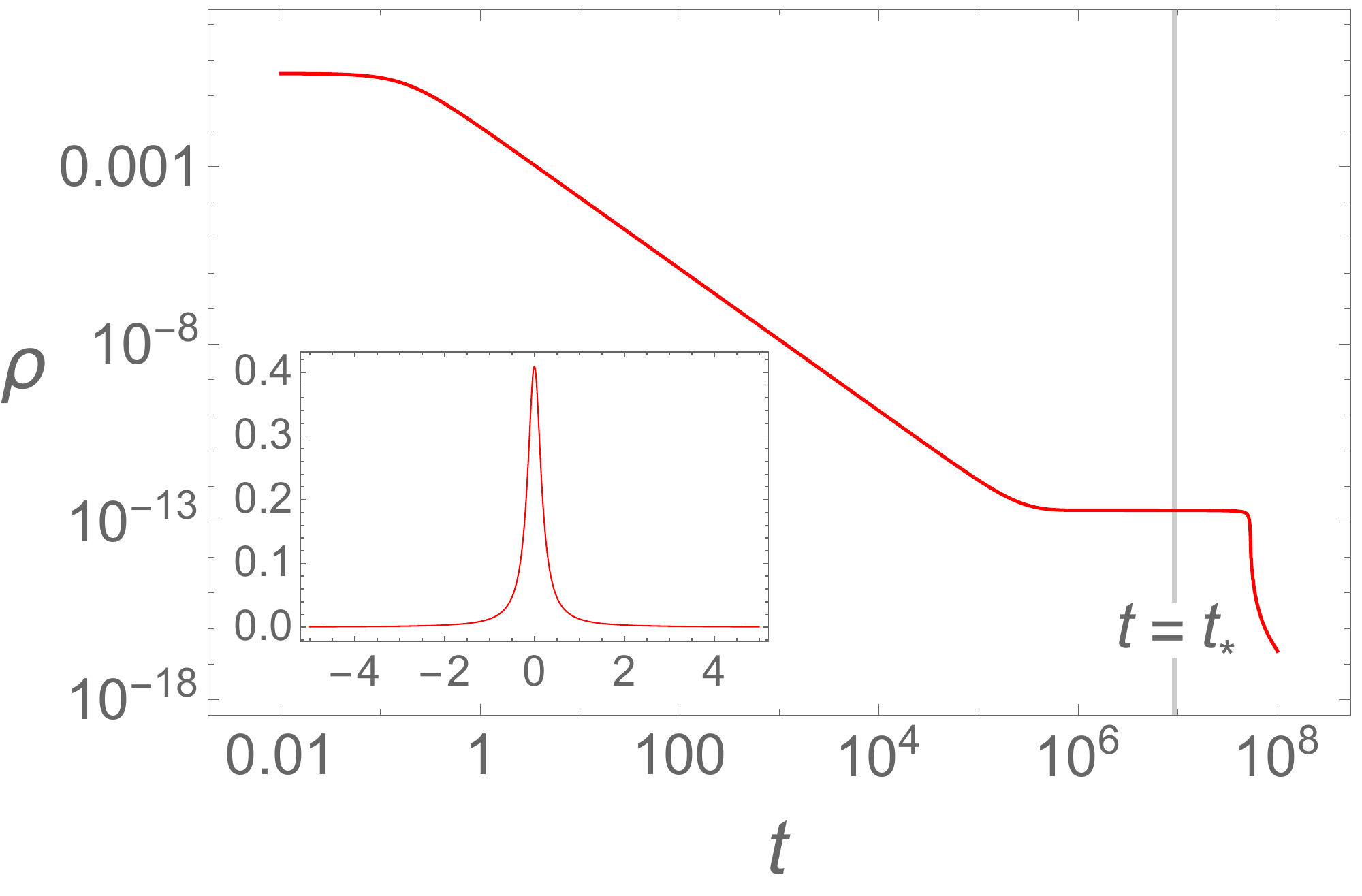}
\caption{{\it Left panel:} Number $N_B$ of \efolds from bounce as a function of the cosmic time in Planck seconds for the Starobinsky potential with initial conditions: $\phib=-1.421~\mpl,~\phid\approx0.9046~\mpl^2$.\\ 
{\it Right panel:} The evolution of energy density from the bounce till the end of inflation.\\ In both figures the evolution near the bounce is shown in insets and $t=t_\star$ marks the instant of time when the pivot mode $k_\star=0.002~\mpc$ exits the curvature radius during slow-roll.}
\label{figs:NbRhoStarobinsky}
\vskip0.6cm
\efig

\bfig
 \ig[width=0.47\textwidth]{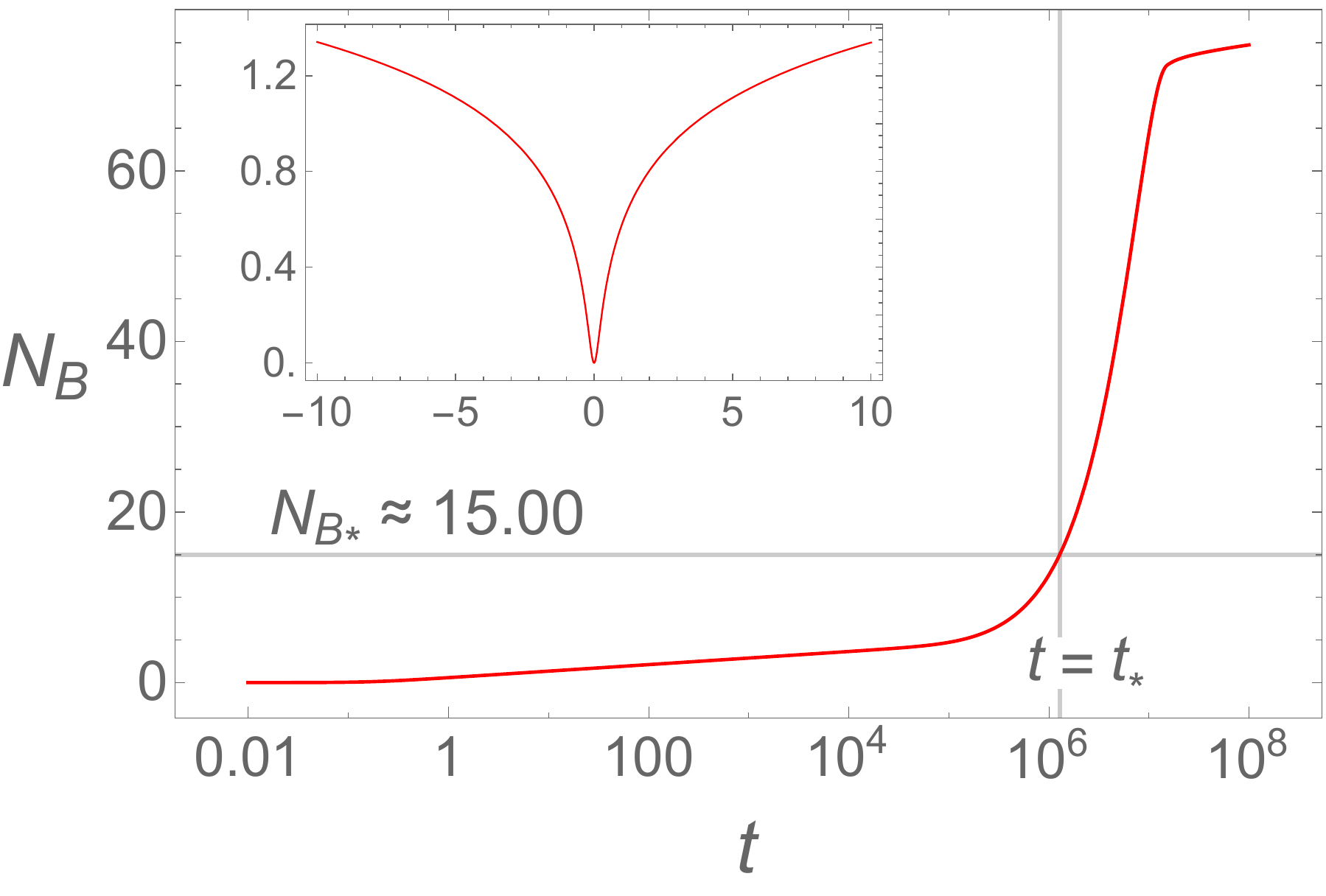}
 \hskip0.5cm
 \ig[width=0.47\textwidth]{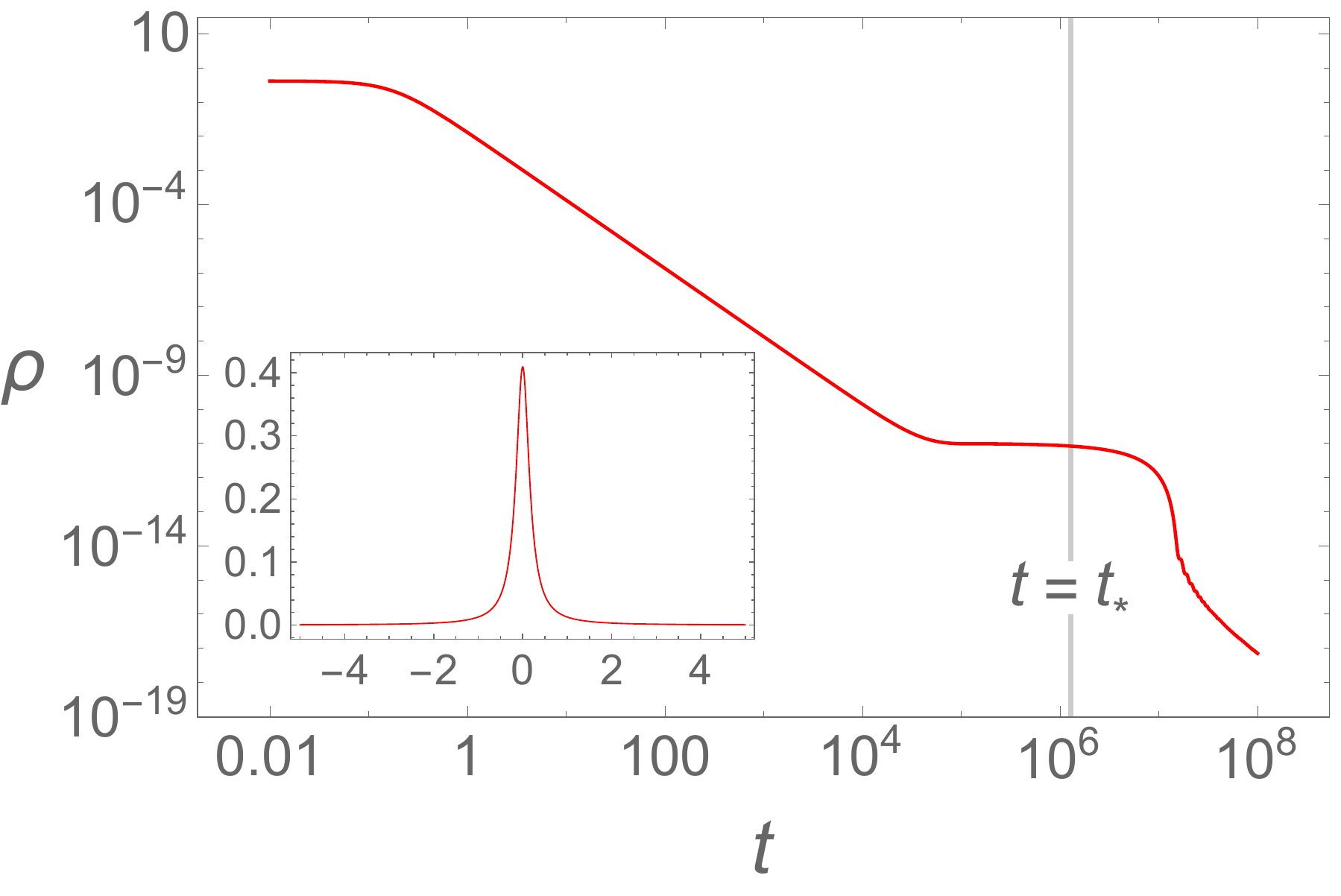}
\caption{{\it Left panel:} Number of \efolds from bounce $N_B$ as a function of the cosmic time in Planck seconds for the quadratic potential with initial conditions: $\phib=1.033~\mpl,~\phid\approx0.9046~\mpl^2$.\\ 
{\it Right panel:} The evolution of energy density from the bounce till the end of inflation.\\ In both figures the evolution near the bounce is shown in insets and $t=t_\star$ marks the instant of time when the pivot mode $k_\star=0.002~\mpc$ exits the curvature radius.}
\label{figs:NbRhoQuadratic}
\efig

The most striking features of this history are the following
\begin{itemize}
\item For both potentials, the bounce is extremely kinetic energy dominated: the potential contribution to $\rhosup$ is only 1 part in $10^{8}$ for the Starobinsky potential and 1 part in $10^{13}$ for the quadratic potential. As a result, in the Planck regime the evolution of the coupled system is almost indistinguishable from that consisting of   a massless scalar field coupled to gravity. This simplifies the technical part our analysis significantly. In particular one can evolve the quantum geometry wave function $\Psi_{o}$ in the Planck regime using the well-developed infra-structure that already exists in LQC \cite{aps1,aps2,aps3}. Results from \cite{ag1} guarantee that the dressed effective metric $\t{g}_{ab}$ satisfies the LQC effective equations (\ref{effeq}) to the desired accuracy (so long as $\Psi_{o}$ is chosen to be sharply peaked, as we do.) Finally, since the LQC evolution is well approximated by GR outside the Planck regime, the entire LQC evolution from $t_{\rm B}$ to $t_{\star}$ can be carried out using effective equations (\ref{effeq}). Note that this simplification arose because of Principle 1. The PLANCK data itself does not require the bounce to be kinetic energy dominated. Indeed, as discussed in, e.g. Refs \cite{asloan_prob2,ck1,LB} compatibility with observations is a very weak constraint on pre-inflationary LQC dynamics. It is Principle 1 that selects from all LQC solutions $(\t{g}_{ab} (t), \phi(t))$ that are compatible with the PLANCK data only two, and the bounce is extremely kinetic energy dominated in both of them.

\item The evolution in the Planck regime for the Starobinsky potential is indistinguishable from that in the quadratic potential. This feature can again be traced back to the fact that the bounce is kinetic energy dominated. The subsequent classical GR dynamics is of course quite different for the two potentials. It is this phase in the evolution that leads to very different values of the slow-roll parameter at time $t_{\star}$ which in turn lead to very different values of the ratio $r$  of tensor to scalar power. 

\item For both potentials, the Planck era lasts a very short time $\sim 11.52 \spl$ or $\sim 1.4$ \efolds (see \tref{tab:timeline}). But, as we will see in section \ref{s4.3}, this short phase suffices to introduce a key difference to the dynamics between the short and long wavelength modes: the long wavelength modes are excited during this phase while the short wavelength modes are not. Consequently, in the end it is this brief phase of \emph{Planck scale dynamics} that is at the root of the key difference between LQC predictions discussed in this paper and the standard predictions based on the BD vacuum at $t_{\star}$.

\item Since in arriving at these solutions for background fields we made use of the data provided by the PLANCK mission, one would expect that for each potential the solution would satisfy the constraints imposed by the data at time $t=t_{\star}$. A comparison of the fourth column of the table with Eqs. (\ref{eq:paramS}) and (\ref{eq:paramQ}) shows that this consistency check is met.

\item For each potential, Principle 1 selects two solutions. As remarked above, to the future of the bounce, the space-time geometry in these two solutions is almost indistinguishable. However, as is clear from Eqs (\ref{eq:iniS}) - (\ref{eq:iniQ2}), the values of the inflaton field in the early phase of the evolution from the bounce is quite different. Nonetheless, in the GR era the values of $\phi$ and $\dot\phi$ in the two solutions also approach one another and by $t=t_{\star}$, when the pivot mode exits the Hubble horizon, dynamics of the inflaton is also indistinguishable. Within the (10 decimal place) accuracy of our simulations, one cannot distinguish between the two solutions at all at, and hence to the future of,  $t=t_{\star}$. This is a reflection of the fact that inflationary solutions are attractors: to see the differences in the two solutions at and to the future of $t=t_{\star}$, one would have to carry out simulations with much higher accuracy. Since, as we saw in section \ref{s2}, the Planck data has error bars, we have to retain both branches to the \emph{past} of $t=t_{\star}$. The permissible solution will be represented by a pencil of solutions around these two, where the radius of the pencil is dictated by the error bars in the data. 
\end{itemize}

To summarize, for any given inflaton potential, the PLANCK data and GR determine the evolution of background fields from $t_{\star}$ to $t_{\rm CMB}$ (and beyond). However, because the inflationary trajectory is an attractor, this solution can be reached starting from `almost any' initial data (within the high accuracy of numerical codes) \cite{asloan_prob2,ck1,LB}. Thus, there is a very large freedom in extending the trajectory back in time from $t=t_{\star}$ to the bounce time $t_{\rm B}$. This freedom is eliminated by Principle 1 since it implies that there is a specific number of \efolds from $t_{\rm B}$ to $t_{\star}$. In particular, we find that the bounce is extremely kinetic energy dominated. Therefore, within desired accuracy, one can use the inflaton $\phi$ as the internal time in the Planck regime to evolve the quantum FLRW  wave function $\Psi_{o}$, extract from it $\t{g}_{ab}$ using (\ref{qcg}) and (\ref{qpara}), and write it in terms of cosmic time of $\t{g}_{ab}$. Outside the Planck regime, the LQC evolution is extremely well-approximated already by GR and we can carry out the evolution directly using proper time $t$. Finally, this evolution shows that while the total number $N_{\rm B-CMB}$ of \efolds between $t_{\star}$ and $t_{\rm CMB}$ is the same for  the starobinsky and quadratic potentials as required by Principle 1, they distribute this number slightly differently between pre-inflationary \efolds $N_{\rm B\,-\,\star}$ and the number $N_{\rm \star\, -\, CMB}$ of post-$t_{\star}$ \efolds.

\subsubsection{Evolution of observable modes in the Planck era}
\label{s4.1.2}

We will now focus on the observable modes with co-moving wave numbers $\sim \, (0.1 k_{\star},\, 300 k_{\star})$. During evolution, these modes experience the curvature in the background metric $\t{g}_{ab}$ only if their \emph{physical} wavelength $\lambda(t) = a(t)/k$ is comparable or larger  than the radius of curvature $\rcurv (t)$ of $\t{g}_{ab}$ at that time. Therefore, we will now discuss the evolution of these two quantities to the past of $t=t_{\star}$ both in standard inflation based on GR and in LQC, compare and contrast the results and comment on their physical implications. This dynamics is shown in  \fref{fig:rcurv}. 

\bfig
 \ig[width=0.45\textwidth]{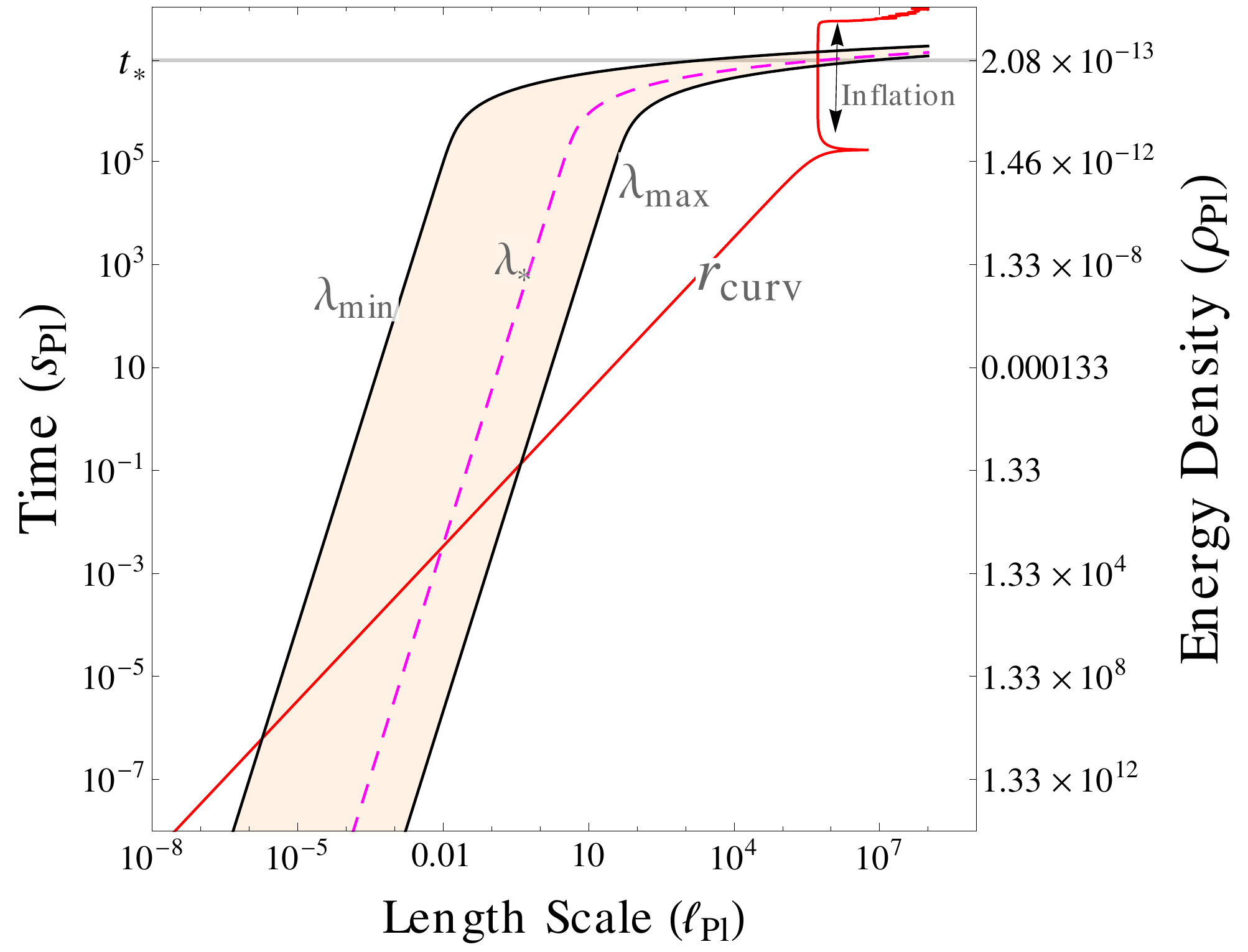}
 \hskip0.7cm
 \ig[width=0.45\textwidth]{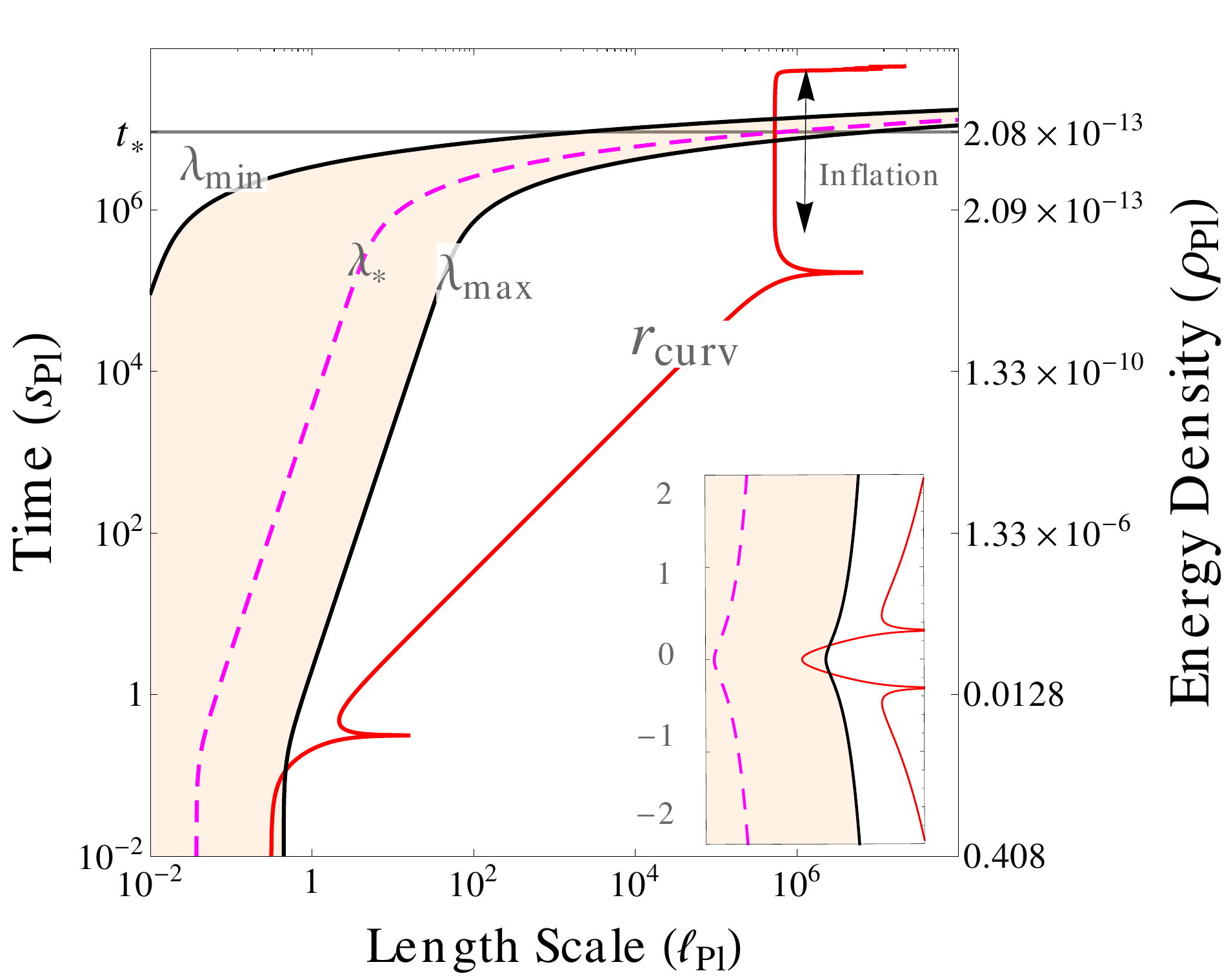}
\caption{Time dependence of the physical wavelengths $\lambda =a/k$ of modes and radius of curvature $\rcurv$ of $\t{g}_{ab}$ in the pre-inflationary era, i.e., prior to $t=t_{\star}$ using Starobinsky potential. The left vertical axis shows cosmic time $t$ and right vertical axis shows the energy density, both in Planck units. The shaded regions represent the wavelengths of observable modes and the dashed line denotes the pivot mode with wavelength $\lambda_\star=a(t)/k_\star$. The red lines represent the evolution of $\rcurv$. The plots of the quadratic potential are very similar.\\
\emph{Left panel: General Relativity.} In the Planck regime near singularity ($t=0$), \emph{all} observable modes exit the curvature radius and are thus excited.\\
\emph{Right panel: LQC.} The bounce occurs at $t=0$. Since all fields are smooth at the bounce, one can continue the evolution to the past of $t=0$. The evolution in a neighborhood of the bounce is shown in the inset. The main figure shows that \emph{only the longest wavelength modes} exit the curvature radius in the pre-inflationary epoch. This is why only the longest wavelength modes are exited and are not in the BD vacuum at $t=t_{\star}$ leading, ultimately, to departures from standard inflation for $\ell \le 30$.} 
\label{fig:rcurv} 
\efig

The main features of this evolution can be summarized as follows. First, in GR, the energy density diverges at the big bang (which corresponds to $t=0$ in the left panel of  Fig. \ref{fig:rcurv}) while in LQC all physical quantities are finite at the bounce (which corresponds to $t=0$ in the right panel of  Fig. \ref{fig:rcurv}). Now, the evolution equation of the modes shows that their dynamics is affected by curvature when the physical wavelength $\lambda (t)$ becomes comparable to or is larger than the curvature radius $\rcurv (t)$ at that time \cite{aan3}.%
\footnote{\label{1} This is most clearly seen through the dynamical equation in conformal time $\eta$ satisfied by $\chi_{k}(\eta)$, related to the Mukhanov-Sasaki mode functions $q_{k}(\eta)$ via $\chi_{k}(\eta) = a(\eta)\, q_{k}(\eta)$:\,\, $\partial^{2}_{\eta} \chi_{k} + a^{2}(\eta) \, \big(\lambda^{-2} - (\sqrt{6}\,\rcurv)^{-2}\big)\, \chi_{k}\, =\, 0$.} 
In both evolutions, the observable modes are well within the curvature radius outside the Planck regime. However, in the Planck regime the situation is different. In GR, all observable modes exit the curvature radius close to the big bang. Therefore, if the Heisenberg state of the perturbation $\h{\Q}_{\vk}$ were chosen to be a `ground state' in the Planck era, all the modes would be excited during evolution (i.e. undergo a non-trivial Bogoliubov transform), whence none of them would be in the BD vacuum a few \efolds before time $t_{\star}$, i.e., at the onset of the slow-roll phase. Put differently, as remarked in section \ref{s1}, the Heisenberg state representing the BD vacuum at the onset of inflation  is an unnatural initial state from the perspective of the Planck regime because it carries non-trivial excitations in the Planck regime. 

In LQC, the situation is different. As the right panel of \fref{fig:rcurv} shows, among the observable modes only those with the longest wavelengths experience curvature during their evolution from $t_{\rm B}$ to $t_{\star}$. Therefore, if the Heisenberg state is chosen to be a `ground state' in the Planck regime, one would find that short wavelength modes are in fact in the BD vacuum at the onset of inflation. It is only the longest wavelength modes that will be excited and hence not in the BD vacuum. Thus, in the LQC geometry $\t{g}_{ab}$ singled out by Principle 1, if one starts in the Planck regime and chooses Heisenberg states $\psi$ using Principle 2, one would find that the correlation functions at the end of inflation will agree with those predicted by standard inflation at small angular scales, but deviate from those predictions at large angular scales. This general conclusion follows already from calculations summarized in the right panel of  \fref{fig:rcurv}. However, further calculations are needed to determine the precise angular scale at which the deviation occurs and the nature of the deviation. These are discussed in section \ref{s4.3}.\\

\emph{Remark:} The difference between GR and LQC captured in \fref{fig:rcurv} brings out a deep interplay between the UV and IR in LQC. As our explicit calculations of section \ref{s4.1.1} show, it is the UV modifications of GR in the Planck regime that tame the  big bang singularity and make all physical quantities finite in the Planck regime of LQC. In particular, because the curvature does not diverge, we now have a non-zero curvature radius at the bounce. This is why, in contrast to the situation in GR, most of the observable modes in LQC have physical wavelengths smaller than the curvature radius throughout the evolution from the bounce to the onset of the slow-roll phase, $t=t_{\star}$ (see the right panel of \fref{fig:rcurv}). Among the observable modes, it is only the ones with longest wavelengths that experience curvature during this phase. The result of this interplay between the UV properties of the background and IR properties of perturbations is that all but the longest wavelength modes are in the BD vacuum at the onset of inflation for the Heisenberg state $\psi$ selected by Principle 2. 
 
\subsection{The preferred Heisenberg state $\psi$ of perturbations}
\label{s4.2}

Let us now consider the scalar mode $\h\Q_{\vk}(t)$ on the LQC space-time $(M, \t{g}_{ab})$. Every Heisenberg state in the collection $\mathcal{C}$ corresponds to a 1-parameter family of basis functions $q_{k}(t)$, normalized via 
\be  a^{3}(t) \big(q_{k}\,\partial_{t} q_{k}^{\star} \, -\, q_{k}^{\star}\,\partial_{t} q_{k}\big) \, =\, i \ee
so that the operator $\h\Q_{\vec{k}}$ can be expanded in terms of creation and annihilation operators:
\be \h{\Q}_{\vk}(t)\, =\, q_{k}(t)\,\h{a}_{\vk}\, +\, q^{\star}_{k}(t)\,\h{a}^{\dag}_{\vk}\,\, . \ee
Let us fix a time $t=t_{0}$ in the interval $I$ defining the Planck regime around the bounce and denote by $|0_{t_{0}}\rangle$ the Heisenberg state satisfying conditions (i)\! -\! (iii) at time $t_{0}$, thereby minimizing the uncertainty $\sigma_{k}^{2}(t_{0})$ (see the first part of section \ref{s3.2}). As shown in \cite{ag2}, the basis functions defined by $|0_{\eta_{0}}\rangle$ are given by
\be
  q_k(t)|_{t=t_0} = \f{1}{a(t_0) \sqrt{2 k}}, \qquad 
  q^\prime_k(t)|_{t=t_0} = \f{-i}{a(t_0)}\,\, \sqrt{\f{k}{2}}. 
\label{eq:state}
\ee
We can evolve the basis functions $q_{k}$ --and hence the operators $\h\Q_{\vec{k}}(t), \h\Pi_{\vec{k}}(t)$-- and calculate the dispersion
\be  \sigma_{k}^{2} ({t_{0}}|{t}) := k\, \langle 0_{t_{0}} |\, \f{1}{2} \big(\h{\Q}_{\vk}\h{\Q}_{\vk}^{\dag}+ \h{\Q}_{\vk}^{\dag} \h{\Q}_{\vk}\big)(t)\, | 0_{t_{0}} \rangle +\f{1}{k}\, \langle 0_{t_{0}} |\, \f{1}{2} \big(\h{\Pi}_{\vk}\h{\Pi}_{\vk}^{\dag}+ \h{\Pi}_{\vk}^{\dag} \h{\Pi}_{\vk}\big)\, |0_{t_{0}} \rangle \ee
for any time $t \in I$. 

As explained in Remark 2 at the end of section \ref{s3.2}, the task of selecting the preferred Heisenberg state can be divided in two steps. In the first, we fix $t$ to the bounce time $t_{\rm B}$ and in the second we allow it to vary. Since we are interested in selecting the preferred state $\psi$ using dynamics of the Planck regime, it suffices to restrict all times to the interval $I$ in which $\rho \in (10^{-4}\rhopl, \rhosup)$. (In the numerical simulations we extended this interval a little to energy densities $\rho \ge  3.315\times 10^{-5} \rhopl$.)

\bfig
 \ig[width=0.47\textwidth]{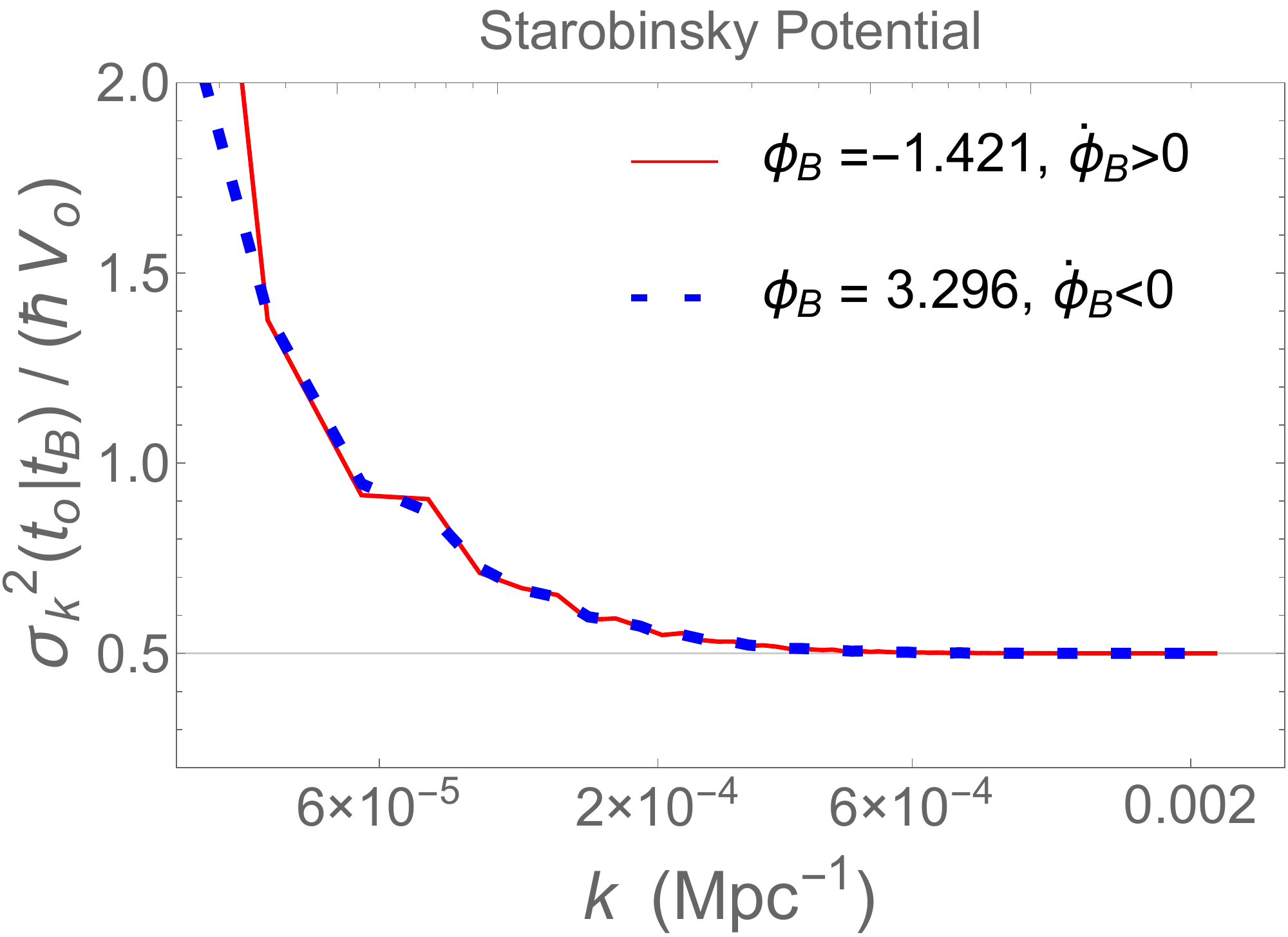}
\hskip0.2cm
 \ig[width=0.47\textwidth]{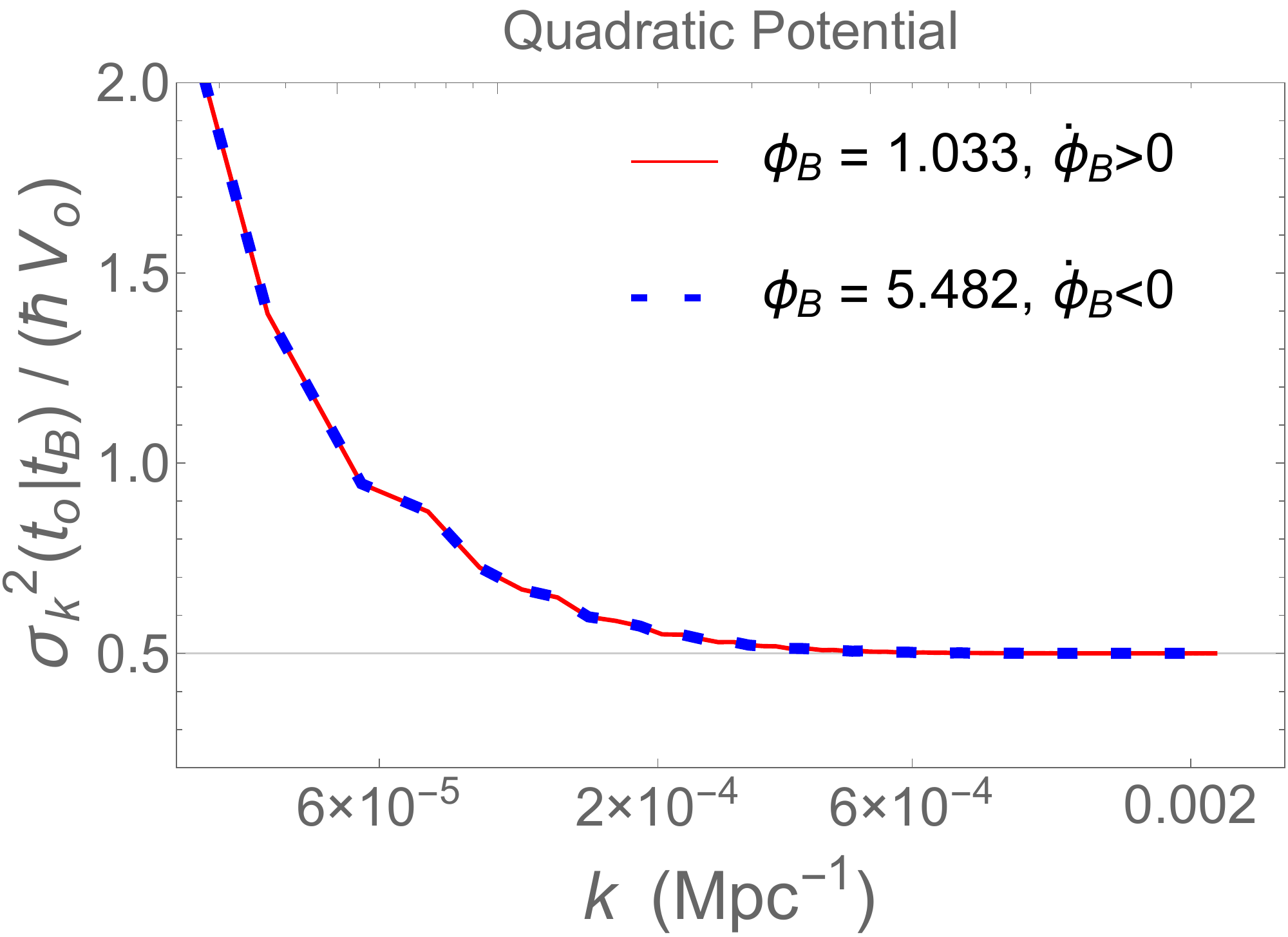}
\caption{The dispersion $\sigma^2_{k} ({t_{o}}|{t_{\rm B}})$ computed at $t=t_{B}$ in the vacuum state $|0_{t_{0}}\rangle$. Here we set $t_{0}=-20~\tpl$ (i.e. before the bounce). The minimum value of this dispersion, $\hbar V_o/2$, is denoted by thin horizontal line. As expected, $\sigma^2_{k} ({t_{o}}|{t_{\rm B}})$ is larger for small $k$ and quickly approaches the minimum value $\hbar V_o$ as $k$ increases.} 
\label{fig:sigBounce}
\efig

\fref{fig:sigBounce} illustrates the  behavior of $\sigma_{k}^{2} ({t_{0}}|{t_{\rm B}})$ for the Starobinsky and quadratic potentials, for initial conditions (\ref{eq:iniS})\,-\,(\ref{eq:iniQ2}). In this simulation we chose $t_{0}\! =\! -20 \spl$, i.e, before the onset of the Planck regime in the pre-bounce phase. For each potential, we see that the dependence on wave number $k$ of $\sigma_{k}^{2} ({t_{0}}|{t_{\rm B}})$ is the same for both solutions selected by Principle 1. Secondly, in the state $|0_{t_{0}}\rangle$ dispersion is minimum, $\hbar V_{o}/2$ at $t=t_{0}$, and increases subsequently for modes which do not remain well within the curvature radius during time evolution. This feature is seen clearly in both figures. In particular, for short wavelength (large $k$) modes, the dispersion remains at the minimum during time evolution.  
\bfig
 \ig[width=0.47\textwidth]{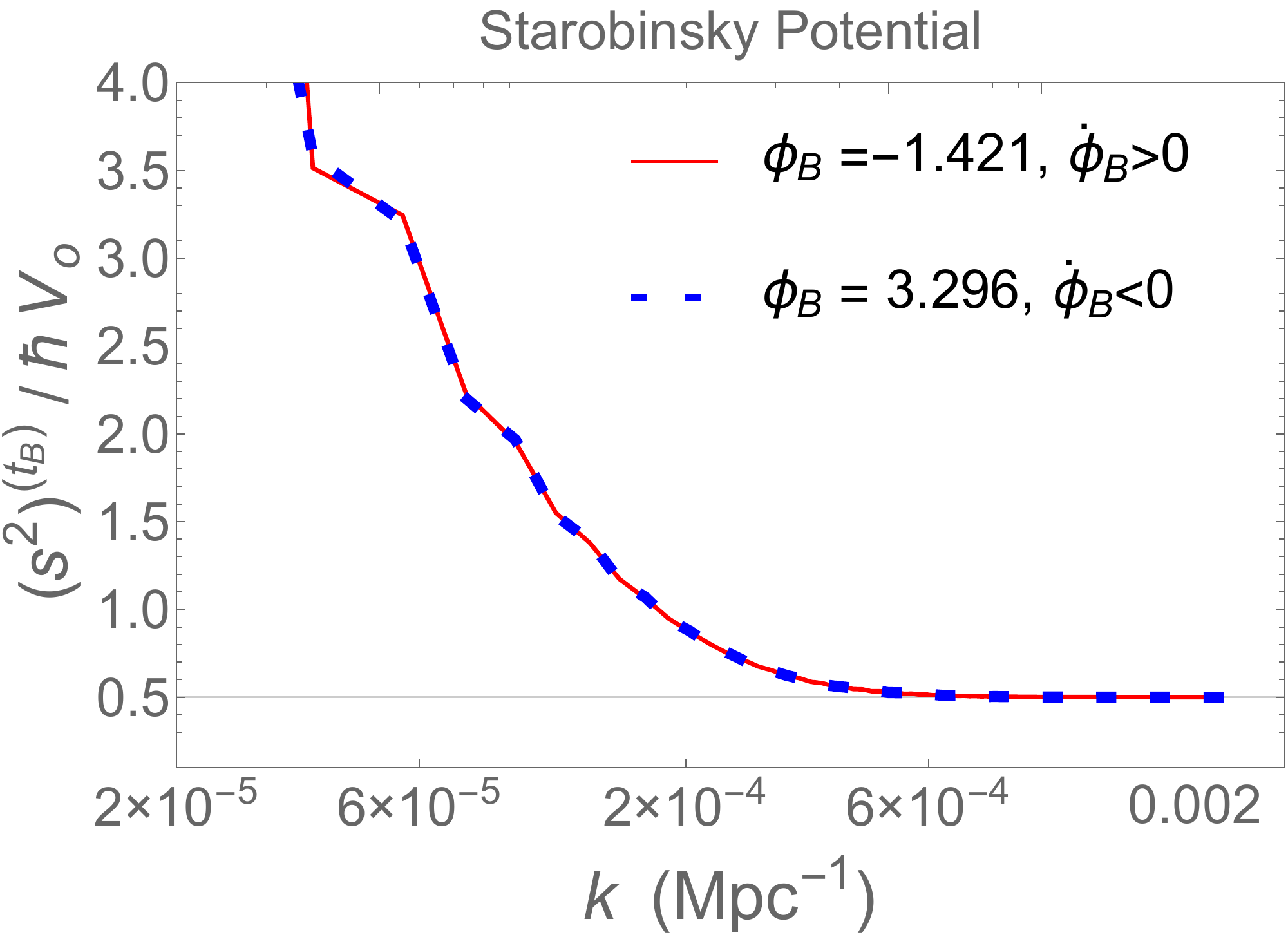}
\hskip0.2cm
 \ig[width=0.47\textwidth]{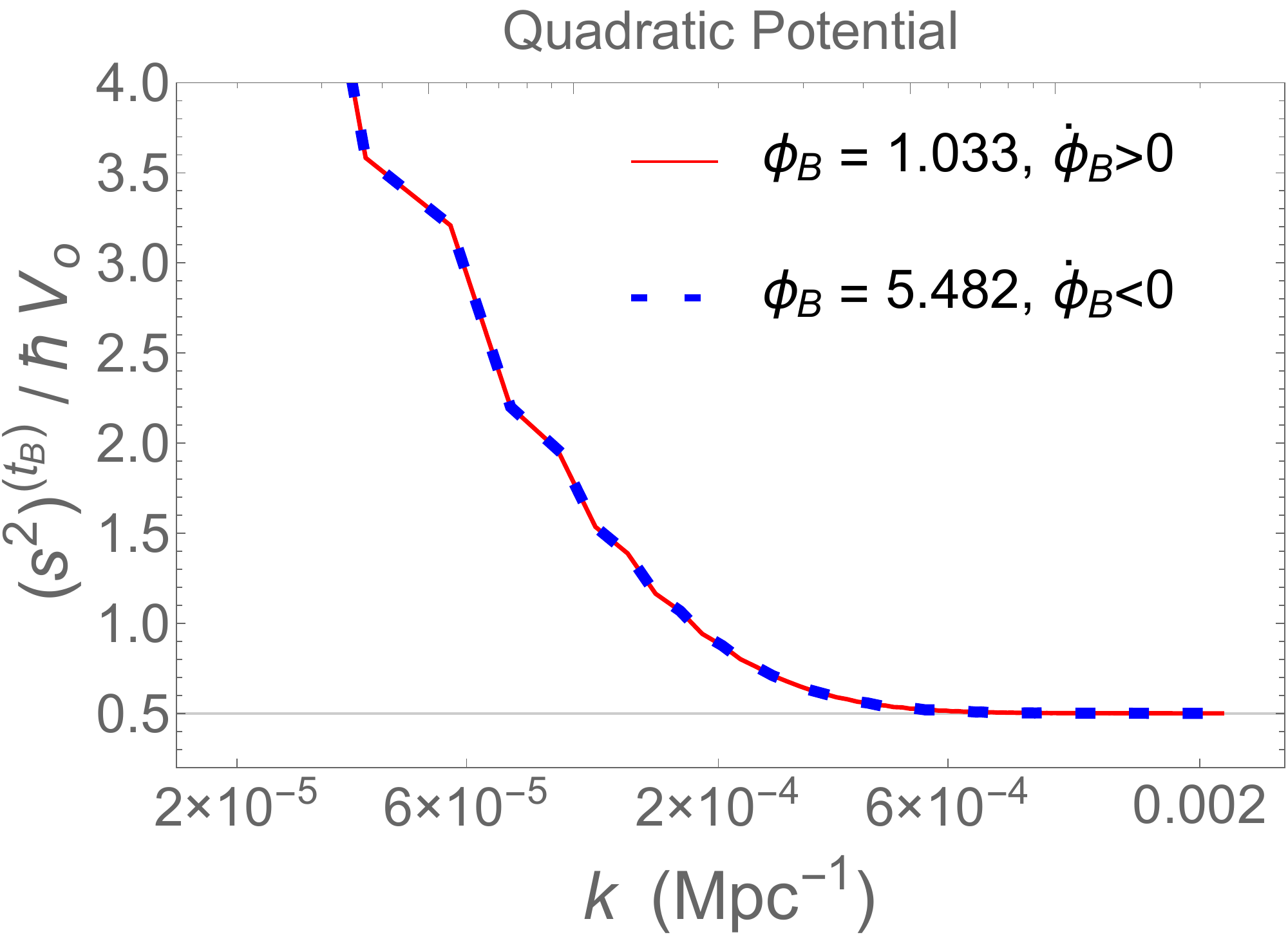}
\caption{The supremum $(s^{2})_{(t_{\rm B})}$ of $\sigma^{2}_{k}(t_{0}|t_{\rm B})$ obtained by varying $t_{0}$ in the interval $(-20 \spl, 20\spl)$ is plotted against the co-moving wave numbers $k$. Again, while for small $k$ the supremum is significantly larger than the absolute minimum ($\hbar/2V_{o}$) set by the uncertainty principle. But as $k$ approaches $k_{\star}$, the supremum rapidly approaches the minimum. }
\label{fig:sigSup}
\efig
We carried out simulations for a large number of values of $t_{0}$ in the interval $[-20\spl,\,\, 20\spl]$ around the bounce ($t_{\rm B}=0$) and verified this behavior. \fref{fig:sigSup} shows the supremum $(s^{2})_{(t_{\rm B})}$ of $\sigma^{2}_{k}(t_{0}|t_{\rm B})$  --i.e., of dispersions evaluated at the bounce in the states $|0_{t_{0}}\rangle$-- by varying $t_{0}$ in the above interval. The left panel shows the behavior for the Starobinsky potential and the right panel for the quadratic potential. Two features of these plots are important. First, while Principle 1 allows two background solutions for each potential, the functions $(s^{2})_{(t_{\rm B})}$ of $k$ are indistinguishable. Second, while $(s^{2})_{(t_{\rm B})}$ is significantly larger for large physical wavelengths (small $k$) than the absolute minimum dictated by Heisenberg uncertainty, it rapidly approaches this limit for small wavelengths. In particular, since $k_{\star} = 0.002\, {\rm Mpc}^{-1}$, departure from the minimum are visible only for modes with $k < k_{\star}$. It is these modes that are affected by the LQC Planck scale dynamics in states selected by the two principles of section \ref{s3}.
 
The next step is to consider the ball $\B_{(t_{\rm B})}$ of states $\t\psi$ in our collection $\mathcal{C}$ that are selected by the supremum $(s^{2})_{(t_{\rm B})}$: 
\be \B_{(t_{\rm B})} := \Big{\{}\,\,\t\psi\in \mathcal{C}\,\Big{|}\, \big[\, k\, \langle \t\psi |\, \f{1}{2} \big(\h{\Q}_{\vk}\h{\Q}_{\vk}^{\dag}+ \h{\Q}_{\vk}^{\dag} \h{\Q}_{\vk}\big)(t_{\rm B})\, |\t\psi \rangle +\f{1}{k}\, \langle \t\psi|\, \f{1}{2} \big(\h{\Pi}_{\vk}\h{\Pi}_{\vk}^{\dag}+ \h{\Pi}_{\vk}^{\dag} \h{\Pi}_{\vk}\big)(t_{\rm B})\, |\t\psi\rangle \,\big] \le (s^{2})_{(t_{\rm B})}\,\Big{\}}\, , \ee
and select a preferred state in this ball by minimizing the uncertainty in $\h\Q_{\vk}$ \emph{at the end of inflation} i.e. $t=t_{\rm end}$. 

This task is facilitated if one exploits the fact that all states in our collection $\mathcal{C}$ --and hence in the ball $\B_{(t_{\rm B})}$\,-- are related by a Bogoliubov transformation. Since the ball is tied to the bounce time, the state $|0_{t_{\rm B}}\rangle$ minimizes the dispersion for all $k$. Therefore it serves as a convenient reference state. The basis functions $\t{q}_{k}$ of a generic state $\t\psi \in \B_{(t_{\rm B})}$ are related to the basis functions $q^{(\rm B)}_{k}$ of $|0_{t_{\rm B}}\rangle$ via 
\be \t{q}_{k}(t) = \alpha_{k}\, q^{(\rm B)}_{k} + \beta_{k}^{\star}\, (q^{(\rm B)}_{k})^{\star} \quad {\rm with}\quad  \alpha_{k} =  (1+ r_{k})^{\f{1}{2}} \,\,\, {\rm and}\,\,\, \beta_{k} = r_{k}\, e^{i\theta_{k}} \, ,\ee
where $r_{k} \ge 0$ and $\theta_{k} \in (-\pi, \pi)$ are real functions. Now, one can show  that the dispersion in the state $\t\psi$ can be expressed in terms of these Bogoliubov coefficients and the dispersion in $|0_{t_{B}}\rangle$ (which equals $\hbar V_{o}/2$ for all $k$)\, \cite{ag2}: 
\ba \label{rsingma} \t\sigma_{k}^2(t_{\rm B}) &:=&  \big[\, k\, \langle \t\psi |\, \f{1}{2} \big(\h{\Q}_{\vk}\h{\Q}_{\vk}^{\dag}+ \h{\Q}_{\vk}^{\dag} \h{\Q}_{\vk}\big)(t_{\rm B})\, |\t\psi \rangle +\f{1}{k}\, \langle \t\psi|\, \f{1}{2} \big(\h{\Pi}_{\vk}\h{\Pi}_{\vk}^{\dag}+ \h{\Pi}_{\vk}^{\dag} \h{\Pi}_{\vk}\big)(t_{\rm B})\, |\t\psi\rangle \,\big]\nonumber\\
& =& (1 + 2r_{k}^{2})\, \f{\hbar}{2} V_{o} \, .\ea
Note that the right side refers only to $r_{k}$ and not to the angular parameter $\theta_{k}$ in the Bogoliubov transformation. Since $\t\psi$ is in the ball $\B_{(t_{\rm B})}$, Eq (\ref{rsingma}) implies that the Bogoliubov parameter $r_{k}$ is bounded above:
\be  r_{k}^{2}\,\,  \le\,\, \f{(s_{k}^{2})^{(t_{\rm B})}}{\hbar V_{o}}  - \f{1}{2}\, =:\, (r_{k}^{2})^{(t_{\rm B})}\, . \ee
It is this fact that enables us to find a unique $\t\psi$ that minimizes the uncertainty in $\h\Q_{\vk}$ at the end of inflation. Without this bound, there is no minimum: In the full collection $\mathcal{C}$ the uncertainty in $\h\Q_{\vk} (t_{\rm end})$ can be made arbitrarily small; it is the restriction to $\B_{(t_{\rm B})}$ that yields a unique $\t\psi$.

\bfig
 \hskip-0.63cm
 \ig[width=0.495\textwidth]{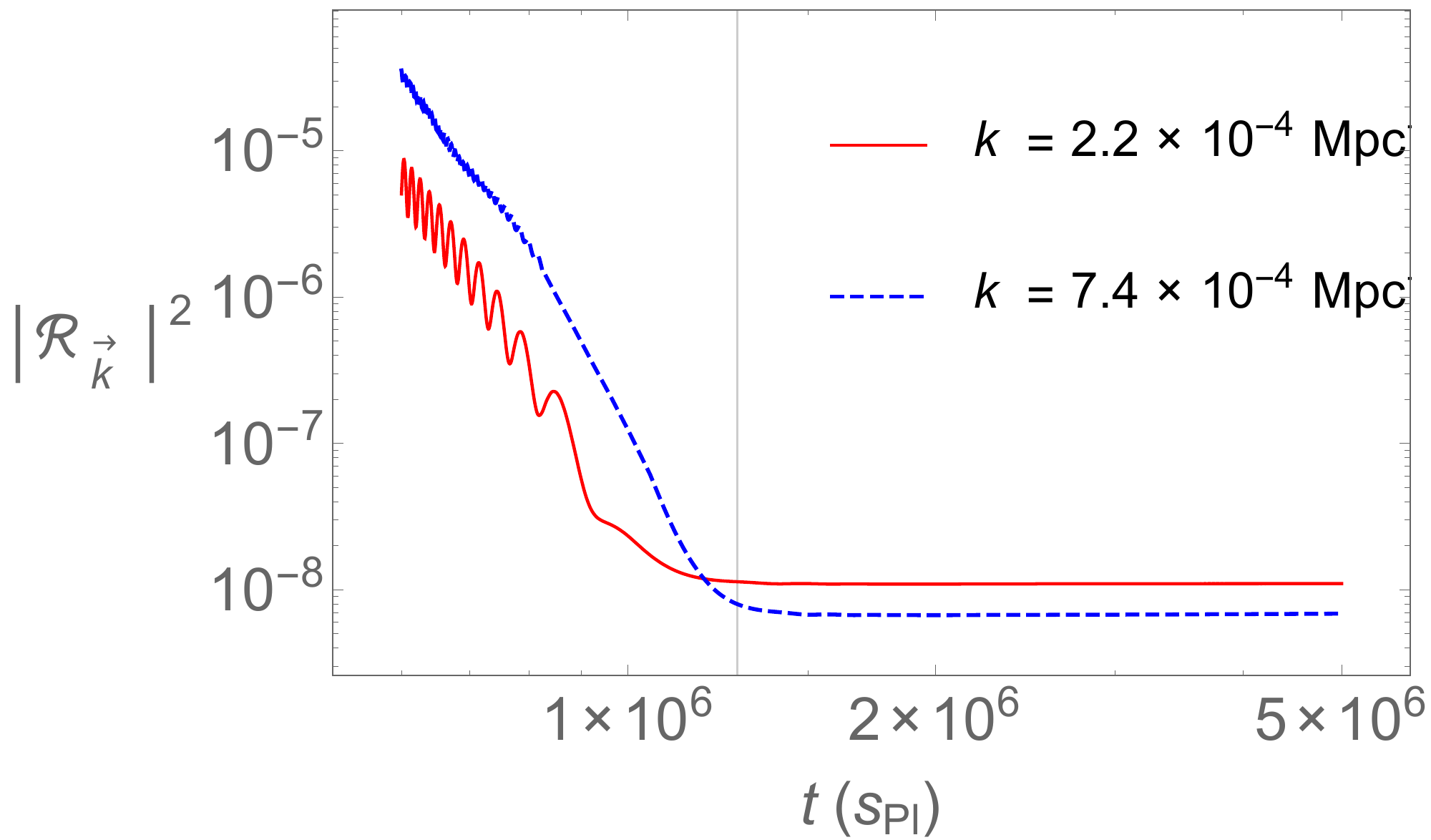}
 \hskip0.5cm
 \ig[width=0.425\textwidth]{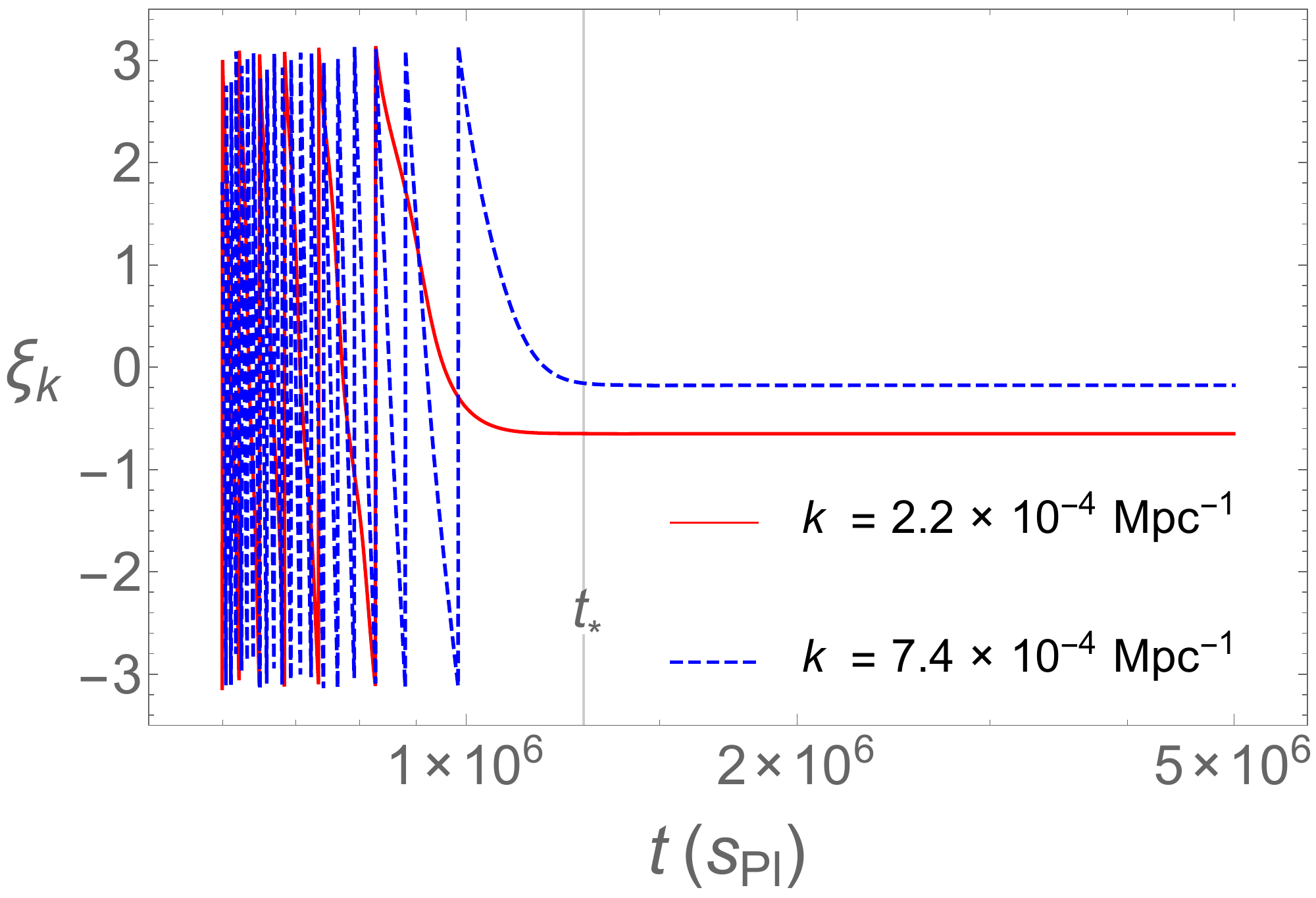}
\caption{Time evolution of the absolute value ({\it left panel}) and the phase ({\it right panel}) of the curvature perturbation $\R_{\vk}$ for two modes. Since $k < k_{\star}$, the LQC dynamics in the Planck regime has observable consequences for both these modes. $\R_{\vk}$ is defined by $\langle 0_{t_{\rm B}}|\, \f{1}{2} \big(\h{\R}_{\vk}\h{\R}_{\vk}^{\dag}+ \h{\R}_{\vk}^{\dag} \h{\R}_{\vk}\big)\, |0_{t_{\rm B}}\rangle = |\R_{\vk}|^{2} \exp {2 i \xi_{k}}$. The thin vertical line in both plot mark the time $t=t_\star$ when pivot mode $k_\star=0.002~\mpc$ crosses the Hubble radius during inflation. Before $t_\star$, when the modes are within the curvature radius in the pre-inflationary era, the modes oscillate. After they exit the curvature radius, the modes are frozen and their phases remain constant during the rest of the inflationary phase. This feature simplifies the numerics.}
\label{fig:modes}
\efig

Computationally, the task of minimization of the uncertainty in the scalar mode is simplified if one evolves $\h\Q_{\vec{k}}(t)$ from the Planck regime until a few \efolds after the mode $k$ exits the Hubble horizon during inflation and \emph{then} passes to the curvature perturbation $\t\R_{\vec{k}}$  
\be \h\R_{\vec{k}}(t) = \frac{H}{\dot{\phi}}\, \h\Q_{\vec{k}}(t) \ee
because $\h{\R}_{\vk}$ remains frozen to the future of this time. (While $\dot\phi$ vanishes in the pre-inflationary evolution when the inflaton undergoes a turn-around, it does not vanish during inflation. So this procedure yields a well-defined $\h\R_{\vk}(t)$.) Since $\h\R_{\vk}$ and $\h{Q}_{\vk}$ are related by a background quantity, a state $\t\psi$ in our ball minimizes the uncertainty in $\h\Q_{\vk}(t_{\rm end})$ if and only if it minimizes the uncertainty in  $\h\R_{\vk}(t_{\rm end})$. Now, at any time $t$, the uncertainty in the state $\t{\psi}$ is related to that in the state $|0_{t_{\rm B}}\rangle$ via:
\ba \label{relation}
 \langle \t\psi|\, |\, \f{1}{2} \big(\h{\R}_{\vk}\h{\R}_{\vk}^{\dag}+ \h{\R}_{\vk}^{\dag} \h{\R}_{\vk}\big)\,|\t\psi \rangle (t) = 
  \Big(\, 1+2r_k^2 +  2~r_k\, (1+r_k^2)^{\f{1}{2}}\,\, \cos\big(\theta_{k} + 2\xi_{k}(t)\big)\Big)\, \times\nonumber\\
  \langle 0_{t_{\rm B}}|\, \f{1}{2} \big(\h{\R}_{\vk}\h{\R}_{\vk}^{\dag}+ \h{\R}_{\vk}^{\dag} \h{\R}_{\vk}\big) (t)\,| 0_{t_{\rm B}} \rangle\, .
\ea
where $2\xi_{k}(t)$ is the phase in the evolution of $\langle 0_{t_{\rm B}} |\,\f{1}{2} \big(\h{\R}_{\vk}\h{\R}_{\vk}^{\dag}+ \h{\R}_{\vk}^{\dag} \h{\R}_{\vk}\big) \, | 0_{t_{\rm B}}\rangle$. The evolution of $\xi_{k}(t)$ is shown in the right panel of \fref{fig:modes} for two modes in the interesting range $k <k_{\star}$ where LQC corrections are non-negligible. Note in particular that this phase remains constant in time once the mode exits the Hubble radius during inflation. This feature simplifies computations considerably.

It follows immediately from (\ref{relation}) that the state $|\t\psi\rangle  \in \B_{(t_{\rm B})}$ which minimizes the uncertainty in $\h\R_{\vk}(t_{\rm end})$ is related to the reference state $|0_{t_{\B}} \rangle$ via  
\be
  \theta_k = \pi -  2 \xi_{k}(t_{\rm end})  \quad {\rm and} \quad 
  r_k \,= \, r_{k}^{(\rm B)}\, .
\label{eq:rkthetakbounce}
\ee
These values of $r_{k} \,{\rm and}\, \theta_{k}$ singles out a unique state from the ball $\B_{(t_{\rm B})}$ precisely because $r_{k}$ is bounded above in our ball.

In the last step, we replace $|0_{t_{\rm B}}\rangle$ with $|0_{t}\rangle$ with $t\in I$, find the state $\t\psi$ that minimizes the uncertainty in $\h{\R}_{\vk}(t_{end})$ for each value of $t \in I$. This provides a 1-parameter family of states $\t{\psi}(t)$. The state $\psi$ singled out by Principle 2 is the one that minimizes the uncertainty in $\h{\R}_{\vk}(t_{\rm end})$ among this 1-parameter family with $t$ in the compact interval $I$.

In summary, through a 2-step procedure we have obtained a unique Heisenberg vacuum state $\psi$ that has minimum uncertainty in $\h\R_{\vk}$   
 -- or, equivalently, $\h{Q}_{k}$-- at the end of inflation among all states contained in the ball $\B$ of Principle 2. In the next subsection we will use the background fields $\t{g}_{ab}, \phi$, provided by Principle 1, and the Heisenberg state $\psi$ of scalar perturbations $\h{\R}_{\vk}$, provided by Principle 2, to calculate the primordial scalar power spectra and discuss their observational consequences.\\

\emph{Remarks:}

(i) In this procedure to narrow down the choice of $\psi$, we started with states $|0_{t_{0}} \rangle$ that minimize the dispersion $\sigma_{k}^{2}$ of (\ref{dispersion}) at a time instant $t_{0}$. Had the geometry been stationary, adapted to the scale factor $a(t_{0})$\, --i.e., if the space-time metric had been $ds^2 = -dt^2 + a^2 (t_0) d\vec{x}^2$--\, then  $|0_{t_{0}} \rangle$ would have been the natural vacuum state. But of course the actual background metric $\t{g}_{ab}$ is time dependent, whence the dispersion does not remain constant, but increases as we evolve away from $t_{0}$. With the goal of using quantum dynamics in the full Planck regime to select the preferred state $\psi$, then, we were naturally led to vary $t_{0}$, consider the 1-parameter family $|0_{t}\rangle$ with $t$ in the Planck regime, and furthermore allow \emph{all} states $|\t{\psi}\rangle \in \B$ which are on the `same footing' as the $|0_{t}\rangle$ as far as the dispersions $\sigma_{k}^{2}$ in the Planck regime are concerned. Thus, it is our emphasis on the \emph{full} Planck regime that led us to a considerably wider class of states than the initial $|0_{t}\rangle$. The state $\psi \in \B$ which minimizes the uncertainty in $\h\Q$ at the end of inflation is in fact \emph{not} one of the $|0_{t}\rangle$ we began with: the fact that we focus on the \emph{full} Planck regime is crucially important for the final result. (For further discussion of the importance of considering the full ball $\B$ in place of the 1-parameter family of states $|0_{t}\rangle$, see \cite{ag2}.)

(ii) Because our aim was to choose initial conditions in the Planck epoch, we began by constructing the ball $\B$ using uncertainties in the Planck regime. However, since the background geometry has a pre-bounce phase, one may want to specify initial conditions in the distant past in the contracting pre-bounce part of space-time. Although this is not our viewpoint, we extended the interval $I$ to distant past in the contracting branch up to the time when the background energy density $\rho\lesssim 10^{-10}\rhopl$. The change in the final observable power spectrum remains less than $0.5\%$. Thus the final results are quite insensitive to this extension. This is because the wavelength of all observable modes remains well within the curvature radius in the extension.

\subsection{Primordial power spectrum and implications for the CMB}
\label{s4.3}

\bfig
 \ig[width=0.7\textwidth]{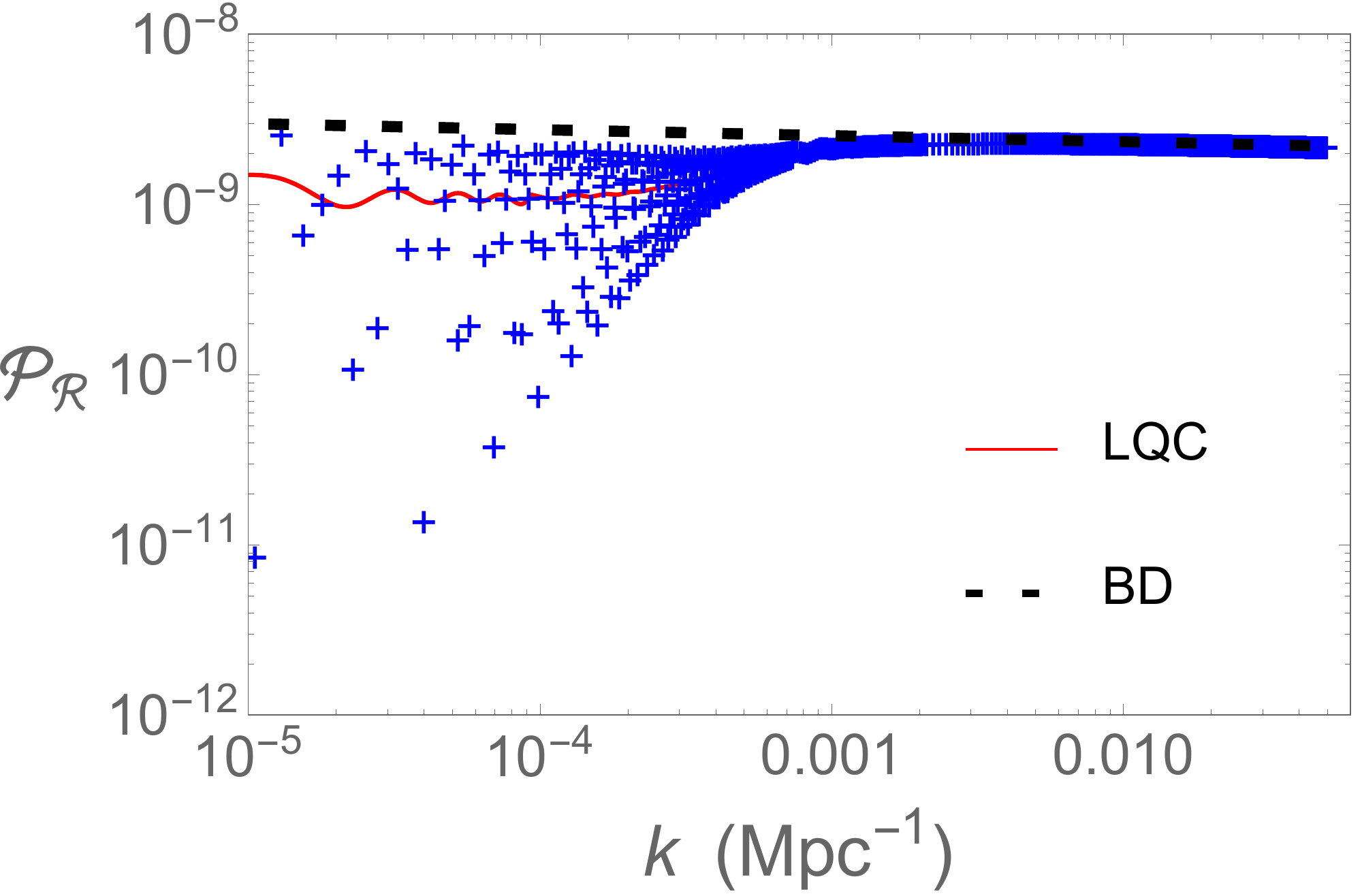}
\caption{The scalar power spectrum $\mathcal P_{\R}$ for the Starobinsky potential with $\phib=-1.421~\mpl$. The points show the power spectrum computed for each mode and the (red) solid curve shows {the average of these points is obtained by binning with the size $5\times10^{-6}~{\rm Mpc}^{-1}$.} The standard power spectrum assuming the BD vacuum at the onset of inflation is shown by the dashed curve. The LQC plot shows deficit of power at $k\lesssim0.002~\mpc$ while there is excellent agreement with the BD power spectrum at large $k$. This behavior is due to the fact that the modes with $k\lesssim0.002~\mpc$ have wavelengths comparable to the the radius of curvature \emph{in the Planck regime}.}
\label{fig:fullpower}
\efig
\fref{fig:fullpower} shows the power spectrum $\mathcal P_\R$ for curvature perturbations at the end of inflation for the Starobinsky potential using initial conditions of Eq. (\ref{eq:iniS}) at the bounce. The dashed curve shows the standard power spectrum using the BD vacuum at the onset of inflation. The scattered points show the LQC power spectrum computed for each mode for the preferred vacuum state obtained in section \ref{s4.2}. {The average of these points is shown with solid (red) curve, where the average is obtained by binning with size of $5\times10^{-6}~{\rm Mpc}^{-1}$.} It is evident that the LQC power spectrum has less power than the BD power spectrum at $k\lesssim0.002~\mpc$ while the two plots agree with each other at large $k$. As discussed before, this is because the two principles of section \ref{s3} turn out to imply that only the modes with $k\lesssim k_{\star} =0.002~\mpc$ have physical wavelengths comparable to the curvature radius during the Planck epoch, leading to a departure from the BD state at the onset of inflation around $t=t_{\star}$. Modes with $k \gg k_{\star}$ remain well within the curvature radius all the way from the bounce to the onset of inflation, whence the state $\psi$ chosen by Principle 2 turns out to be indistinguishable from the BD vacuum at $t=t_{\star}$. Finally, the power in each mode, shown by points in \fref{fig:fullpower}, oscillates very rapidly with $k$. But these oscillations are averaged out when one computes the angular temperature anisotropy in the CMB because each $\ell$ mode in the spherical decomposition receives contributions from a small interval of $k$-modes. Therefore what is relevant for observations is the averaged or binned power spectrum, shown by the solid (red) curve. From now on we will focus on this curve. 

\fref{fig:supsupPower} shows the ratio of the averaged power spectra with respect to the standard Bunch-Davies power spectrum in the standard inflationary scenario for the Starobinsky and quadratic potentials. The solid (red) and dashed (blue) curves correspond, respectively, to positive and negative $\phidb$. In all cases, there is suppression of power at $k\lesssim0.002~\mpc$ with respect to the standard BD prediction. Note that the form of this suppression is the same for both potentials and the two choices of $\phidb$. This property can be traced back to the fact that the underlying mechanism has its origin in the dynamics of perturbations in the well behaved Planck regime of LQC and Principle 1 implies that the bounce is highly kinetic energy dominated. Therefore the Planck scale dynamics is quite insensitive to the details of the potential. In this sense, the power suppression at
$k\lesssim k_{\star} =0.002~\mpc$ is a robust feature of the Planck scale dynamics of LQC, supplemented with the two principles that provided us with the preferred background fields $\t{g}_{ab}, \, \phi$ and the quantum state $\psi$ of scalar perturbations. 
\bfig
 \ig[width=0.47\textwidth]{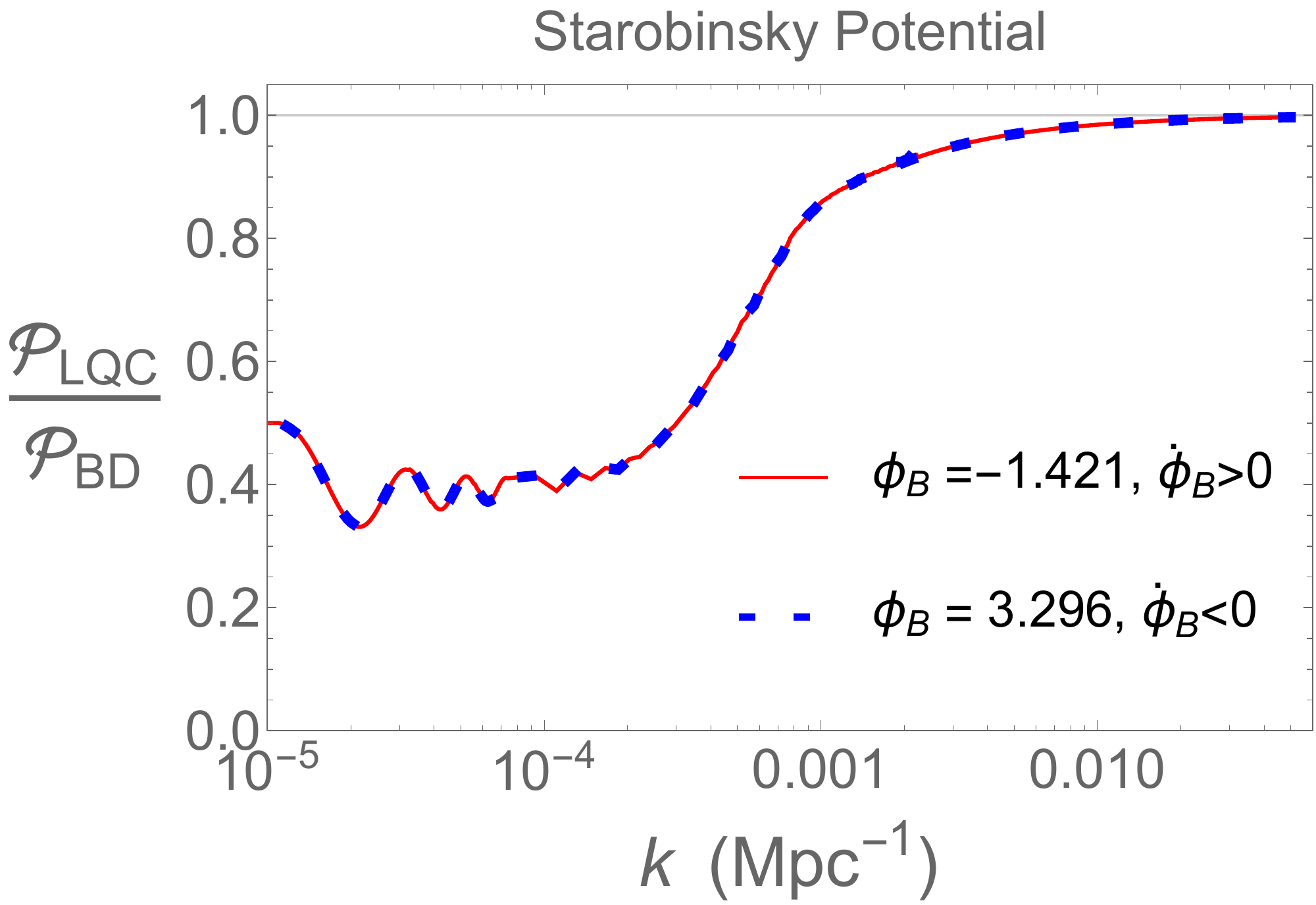}
 \hskip0.2cm
 \ig[width=0.47\textwidth]{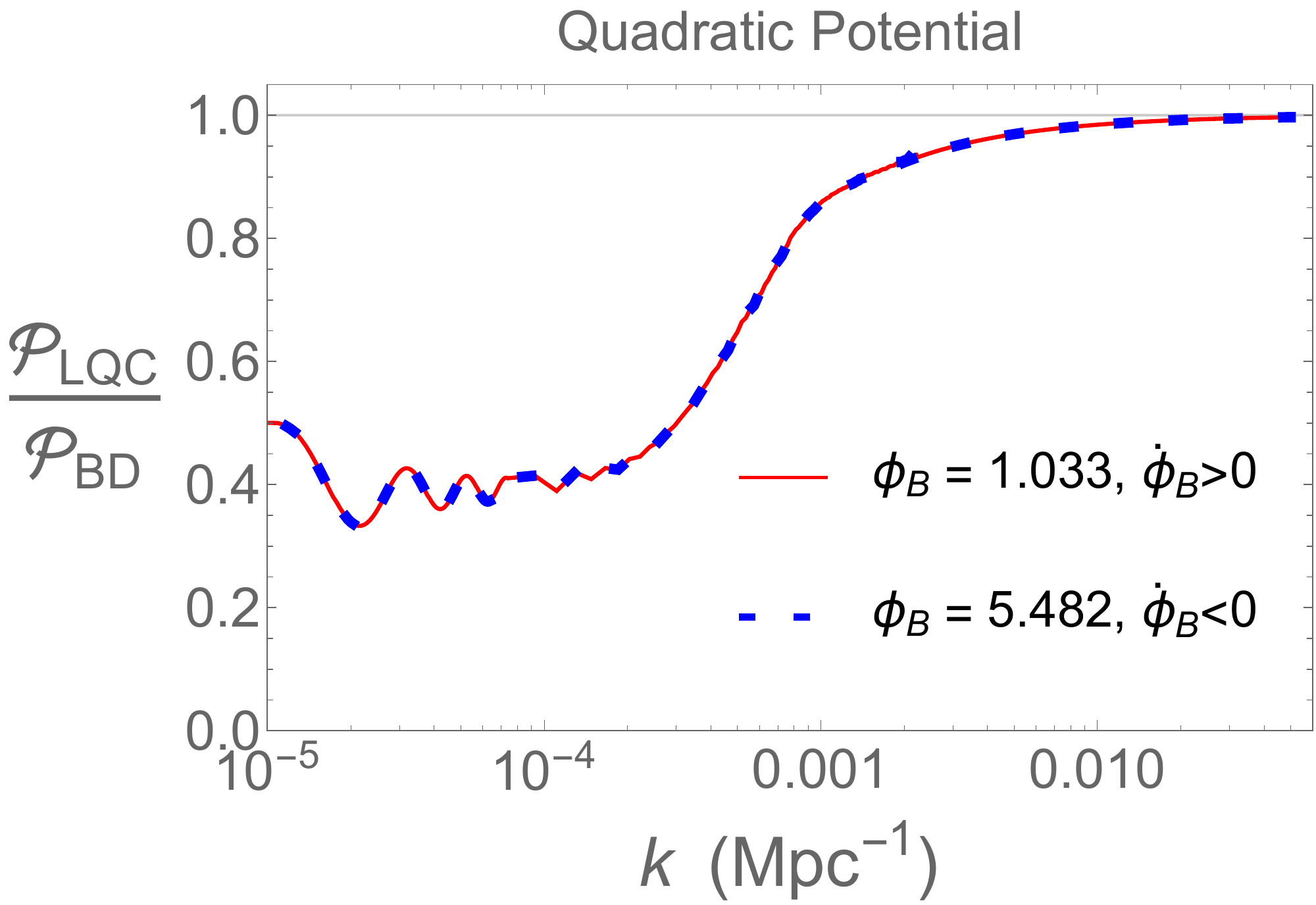}
\caption{Ratio of the averaged LQC curvature perturbation power spectrum ($\mathcal P_{\rm LQC}$) to that in the standard inflationary scenario ($\mathcal P_{\rm BD}$) for the Starobinsky potential \emph{(left panel)} and the quadratic potential \emph{(right panel)}, with inflaton initial conditions \eref{eq:iniS}--(\ref{eq:iniQ2}) selected Principle 1. For both potentials, and for both initial conditions at $t_{\rm B}$ selected by Principle 1 for each potential, there is suppression of power at $k\lesssim k_{\star} =0.002~\mpc$.}
\label{fig:supsupPower}
\efig

As is well-known, the recent data from Planck and WMAP reveal that there is suppression of power in the temperature anisotropy spectrum $\celltt$ at angular scales $\ell\lesssim 30$, with statistical significance of approximately $2$-$3\sigma$ \cite{planck15xx,Copi:2013cya,Schwarz:2015cma,Hunt:2015iua}. The key question then is whether the theoretical power spectra of \fref{fig:supsupPower} generate a suppression of power in $\celltt$ consistent with the data. In order for this to happen two
non-trivial requirements need to be fulfilled: The scale of the quantum gravity corrections in the power spectrum ought to correspond to $\ell \lesssim 30$ and the amount of suppression has to be compatible with the data. As discussed below (in Remark (iii)) these requirements are met as a direct consequence of the two principles. These are non-trivial tests for the theoretical framework proposed here. 

In order to compute the predicted temperature anisotropy $\celltt$ in the CMB we need to solve the Boltzmann equations which govern the evolution of density perturbations in the post-inflationary era. The primordial power spectrum (\fref{fig:supsupPower}) provides the initial conditions for the Boltzmann equations. The corresponding solution was obtained  using the publically available code \camb \cite{camb}. The resulting $\celltt$ is shown in \fref{fig:celltt}. It evident that in LQC there is power suppression for $\ell\lesssim30$ compared to the standard inflationary prediction. Moreover, the LQC predictions agree better with the Planck data than the standard inflationary scenario with
\be
\Delta\chi^2 := \chi^2_{\rm BD} - \chi^2_{\rm LQC}=3.15~.
\label{eq:chisq1}
\ee
Note that, \emph{we have not added any new parameter}. The suppression in the primordial power also affects the E-mode polarization spectrum characterized by the EE and TE correlations in the CMB. As shown in \fref{fig:celltt}, we find that $\cellee$ and $\celltt$ also show suppression at the scale $\ell\lesssim30$. These predictions will be tested over the coming year when the PLANCK team releases the data for TE and EE power spectra for $\ell <30$. If they are in clear conflict with the data, at least one of the two principles introduced in section \ref{s3} will have to be abandoned. Thus, there is synergistic interplay between fundamental theory and observations. \\

\bfig
 \ig[width=0.9\textwidth, height=3.5in]{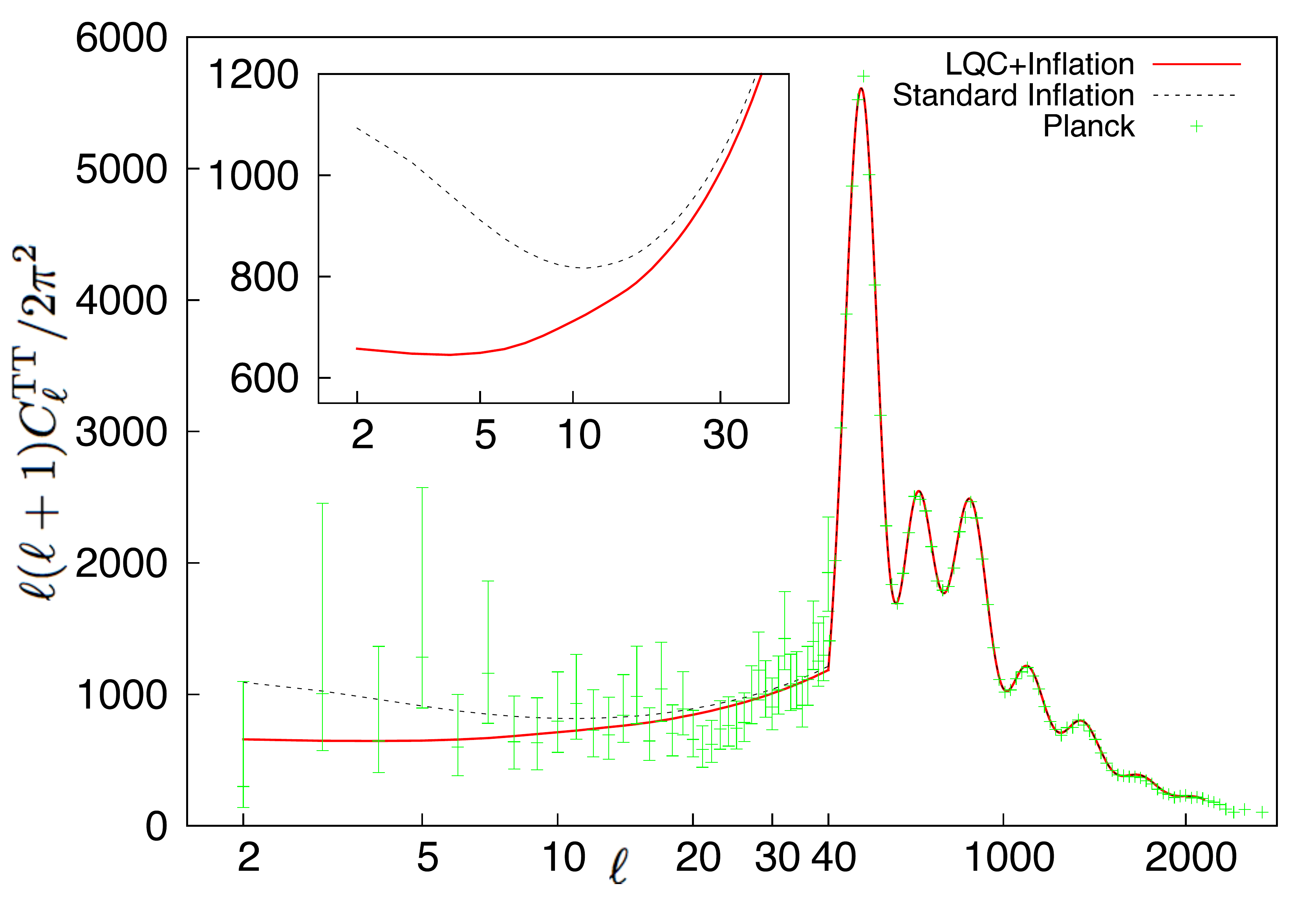}
\caption{Temperature anisotropy spectrum $\celltt$ in the CMB. The LQC prediction for the primordial power spectrum of \fref{fig:supsupPower}  translates to the solid (red) curve. The results for the standard inflation with Bunch-Davies initial conditions is shown by black dashed curve and the PLANCK data is shown by green points with errorbars. Clearly, the LQC curve shows suppression for $\ell\lesssim30$. At large $\ell$, both LQC and the standard spectra agree well with the PLANCK data. The LQC curve provides a better fit to the Planck data with $\Delta \chi^2 = 3.15$ (see \eref{eq:chisq1}). We have used logarithmic scale on the x-axis for $\ell >40$ because the non-trivial effects occur for $\ell <30$.
}
\label{fig:celltt}
\efig

\bfig
 \ig[width=0.47\textwidth]{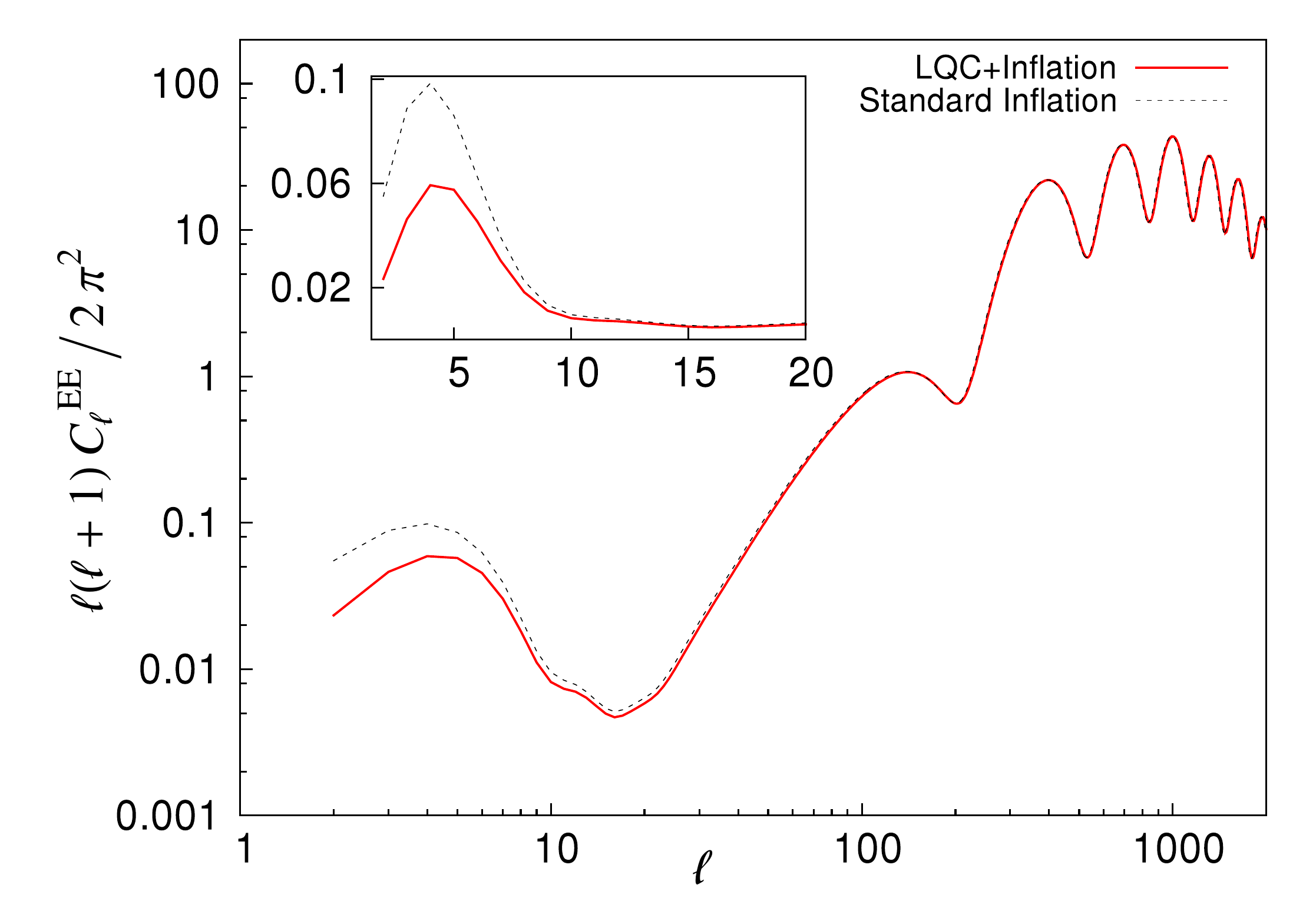}
 \hskip0.2cm
 \ig[width=0.47\textwidth]{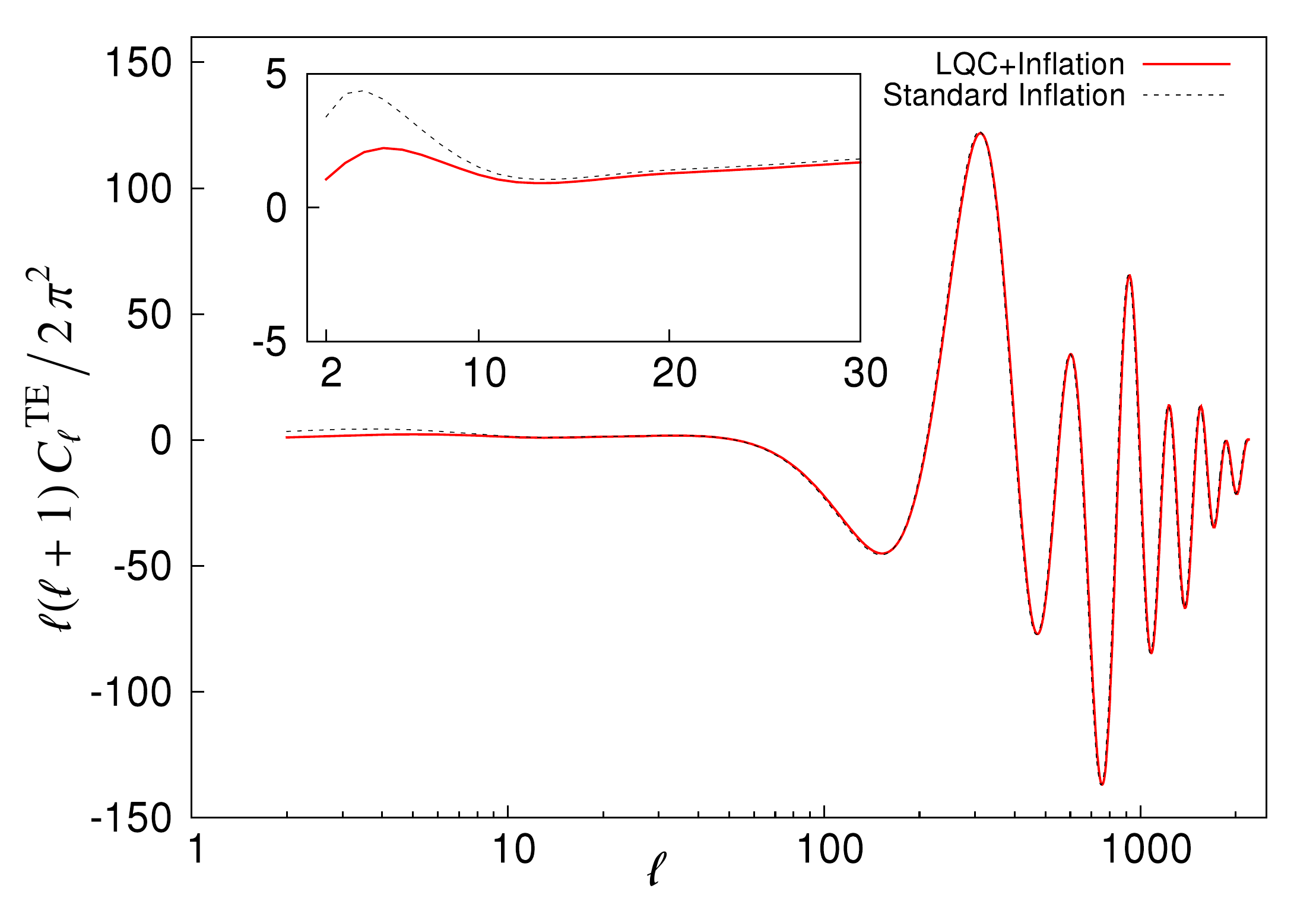}
\caption{Predictions for E-mode polarization (left panel) and T-E cross correlation (right panel) anisotropy spectrum in the CMB. Again, the LQC predictions are shown by solid (red) curve and the predictions of the standard inflationary scenario are shown by black (dashed) curve. The LQC curves show suppression at the angular scales of $\ell\lesssim30$.}
\label{fig:cellee}
\efig

{\it Remarks:}
 
 
(i) The power spectrum predicted by standard inflation has a red tilt because the Hubble parameter decreases during the slow-roll. In GR, there could be a very long slow-roll phase prior to $t=t_{\star}$. If this occurs, the prediction of red tilt would hold for modes with $k \ll k_{\star}$. Although this part of the growth in the power spectrum refers to modes way beyond the observable range, it has some conceptually interesting ramifications. In particular, it served as the primary motivation behind the first discussions of multiverse \cite{linde}. As \fref{fig:fullpower} shows, the situation is quite different in LQC. There is a turn-around in the power spectrum near $k=k_{\star}$ and we now have $\mathcal{P}_{\R}(k) < \mathcal{P}_{\R}(k_{\star})$ fro $k <k_{\star}$. Since the power does \emph{not} continue to grow for $k < k_{\star}$, we are no longer driven to multiverse scenarios. Consequently concerns that have been raised about plateau type potentials stemming from the inevitability of multiverses in standard inflation (e.g. in \cite{steinhardt}) are not applicable to the LQC analysis, supplemented by the two Principles.

(ii) Departure from predictions of standard inflation can be traced back directly to the evolution of modes in the Planck regime: 
As the right panel of \fref{fig:rcurv} shows, \emph{all} observable modes have wavelengths much less than the curvature radius between the time they exit the Planck regime and the onset of slow roll inflation.
On the other hand in the literature there is discussion of robustness of inflationary perturbation spectrum with respect to the Planck scale physics (see especially \cite{starobinsky}). How do we reconcile those results with ours? The focus of those investigations was on the changes in  dispersion relations when the modes become trans-Planckian because of \emph{ad hoc} modifications of special and general relativity that lead to a violation of Lorentz invariance. It was shown that for a large class of modifications for which the WKB approximation holds, predictions of standard inflation remain unaffected. Our analysis is entirely different: Local Lorentz invariance is not violated; WKB approximation is not used, and the departure from standard inflation is not due to modification of dispersion relations. Rather, as discussed in section \ref{s4.1}, it is because of a competition between the physical wave length of modes and the radius of curvature in a specific LQC quantum geometry picked out by Principle 1.

(iii) It is highly non-trivial that the power spectrum agrees with that of standard inflation for $\ell \gtrsim 30$ but is suppressed for $\ell \lesssim 30$. What is responsible for the scale $\ell \approx 30$? And why is there suppression of power rather than enhancement? The answer to the first question lies in Principle 1, and, to the second, in Principle 2. 

If we were to drop Principle 1, we would obtain a very large class of background solutions $(\t{g}_{ab},\, \phi)$ all of which lead to the Cauchy data $(\phi_{\star}, \dot\phi_{\star}, a_{\star}, H_{\star})$ at $t=t_{\star}$ that are compatible with the constraints\, --discussed in section \ref{s2.2}--\, imposed by the PLANCK data within error bars. Therefore, a priori they are all viable past extensions of the `correct' space-time geometry to the future of $t=t_{\star}$. The number $N_{\rm{B}\,-\,\star}$ of \efolds between $t_{\rm B}$ and $t_{\star}$ varies from one space-time to another in this class. If we chose a solution with large $N_{\rm{B}\,-\,\star}$, we would find no power suppression at all relative to standard inflation in any of the the observable modes. For small $N_{\rm{B}\,-\,\star}$ there would be power suppression for modes with $\ell \gg 30$ which is ruled out by observations (even if we used the $\psi$ given by Principle 2). It is the value of $N_{\rm{B}\,-\,\star}$\, dictated by Principle 1 (16.83  for the Starobinsky potential, and 15.00 for the quadratic) and pre-inflationary LQC dynamics that imply there would be departures from predictions of standard inflation precisely for $\ell \lesssim 30$. 

But a priori these departures can be in either direction: power could suppressed or enhanced. Indeed, our ball $\B$ does contain states\, --e.g., $|0_{t_{0}}\rangle$-- that lead to an  enhancement of power at the end of inflation (again for $\ell \lesssim 30$, assuming we used Principle 1 to select $\t{g}_{ab}$). The second part of Principle 2  --the `final condition' which minimizes the uncertainty in $\h\Q$ at the end of inflation-- suppresses power relative to $|0_{t_{0}}\rangle$. But even with this condition, a priori, power may not have been suppressed 
relative to standard inflation, or it could have been suppressed so much as to be ruled out by the PLANCK observations. Therefore, the fact that the two principles together provide a suppression precisely for modes with $\ell \lesssim 30$, and that the result is a better fit to observations than standard inflation for all observable modes is highly non-trivial, especially given that we start dynamics in the Planck regime.

(iv) Principle 1 sets the scale at which modes get excited during the Planck era \emph{irrespective of the potential}: $k \lesssim e^{-N_{\rm B-0}}/ (\sqrt{6} r_{\rm B})$.
As remarked earlier, it is this evolution in the Planck regime that causes LQC predictions to depart from those of standard inflation. (Here $N_{\rm B-0}$ is the number of \efolds from the bounce to today and $r_{\rm B}$ is the universal radius of curvature at the bounce.) But in general the effect of these excitations on the final power spectrum would depend on details of dynamics, and hence the potential. Why are the results of \fref{fig:supsupPower} for the Starobinsky and quadratic potentials essentially indistinguishable? It is because for both these potentials
the bounce turned out to be kinetic energy dominated. Therefore the evolution of modes in the Planck regime is insensitive to the differences between these potentials. 
 
(v) Suppression in $\celltt$ can also be generated by late time phenomena such as the integrated Sachs-Wolfe effect \cite{isw1,isw2}. However, such effects do not produce any suppression in $\cellee$ and produce a very small suppression in $\cellte$. Therefore, measuring $\cellee$ can distinguish between the late time effects and primordial mechanisms of generating the power suppression at large scales. If the future CMB observations reveal a suppression in both TE and EE polarization correlations, it would be a strong indication that the suppression is due to primordial physics rather than an effect of late time evolution of the universe.

(vi) One can calculate the tensor power spectrum in the LQC paradigm provided by the two principles. For each of the two potentials we analyzed in detail, the ratio $r$ of tensor to scalar power is the same as in standard inflation because the differences from standard inflation can be traced back to the dynamics of modes within the Planck regime and, for all observable modes, the potential $\t{\mathcal{U}}$ that enters in the evolution equation of scalar modes is negligible in the Planck regime. Thus, $r$ is much lower for the Starobinsky potential than for the quadratic and the difference can be traced entirely to dynamics in the GR epoch.
 
\subsection{\cosmomc analysis and robustness of Principle 1} \label{s4.4}

In this section, \ref{s4}, we started with the two theoretical principles in conjunction with certain observational facts and derived predictions for various power spectra for the Starobinsky and quadratic potentials. Principle 1 made a crucial use of the area eigenvalues of the LQG quantum geometry. Specifically, we started with the largest sphere $\mathbb{S}_{\rm CMB}$ within the cosmological horizon at $t_{\rm CMB}$ and asked that when evolved back in time to the bounce surface $t_{\rm B}$, it yield the `elementary' 2-sphere $\mathbb{S}^{2}_{\rm B}$ with area $6 \Delta\,\, \lp^{2}$ in the underlying LQG quantum geometry. 
In the geometry defined by the dressed effective metric $\t{g}_{ab}$,\, $\mathbb{S}^{2}_{\rm B}$ has radius $\mathring\RB \approx 1.57~\lp$. This value emerged from purely theoretical considerations involving LQG quantum geometry and then played a key role in our subsequent analysis of sections \ref{s4.1}\, -\,\ref{s4.3}. We can now turn the issue around and ask what \emph{observations tell us about the radius of the 2-sphere at the bounce.} Suppose the metric $\t{g}_{ab}$ were such that when we evolve the 2-sphere $\mathbb{S}_{\rm CMB}$ back in time, it has a radius $\RB \not= \mathring\RB$ at the bounce surface. Setting the theoretical considerations aside, do observations favor a metric $\t{g}_{ab}$ for which the value of $\RB$ is very different from $\mathring\RB$? In other words, can we test the `goodness' of Principle 1 from observational perspective? 

\bfig
 \ig[width=0.7\textwidth]{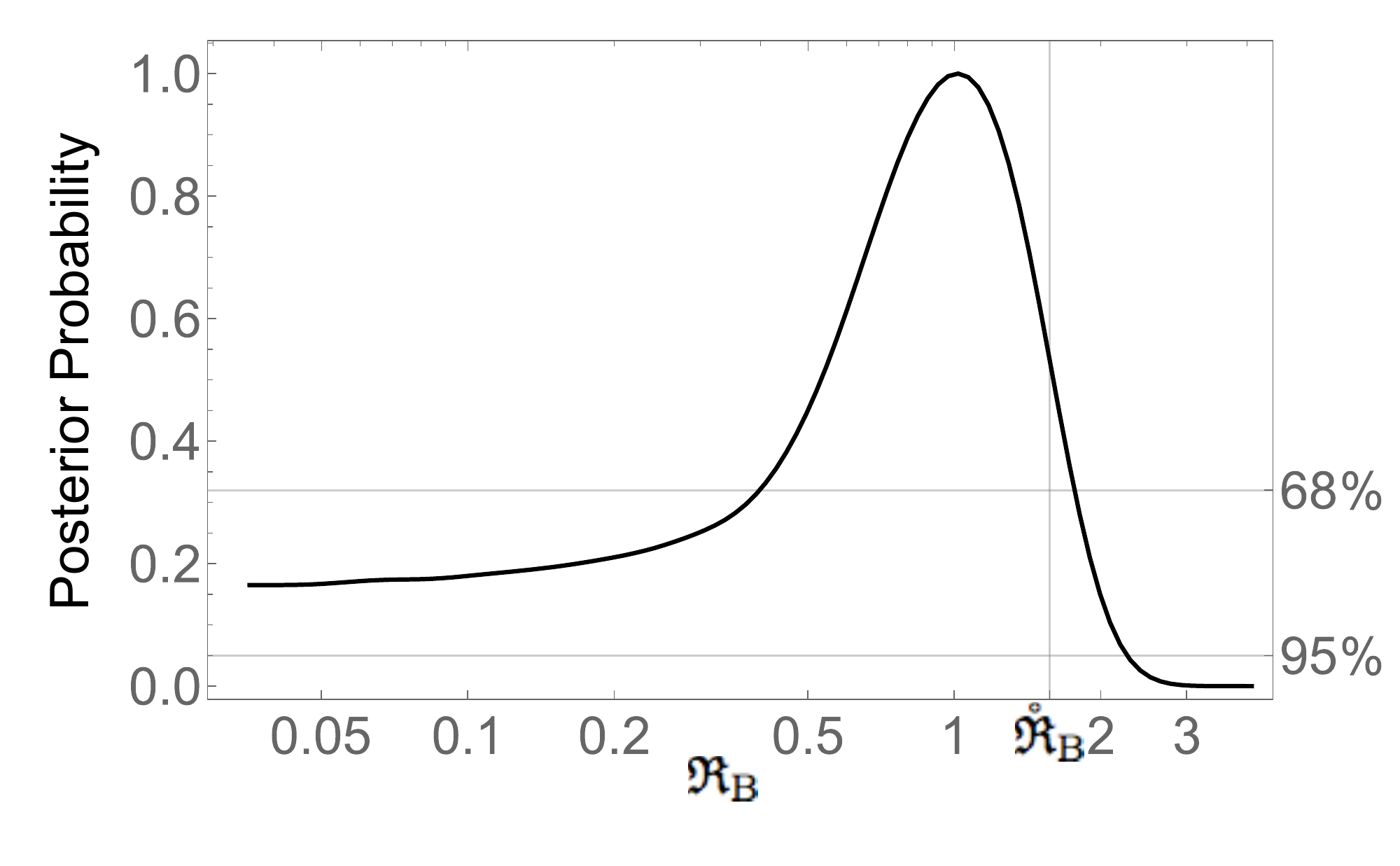}
\caption{One dimensional posterior probability distribution for $\RB$. 
The vertical line denotes the value of $\RB=\mathring{\RB} = 1.57~\lp$ selected by Principle 1 in \sref{s3}. Clearly, $\mathring{\RB}$ is within $68\%$ confidence level of the peak value. The distribution has a sharp cutoff for large values of $\RB$ which would have led to too few \efolds between the bounce and onset of slow-roll, resulting in a large suppression of power also at scales $\ell > 30$. On the other hand the distribution flattens out for small values of $\RB$. In this regime, there are too many \efolds from the bounce to the onset of the slow-roll phase and all the Planck scale corrections to the power spectrum is shifted towards super-horizon modes. Since these are not among observable modes, PLANCK data does not constrain lower values of $\RB$ as tightly.}
\label{fig:RBlikelihood}
\efig

To answer this purely phenomenological question, we dropped principle 1, treated $\RB$ as a free parameter, calculated the LQC predictions for $\celltt$ using this general value of $\RB$, and arrived at the value that fits the PLANCK data the best. At the end we compared this best fit value with $\mathring\RB=1.57~\lp$. (To have a meaningful evaluation of Principle 1  we retained Principle 2 to select $\psi$ in these calculations.)  To carry out this task, we performed Markov-Chain-Monte-Carlo (MCMC) parameter estimation using the publically available code \cosmomc \cite{cosmomc}, using a flat prior on $\RB$. \fref{fig:RBlikelihood} shows the one-dimensional marginalized posterior distribution of $\RB$. Remarkably, PLANCK data does prefer a small range of values. There is a peak in the distribution at:
\be
  \RB^{\rm (peak)} = 1.02~\lp, \quad {\rm for~which}\quad \Delta\chi^2 =
3.78.
\label{eq:chisq2}
\ee
On the right side of the peak, there is sharp cut-off for large values of $\RB$ while on the left side the distribution flattens out for smaller values. This puts the following constraints on the parameter $\RB$:
\ba
  0.40~\lp< \RB &<& 1.77~\lp \quad {\rm (at~68\%~confidence~level)}.
\ea

To summarize, the analysis reveals three points that could not be anticipated a priori: 

\noindent (i) There exists a peak in the posterior distribution of $\RB$. \\
(ii) The value $\mathring\RB=1.57~\lp$ of Principle 1 is well within the $68\%$ confidence level of the best fit value.\\
(iii) The value $\RB =3$ --which would replace the elementary 2-sphere with one whose area is just $\sim\, 5$ times larger at the bounce-- is ruled out at 95\% confidence level!\\
As pointed out in section \ref{s2.1}, in our analysis we worked with only the average values reported by PLANCK and did not keep track of the 68\% confidence level error bars in the data (\ref{parameters}). Therefore, (ii) brings out a rather remarkable confluence between theory and observations: the value $\RB = \mathring\RB$ arrived at from purely theoretical considerations of quantum geometry in the Planck regime is within the 68\% confidence level of the best fit value produced by the data. Furthermore, even slightly larger values are strongly disfavored. Interestingly (as pointed out in Remark 3 at the end of section \ref{s3.1}) if the LQG quantum geometry at the bounce surface were constructed using a graph adapted to the more commonly used  tetrahedral simplicial decomposition, rather than a cubical one used here, the theoretical value of $\mathring\RB$ would have been $1.28~\lp$, even closer to the peak value. \\ 

{\it Remark:} Recall that in standard cosmology the primordial scalar power spectrum is usually assumed to be given by a power law,  parameterized by two parameters $A_s$ and $n_s$, the amplitude of the power spectrum at the pivot scale $k_\star$ and the spectral index which is usually assumed to be constant. Deviation from this ansatz is often modeled by introducing a new parameter, the running in the spectral index, 
\be n_{\rm run} = \f{{\dd}n_s}{{\dd}\ln(k)}\, ,\ee 
assumed to be a constant, and using the modified ansatz \cite{planck15xx}
\be \label{modified}
    \mathcal P_{\mathcal R}(k) = A_s\,\, \Big(\f{k}{k_\star}\Big)^
                 {\big(n_s-1 + \f{1}{2} n_{\rm run}\ln(k/k_\star)\big)} 
                 \,.
\ee
In this ansatz the spectral itself index is no longer constant but allowed to vary with $k$. PLANCK results provide strong constraints on this running: $n_{\rm run}=-0.0057\pm0.0071$, so it is consistent to set $n_{\rm run}$ to zero \cite{planck15xx}. 

Now, in the LQC power spectrum obtained in section \ref{s4.3}, the red tilt does not persist for $k <k_{\star}$. Since the LQC $n_{s}$ is no longer constant on the full range of observable modes, it is natural to revisit the issue of running of the spectral index and inquire if the introduction of a non-zero $n_{\rm run}$ is statistically degenerate with LQC corrections. That is, is there a statistical correlation between introducing non-zero $n_{\rm run}$ and LQC corrections? In order to investigate this issue we further modified the ansatz to
\be
    \mathcal P_{\mathcal R}(k) = A_s\,\, \Big(\f{k}{k_\star}\Big)^
                 {\big(n_s-1 + \f{1}{2} n_{\rm run}\ln(k/k_\star)\big)} 
                 \,\,\,\, \f{\mathcal P_{\mathcal R\rm }^{\rm LQC}}                 {\mathcal P_{\mathcal R}^{\rm BD}}(k)\, ,
\ee
where ${\mathcal P_{\mathcal R\rm }^{\rm LQC}}/{\mathcal P_{\mathcal R}^{\rm BD}}$ is the ratio of the LQC power spectrum to the standard BD power spectrum at the end of inflation (shown in \fref{fig:supsupPower}). We performed an MCMC parameter estimation by keeping by keeping $A_s$ and $n_s$ fixed to their best fit values
(\eref{eq:asns}) and varying $n_{\rm run}$ and $\RB$. We worked with $\ln(\RB/\mathring{\RB})$ rather than $\RB$ in order to explore large parameter space. To test for possible degeneracy between $n_{\rm run}$ and $\ln(\RB/\mathring{\RB})$, one has to calculate the correlation coefficient between the parameter $n_{\rm run}$ and $\RB$. It can range between $-1$ and $1$, the value $-1$ showing complete anti-correlation and $1$ showing complete correlation. Our analysis shows that the correlation coefficient between $\ln(\RB/\mathring{\RB})$ and $n_{\rm run}$ is 0.4582, whence the statistical correlation is weak. Therefore, $n_{\rm run}$ and $\ln(\RB/\mathring{\RB})$ are not degenerate. The suppression of power at scales $k<k_\star$ in LQC cannot be accounted for by introducing a non-zero $n_{\rm run}$. (The reason why the correlation coefficient is not much smaller is that, since for $\ell >30$ the LQC corrections are small and rapidly go to zero with increasing $\ell$, there is some correlation with a small non-zero $n_{\rm run}$ allowed by the PLANCK data and LQC corrections at the small angular scales $\ell >30$.)


\section{Summary and outlook}
\label{s5}

The anomalies discovered by the WMAP and PLANCK teams at large angular scales are only at a 2-3 $\sigma$ level and could well be artifacts of cosmic variance, or, as we pointed out in section \ref{s4.3}, they could arise from a feature of the late time evolution of the universe such as the integrated Sachs-Wolf effect \cite{isw1,isw2}. But, as the PLANCK team pointed out in \cite{planckxvi}, the anomalies could also have a primordial origin, i.e., could be  \emph{``the visible traces of fundamental physical processes occurring in the early universe''}. 
Given the dearth of observational data at the Planck scale, it behooves every serious approach to quantum gravity to seize on this opportunity by examining the possibility seriously. A number of investigations in LQC have done just that (see, e.g., \cite{aan1,aan2,aan3,agullomorris,agulloassym}). 

As explained in section \ref{s1}, in the Planck regime, one needs a wave function $\Psi_{o}$ to describe the background FLRW quantum geometry in which all physical observable remain finite. The `trans-Planckian issues' \cite{brandentrans} can be faced head-on if one replaces the classical FLRW metric with this $\Psi_{o}$. In particular, then, the scalar and tensor perturbations $\h{Q}, \h{\T}^{I}$ propagate on the quantum geometry $\Psi_{o}$ in LQC. Now, it turns out that this propagation can be described \emph{exactly} by replacing $\Psi_{o}$ with a dressed, effective metric $\t{g}_{ab}$ which incorporates all the quantum fluctuations that $\h{Q}, \h{\T}^{I}$ experience during evolution \cite{akl,aan2,aan3}. This is the strategy used in the LQC papers mentioned above.  However, in these investigations, $\t{g}_{ab}$, was left unspecified, and plausible choices of the Heisenberg state $\psi$ of perturbations were made. 

In this paper, we narrowed down the choices of $\t{g}_{ab}$ and $\psi$ severely by introducing two principles in section \ref{s3}, and worked out the consequences for the T-T, T-E and E-E correlation functions for scalar perturbations in section \ref{s4}. The principles restrict the initial conditions in the deep Planck regime when dynamics is very different from that on de Sitter space-time, while in standard inflation initial conditions are chosen a few \efolds before the onset of slow roll, when GR is an excellent approximation and dynamics is well-modeled by the de Sitter metric. Yet we found that at small angular scales, $\ell \gtrsim 30$, the predicted correlation functions are indistinguishable from those of standard inflation, and hence in excellent agreement with observations. However, at large angular scales, $\ell \lesssim 30$, the predictions are quite different. Specifically, in LQC the power is suppressed, and the LQC prediction provides a better fit to the PLANCK data for the entire range of observable modes (with $\Delta \chi^2 = 3.15$). 

The difference from standard inflation can be traced back largely to the specific features of dynamics of the scalar mode $\h\Q$ \emph{in the Planck regime}, which are sensitive to the values of the LQC radius of curvature $r_{\rm curv}$ near the bounce. In  GR, curvature diverges at the big bang while in LQC it remains finite at the bounce, whence $r_{\rm curv}$ is finite and non-zero there. It provides a new scale. Power is suppressed only in those modes whose physical wavelength is comparable to, or exceeds, $r_{\rm curv}$ in the Planck regime. And these are precisely the longest wave length modes that appear at the largest angular scale in the sky. Principle 1 determines the scale that distinguishes the two sets, with the boundary around $\ell \sim 30$. Thus there is a fascinating interplay between the UV properties of the background quantum geometry that determines $r_{\rm curv}$ in the Planck regime, and the IR properties of quantum perturbations at large angular scales $\ell \lesssim 30$.%
\footnote{This interplay was first noted in \cite{aan1,aan3}. It is brought to a sharper focus in the present discussion because of the specific $\t{g}_{ab}, \psi$ selected by Principles 1 and 2 respectively. Finally, note that some care is needed in comparing the results presented here and those of \cite{aan1,aan3} because in that  work the scale factor is set to one at the bounce, $a_{\rm B}=1$, while in this paper we have followed the standard cosmology convention and set the scale factor today to be one, $a_{0}=1$.}
What is even more surprising is that for the two potentials we analyzed in detail, the Planck regime lasts only $\sim 1.4$ \efolds (see Table \ref{tab:timeline}). Yet this short phase is sufficient to make an observable difference in the power spectrum at the end of inflation! Finally, the scale $\ell \approx 30$ is set almost directly by the fact that Principle 1 restricts the number $N_{\rm B \,-\, \star}$ of pre-inflationary \efolds to a  small value ($N_{\rm B \,-\, \star} \approx 16.83$ for the Starobinsky potential and $N_{\rm B \,-\, \star} \approx 15.00$ for the quadratic). In particular, it rules out the possibility of eternal inflation in LQC.

Other mechanisms have been proposed to account for power suppression at large angular scales. We already mentioned the possibility of using the integrated Sachs-Wolf effect \cite{isw1,isw2} with appropriate fine tuning of the late-time dynamics of the universe. By contrast, our mechanism is primordial and leads to power suppression also in the EE power spectrum. Other primordial mechanisms have also been investigated within standard inflation. In particular it was suggested in \cite{ContaldiFastRoll} that the addition of a transient fast-roll phase
preceding the slow-roll would lead to power suppression and the idea was further developed in \cite{ClineFastRoll,JainFastRoll,
PedroFastRoll,LelloFastRoll,cai3}, e.g., by modifying the inflaton potential, or by simply postulating an appropriate fast-roll phase. Finally, power suppression at large angular scales has been studied in the context of other bouncing models where the bounce is either assumed \cite{cai1}, or obtained by adding a scalar field with a negative kinetic term which violates standard energy conditions \cite{cai2}. In all these approaches the resulting power spectrum usually contains additional parameters which are then fixed using the data. Our approach is different in that: (i) the bounce is a \emph{prediction} of LQC; (ii) since the mechanism has its roots in quantum geometry effects underlying LQG, additional scalar fields or violations of energy conditions are not involved; (iii) the standard inflationary potentials are used without adjustments to provide a fast roll phase; (iv) the two Principles lead us to unique power spectrum; there are no additional parameters; 
and, most importantly, (v) our goal is to investigate whether the CMB observations can inform quantum gravity and vice versa.

Since the detailed results presented in section \ref{s4} on power spectra make a crucial use of the two principles, we will now discuss them briefly. Both principles serve to bridge some fundamental tenets of the Planck scale physics with observed facts about the late time universe. The motivation behind them can be summarized as follows. Observations (together with GR) have determined the geometry of the universe quite accurately to the future of the CMB time $t_{\rm CMB}$ (see section \ref{s2.1}). Since CMB or other LSS observations will not bring us direct information about the structure of the universe at earlier epochs, it is natural to invoke theoretical ideas to construct the history of the earlier epoch. Given an inflationary model, we can indeed extend the history all the way back to $t=t_{\star}$ when the pivot mode $k_{\star}$ exited the horizon during inflation (see section \ref{s3}). To extend the history further back in time is difficult because of two reasons. First, because inflationary trajectories are attractors, there is a very large class of LQC histories to the past of $t=t_{\star}$, all of which enter the small neighborhood of the initial data at $t=t_{\star}$ picked out by observations. Therefore, one needs additional input to restrict this phase of evolution to a small class of preferred histories. The second non-triviality is that as we move to the past of $t_{\star}$ we quickly reach the Planck epoch. Therefore the additional conditions have to be specified in the quantum gravity regime. Principles proposed in section \ref{s3} address both these difficult issues and provide us, as desired, with a very restricted class of extensions from $t_{\star}$ to the bounce time. In section \ref{s4}, we showed that observations provide considerable support for the initial conditions thus selected. 

However, precisely because the principles go back and forth between the Planck regime and late time, they mix fundamental quantum concepts (properties of the area operator in quantum geometry and Heisenberg uncertainties in the Planck regime) and semi-classical ideas (the dressed effective metric $\t{g}_{ab}$ and emergence of classical behavior at the end of inflation). Also Principle 1 makes crucial use of the CMB surface because the universe to its past is not directly observable. Indeed, even the proposed missions to detect primordial gravitational waves are based on the B-mode polarization of CMB photons. But from a fundamental perspective of quantum gravity the emphasis on the CMB surface seems unsatisfactory. Our view can be illustrated by the example of the Bohr model of the hydrogen atom. Bohr retained the classical `planetary model' of an electron in orbit around the proton but added an ad-hoc assumption that angular momentum is quantized. But this unnatural strategy was vindicated by the fact that it led to energy levels that were in agreement with observed of spectral lines of hydrogen. The final quantum mechanical treatments of the hydrogen atom by Pauli and Schr\"odinger are vastly different from Bohr's. Yet, the Bohr model was very helpful because it captured a kernel of truth. In the same vein, we hope that the principles introduced here will serve as helpful guidelines, although the final formulation of initial conditions in quantum gravity is likely to be quite different.

Our analysis suggests several directions for future work. We will conclude by discussing three examples.

$\bullet\,\,$ \emph{Robustness:} Our investigation is confined to the inflationary scenario and is furthermore restricted to just two inflation potentials. The first open issue is whether the two principles and LQC dynamics continue to yield a better fit to the PLANCK data for other potentials that are observationally viable in the standard scenario based on the BD vacuum.   
%
%
Next, LQC has also been used in non-inflationary contexts (see, e.g., \cite{csv,cai-wilson}). It would be interesting to investigate whether the two principles can be modified appropriately in these contexts to narrow down initial conditions and, if so, what the observational predictions are.

$\bullet\,\,$ \emph{UV issues:} In this paper we worked entirely within the LQC framework. However, for a deeper understanding of quantum gravity issues, it would be interesting to investigate more carefully the implications of the natural UV cut-offs of full LQG on quantum perturbations $\h\Q$ and $\h\T^{I}$. Presumably these imply that the physical wavelengths of modes cannot be arbitrarily small. In the scenarios considered in this paper, these cut-offs will refer to modes at angular scales which are too small to be cosmologically significant, i.e. at scales where local astrophysics will dominate (see Remark 3 in section \ref{s3.1}). But the issue of whether the modes with too small a wavelength are simply absent because of LQG quantum geometry is conceptually important. It would also be practically important if one were to remove Principle 1 and allow a very large number of pre-inflationary \efolds $N_{\rm B\, -\, \star}$.

$\bullet\,\,$ \emph{Arrow of time:} The background solutions selected by Principle 1 are highly time asymmetric. For example, in the solution (with initial conditions (\ref{eq:iniS})) discussed in detail in this paper, there are very few \efolds during deflation before the bounce. As a result, detailed considerations show that if we evolve the perturbations \emph{in the past direction} starting from the bounce, they would not lead to structure formation. Thus, in this solution   the pre-bounce phase contains only the background with weak perturbations; structure formation and interesting dynamics occurs only in the post-bounce phase. Heuristically, then, it would seem that the universe is in a low entropy state throughout the contracting phase. This scenario fits well with our overall strategy since the motivation behind the two principles was to introduce very special initial conditions, as in Penrose's Weyl curvature hypothesis \cite{rp-weyl}. It would be very interesting to develop these ideas further in greater detail.

\vskip0.04cm
\acknowledgements{We would like to thank Nishant Agarwal, Ivan Agullo, Fernando Barbero, Eugenio Bianchi, B\'{e}atrice Bonga, Robert Brandenberger, Yi-Fu Cai, Aruna Kesavan, Luis Lehner, Yuexing Li, Andrew Liddle, Peter Saulson, Shiv Sethi, William Unruh, Madhavan Varadarajan, and Robert Wald for discussions and correspondence. This work was supported in part by the NSF grant PHY-1505411, the Eberly research funds of Penn State and Perimeter Institute for Theoretical Physics. Research at Perimeter Institute is supported by Government of Canada through the Department of Innovation, Science and Economic Development and by the Province of Ontario through the Ministry of Research, Innovation and Science. This research used the Extreme Science and Engineering Discovery Environment (XSEDE), which is supported by National Science Foundation grant number ACI-1053575.}

\end{document}